\documentclass[11pt]{article}
\usepackage[pass,letterpaper,margin=1in]{geometry}
\usepackage{fullpage}
\usepackage{amsfonts}
\usepackage{amsmath}
\usepackage{amssymb}
\usepackage{fixltx2e}
\usepackage{mathrsfs}
\usepackage{marvosym}
\usepackage{nicefrac}
\usepackage{array,multirow}
\usepackage{color}
\usepackage{authblk}
\usepackage{ifpdf}
\usepackage{tikz}
\usetikzlibrary{arrows,automata,shapes,decorations,calc,matrix,decorations.pathmorphing}
\ifpdf
 \usepackage[pdftex]{hyperref}
\else
 \newcommand{\href}[2]{#2}
 \newcommand{\texorpdfstring}[2]{#1}
\fi
\usepackage{amsthm} % moved here to remove some of the cleveref-introduced warnings
\usepackage[capitalise]{cleveref}

\definecolor{light-gray}{gray}{0.75}

\theoremstyle{plain}
\newtheorem{theorem}{Theorem}%[section]
\newtheorem{lemma}[theorem]{Lemma}
\newtheorem{proposition}[theorem]{Proposition}
\newtheorem{corollary}[theorem]{Corollary}
\newtheorem{claim}[theorem]{Claim}
\newtheorem{observation}{Observation}

\theoremstyle{definition}

\newtheorem{assumption}{Assumption}

\theoremstyle{remark}
\newtheorem*{remark}{Remark}

\newcommand{\transpose}{\ensuremath{\mathsf{T}}}
\DeclareMathOperator{\val}{v}
\DeclareMathOperator{\sgn}{sgn}

\newcommand{\NN}{\ensuremath{\mathbb{N}}}
\newcommand{\ZZ}{\ensuremath{\mathbb{Z}}}

\newcommand{\RR}{\ensuremath{\mathbb{R}}}

\newcommand{\abs}[1]{\ensuremath{\mathopen\lvert #1 \mathclose\rvert}}
\newcommand{\norm}[1]{\ensuremath{\mathopen\lVert #1 \mathclose\rVert}}
\newcommand{\Abs}[1]{\ensuremath{\left| #1 \right|}}

\newcommand{\eps}{\varepsilon}
\newcommand{\Dist}{\ensuremath{\Delta}}
\newcommand{\Exp}{\ensuremath{{\mathrm E}}}

\newcommand{\margval}[2]{\ensuremath{\frac{\partial\val(#1)}{\partial #2}}}
\newcommand{\absorb}[1]{\ensuremath{ #1{\mbox{*}}}}
\newcommand{\nonabsorb}[1]{\ensuremath{ #1\phantom{\mbox{*}}}}
\newcommand{\lef}{\ensuremath{\text{\bfseries L}}}
\newcommand{\rig}{\ensuremath{\text{\bfseries R}}}
\newcommand{\payoff}{\ensuremath{\pi}}
\newcommand{\stopprob}{\ensuremath{\omega}}
\newcommand{\derivedgame}[1]{\ensuremath{\widetilde{#1}}}

\DeclareMathOperator{\dens}{dens}
\DeclareMathOperator{\gdens}{gdens}
\newcommand{\calM}{\ensuremath{\mathcal{M}}}

\newcommand{\uinf}{\ensuremath{u_{\operatorname{inf}}}}
\newcommand{\usup}{\ensuremath{u_{\operatorname{sup}}}}

\newcommand{\lowvinf}{\ensuremath{\underline{\val}_{\operatorname{inf}}}}
\newcommand{\lowvsup}{\ensuremath{\underline{\val}_{\operatorname{sup}}}}
\newcommand{\uppvinf}{\ensuremath{\overline{\val}_{\operatorname{inf}}}}
\newcommand{\uppvsup}{\ensuremath{\overline{\val}_{\operatorname{sup}}}}

\newcommand{\ov}{\overline}
\def\ploss{p_{\mathrm{loss}}}
\def\pwin{p_{\mathrm{win}}}
\def\lft{\lef}
\def\rgt{\rig}
\newcommand{\tail}{\ensuremath{\text{tail}}}
\crefname{figure}{Figure}{Figure}
\crefname{equation}{Equation}{Equation}

\title{The Big Match in Small Space}
\author[1]{\mbox{Kristoffer Arnsfelt Hansen}\thanks{The author acknowledges support from the Danish National Research Foundation and The National Science Foundation of China (under the grant 61361136003) for the Sino-Danish Center for the Theory of Interactive Computation and from the Center for Research in Foundations of Electronic Markets (CFEM), supported by the Danish Strategic Research Council.}}
\author[2]{\mbox{Rasmus Ibsen-Jensen}}
\author[3]{\mbox{Michal Kouck\'y}\thanks{The research leading to these results has received funding from the European Research Council under the European Union's Seventh Framework Programme (FP/2007-2013) / ERC Grant Agreement n. 616787. Supported in part by grant from Neuron Fund for Support of Science, and by the Center of Excellence CE-ITI under the grant P202/12/G061 of GA \v{C}R.}}
\affil[1]{Aarhus University}
\affil[2]{IST Austria}
\affil[3]{Charles University, Prague}

\begin{document}
\maketitle
\begin{abstract}
  In this paper we study how to play (stochastic) games optimally
  using little space.  We focus on repeated games with absorbing
  states, a type of two-player, zero-sum concurrent mean-payoff games.
  The prototypical example of these games is the well known Big Match
  of Gillete (1957). These games may not allow {\em optimal}
  strategies but they always have $\eps$-optimal strategies. In this
  paper we design $\eps$-optimal strategies for Player 1 in these
  games that use only $O(\log\log T)$ space. Furthermore, we construct strategies for
  Player 1 that use space $s(T)$, for an arbitrary small unbounded
  non-decreasing function $s$, and which guarantee an $\eps$-optimal
  value for Player 1 in the limit superior sense. The previously known
  strategies use space $\Omega(\log T)$ and it was known that no strategy
  can use constant space if it is $\eps$-optimal even in the limit superior sense. 
  We also give a complementary lower bound.
  
 Furthermore, we also show that no Markov strategy, even extended with finite memory, can ensure value greater than~0 in the Big Match, answering a question posed by Abraham Neyman.
%
%
%  We consider two-player, zero-sum concurrent mean-payoff games. We
%  focus on the well-known game the Big Match, in which Player~1 can
%  choose go or stop and Player~2 safe and unsafe. If Player~1 ever
%  plays stop, then Player~1 wins~1 iff Player~2 played unsafe in that
%  round, otherwise Player~1 gets the limit-average of the number of
%  times Player~2 played safe (either lim-sup or lim-inf depending on
%  variant). This game has value~$\frac{1}{2}$. It has been shown that
%  no finite-memory $\eps$-optimal strategy for Player~1 exists for
%  $\eps<\frac{1}{2}$ and either variant. We show that this is
%  optimal for the lim-sup variant, by showing that for all unbounded
%  functions $f$, there is a strategy which uses at most $f(T)$ space
%  upto round $T$ with arbitrarily high probability for all
%  sufficiently large $T$'s. We also give a strategy that uses at most
%  $O(\log \log T)$ space upto round $T$ with arbitrarily high
%  probability for all sufficiently large $T$'s, in the lim-inf
%  variant.  Afterwards we show how to extend our results to all
%  two-player, zero-sum concurrent mean-payoff games, in which all but
%  one state are absorbing.
\end{abstract}

\newpage
\tableofcontents 
\newpage

\section{Introduction}

In game theory there has been considerable interest in studying the
complexity of strategies in infinitely repeated games. A 
natural way how to measure the complexity of a strategy is by the number of
states of a finite automaton implementing the strategy. A common theme
is to consider what happens when some or all players are restricted to
play using a strategy given by an automaton of a certain bounded
complexity. 
% In this work we shall be concerned only with two-player
% zero-sum repeated games. In such settings a natural question is to
% understand the complexity of optimal or near-optimal play. 
% Here

\smallskip\noindent{\bf Asymptotic view.}
Previous works have mostly been limited to dichotomy results: either
there is a good strategy implementable by finite automaton or there is
no such strategy. Our goal here is to refine this picture. We do this
by taking the asymptotic view: measuring the complexity as a function
of the number of rounds played in the game. Now when the strategy no
longer depends just on a finite amount of information about the
history of the game it could even be a computationally difficult
problem to decide the next move of the strategy. But we focus on
investigating how much information a good strategy must store about
the play so far to decide on the next move; in other words, we study
how much space the strategy needs.

\smallskip\noindent{\bf Game classes.}
The class of games we study is that of repeated zero-sum games with
absorbing states. These form a special case of undiscounted stochastic
games.  Stochastic games were introduced by
Shapley~\cite{PNAS:Shapley53}, and they constitute a very general model of
games proceeding in rounds. We consider the basic version of
two-player zero-sum stochastic games with a constant number of states
and a constant number of actions. In a given round $t$ the two players
simultaneously choose among a number of different \emph{actions}
depending on the current \emph{state}. Based on the choice of the pair
$(i,j)$ of actions as well as the current state $k$, Player~1 receives
a \emph{reward} $r_t=a_{ij}^k$ from Player~2, and the game proceeds to
the next state $\ell$ according to probabilities $p_{ij}^{k\ell}$.

\smallskip\noindent{\bf Limit-average rewards.}
In Shapley's model, in every round the game stops with non-zero
probability, and the payoff assigned to Player~1 by a play is simply
the sum of rewards $r_i$. The stopping might be viewed as
\emph{discounting} later rewards by a discounting factor
$0<\beta<1$. Gillette~\cite{AMS:Gillette1957} considered the more
general model of undiscounted stochastic games where all plays are
infinite. He is interested in the average reward $\frac{1}{T}
\sum_{t=1}^T r_t$ to Player~1 as $T$ tends to infinity. As the limit
may not exist one needs to consider $\liminf$, $\limsup$, or some
Banach limit~\cite{Sorin-chapter92} of the sums. In many cases the
particular choice of the limit does not matter much, but it turns out that
for our results it has interesting consequences. For this reason we
consider both $\liminf_{T \rightarrow \infty} \frac{1}{T} \sum_{t=1}^T
r_t$ and $\limsup_{T \rightarrow \infty} \frac{1}{T} \sum_{t=1}^T
r_t$. 

Note that both these notions have natural interpretations. For
instance, the $\liminf$ notion suits the setup where the infinite
repeated game actually models a game played repeatedly for an
unspecified (but large) number of rounds, where one thus desires a
guarantee on the average reward after a certain number of rounds.  The
$\limsup$ notion on the other hand models the ability to always
recover from arbitrary losing streaks in the repeated game.

\smallskip\noindent{\bf The Big Match.}  A prototypical example of an
undiscounted stochastic game is the well-known Big Match of
Gillette~\cite{AMS:Gillette1957} (see \cref{fig:big-match2} for an
illustration of the Big Match). This game fits also into an important
special subclass of undiscounted stochastic games: the \emph{repeated
  games with absorbing states}, defined by
Kohlberg~\cite{AS:Kohlberg1974}.  In a repeated game with absorbing
states there is only one state that can be left; all the other states
are \emph{absorbing}, i.e., the probability of leaving them is zero
regardless of the actions of the players. Even in these games, as for
general undiscounted stochastic games, there might not be an optimal
strategy for the players~\cite{AMS:Gillette1957}. On the other hand
there always exist $\eps$-optimal strategies~\cite{AS:Kohlberg1974},
which are strategies guaranteeing the value of the game up to an
additive term $\eps$. The Big Match provides such an example: the
value of the game is $1/2$, but Player~1 does not have an optimal
strategy, and must settle for an $\eps$-optimal
strategy~\cite{AMS:BlackwellFerguson1968}.  On the other hand, it is
known that such $\eps$-optimal strategies in the Big Match must have a
certain level of complexity. More precisely, for any
$\eps<\frac{1}{2}$, an $\eps$-optimal strategy can neither be
implemented by a finite automaton nor take the form of a Markov
strategy (a strategy whose only dependence on the history is the
number of rounds played) \cite{Sorin-FirstCourse2002}.

In this paper we consider the Big Match in particular
and then generalize our results to general repeated games with
absorbing states.

\smallskip\noindent {\bf The model under consideration.}  We are
interested in the \emph{space complexity} of $\eps$-optimal strategies
in repeated games with absorbing states. A general strategy of a
player in a game might depend on the whole history of the play up to
the current time step.  Moreover the decision about the next move
might depend arbitrarily on the history.  This provides the strategies
with lots of power. There are two natural ways how to restrict the
strategies: one can put computational restrictions on how the next
move is decided based on the history of the play, or one can put a
limit on how much information can the strategy remember about the
history. One can also combine both types of restrictions, which leads
to an interactive Turing machine based model, modelling a dynamic
algorithm.

In this paper we mainly focus on restricting the amount of information
the strategy can remember. This restriction is usually studied in the
form of how \emph{large size a finite automaton} (transducer) for the
strategy has to be, and we follow this convention. By the size of a
finite automaton we mean the number of states. The automatons we
consider can make use of probabilistic transitions, and we will not consider
the describtion of these probabilities as part of the size of the
automaton. We do address these separately, however.

\smallskip\noindent{\bf History of the model.}
The idea of measuring complexity of strategies in repeated games in
terms of automata was proposed by Aumann~\cite{AumannSurvey1981}. The
survey by Kalai~\cite{KalaiSurvey1990} further discuss the idea in
several settings of repeated games. However in this line of research
the finite automata is assumed to be fixed for the duration of the
game. This represents a considerable restriction as for many games
there is no good strategy that could be described in this
setting. Hence we consider strategies in which the automata can grow
with time. To be more precise we consider infinite automata and
measure how many different states we could have visited during the first
$T$ steps of the play. The logarithm of this number corresponds to
the amount of space one would need to keep track of the current state
of the automaton. We are interested in how this space grows with the
number of rounds of the play.

\smallskip\noindent{\bf Comparison of our model with a Turing machine
  based model.}  To impose also computational restrictions on the
model, one can consider the usual Turing machine with one-way input
and output tapes that work in lock-step and that record the play:
whenever the machine writes its next action on the output tape it
advances the input head to see the corresponding move of the other
player. The space usage of the model is then the work space used by
the machine, growing with the number of actions processed. The Turing
machine can be randomized to allow for randomized strategies. The main
differences between this model and the automaton based model we
focus on in this paper is that in the case of infinite automata the
strategy can be {\em non-uniform} and use {\em arbitrary}
probabilities on its transitions whereas the Turing machine is {\em
  uniform} in the sense that it has a finite program that is fixed for
the duration play and in particular, all transition probabilities are
explicitly generated by the machine.

\smallskip\noindent{\bf Bounds for strategies with deterministic update.}
Trivially, any strategy needs space at most $O(T)$, since such memory
would suffice to remember the whole history of the play. It is not hard to see
(cf. \cite[Chap. 3.2.1]{RasmusThesis}) that if a strategy is not
restricted to a finite number of states, then the number of reachable
states by round $T$ must be at least $T$.  This means that the space
needed by any such strategy is $\Omega(\log T)$.  However this
provides only worst-case answer to our question, since for randomized
strategies it might happen that only negligible fraction of the states
can be reached with reasonable probability. Indeed, it might be that
with probability close to 1 the strategy reaches only a very limited
number of states. This is the setup we are interested in.  As we will
see in a moment the strategies we consider use substantially less
space than $O(\log T)$ with high probability (and $O(\log T)$
space in the worst case). 

% For similarity with standard memory usage on a computer we say that
% a strategy that uses $k$ distinct memory states uses $\log k$ space.
\smallskip\noindent{\bf Relationship to data streaming}
We find that our question is naturally related to algorithmic
questions in data streaming. In data streaming one tries to estimate
on-line various properties of a data stream while minimizing the
amount of information stored about the stream. As we will see our
solutions borrow ideas from data streaming in particular, we use
sampling to estimate properties of the play so far. It is rather
interesting that this is sufficient for a large class of games.

%% COMMENTED: This seems too technical for the introduction?
% The relation is also going in the other direction: That is, all known
% $\eps$-optimal strategies for the Big Match can be viewed as
% algorithms solving the following datastreaming problem, where one
% tries to predict numbers in arbitrary bit-streams: The wanted
% algorithm considers a infinite bit-stream and has a operator \predict,
% which, if invoked at position $T$ in the stream, must ensure that the
% bit $T+1$ is 0 with probability at least slightly below
% $\frac{1}{2}$. Also, if the limit-average (either limit supremum or
% limit infimum) is strictly below $\frac{1}{2}$ then \predict must be
% invoked at some point with probability~1.  Our strategy for
% limit-infimum can be viewed as such an algorithm using $O(\log \log
% T)$ space at step $T$ with probability nearly~1. Also, our strategy
% for limit-supremum can be viewed as such an algorithm using $O(f(T))$
% space at step $T$ with probability nearly~1, for any given unbounded
% function $f$.

\subsection{Our results}
%\smallskip\noindent {\bf Our results.} 
We provide two types of results. We show
that there are $\eps$-optimal strategies for repeated games with
absorbing states, and we also show that there are limits on how small
space such strategies could possibly use. Our strategies are first
constructed for the Big Match. Then, following
Kohlberg~\cite{AS:Kohlberg1974} these strategies are extended to
general repeated games with absorbing states.

\smallskip\noindent {\bf Upper bounds on space usage.}  Our first results concern
the Big Match. We show that for all $\eps>0$, there exists an
$\eps$-optimal strategy that uses $O(\log \log T)$ space with
probability~$1-\delta$ for any $\delta>0$. We note that the previous
constructed strategies of Blackwell and
Ferguson~\cite{AMS:BlackwellFerguson1968} and
Kohlberg~\cite{AS:Kohlberg1974} uses space $\Theta(\log T)$.

\begin{theorem}
  For all $\eps>0$, there is an $\eps$-optimal strategy $\sigma_1$ for
  Player~1 in the Big Match such that for any $\delta>0$ with
  probability at least $1-\delta$, the strategy $\sigma_1$ uses
  $O(\log \log T)$ space in round $T$.
\end{theorem}

\begin{remark}
  We would like to stress the order of quantification above and their
  impact on the big-O notation used above for conciseness. The
  strategy we build depends on the choice of $\eps$, but only for the
  actions made -- the memory updates are independent thereof, and thus
  likewise is the space usage. The dependence of $\delta$ is also very
  benign. More precisely, there exists a constant $C>0$ independent
  of $\eps$ and $\delta$, and an integer $T_0$ depending on $\delta$,
 but independent of $\eps$, in such a way that with probability at
  least $1-\delta$, the strategy $\sigma_1$ uses at most space $C
  \log\log T$, for all $T \geq T_0$. The same remark holds elsewhere
  in our statements.
\end{remark}

\smallskip\noindent{\bf Our results translated to the Turing based model.}
After a slight modification our $\eps$-optimal strategy can be
implemented by a Turing machine so that (1)~it processes $T$ actions
in time $O(T)$; and (2)~each time it processes an action, all
randomness used comes from at most 1 unbiased coin flip; and (3)~it,
for all $\delta>0$, uses $O(\log \log T+\log \log \eps^{-1})$ space
with probability $1-\delta$, before round $T$. See
Corollary~\ref{cor:tm}.

\smallskip\noindent{\bf Arbitrary small, but growing space for $\limsup$.}
For the case of $\limsup$ evaluation of the average rewards we can
design strategies that uses even less space, in fact arbitrarily
small, but growing, space.

\begin{theorem}
  For any non-decreasing \emph{unbounded} function $s$, there exists
  an $\eps$-supremum-optimal strategy $\sigma_1$ for Player~1 in the
  Big Match such that for each $\delta>0$, with probability at least
  $1-\delta$,  strategy $\sigma_1$ uses $O(s(T))$ space in round
  $T$.
\end{theorem}
We may for instance think of $s$ as the inverse of the Ackermann
function. Although the strategy from this theorem has very uniform
description it might not always be implementable by a Turing machine
running in the same space since the machine needs to sample
probabilistic events comparable to $1/T$ or smaller. That might not be
achievable in small space using just a fair coin.

Our strategy that is $\eps$-optimal is actually an instantiation of
the $\eps$-supremum-optimal strategy to the setting of $O(\log\log T)$
space. We are unable to achieve $\eps$-optimality in less space, and
this seems to be inherent to our techniques.

\smallskip\noindent{\bf Generalization to repeated games with absorbing states.}
We can generalize the above statements to the case of general repeated
games with absorbing states.
\begin{theorem}For all $\eps>0$ and any repeated game with absorbing
  states $G$, there is an $\eps$-optimal strategy $\sigma_k$ for
  Player~$k$ in $G$ such that, for each $\delta>0$, with probability
  at least $1-\delta$, the strategy $\sigma_k$ uses $O(\log \log T +
  \log 1/\eps \cdot \mathrm{poly}(\abs{G}))$ space in round~$T$.
\end{theorem}
\begin{theorem}
  For all $\eps>0$, any repeated game with absorbing states $G$, and
  any non-decreasing unbounded function $s$, there exists an
  $\eps$-supremum optimal strategy $\sigma_k$ for Player~$k$ in $G$
  such that for each $\delta>0$, with probability at least $1-\delta$,
  the strategy $\sigma_1$ uses $O(s(T) + \log 1/\eps \cdot
  \mathrm{poly}(\abs{G}))$ space in round $T$.
\end{theorem}
These strategies are obtained by reducing to a special simple case of
repeated games with absorbing states, generalized Big Match games, to
which our Big Match strategies can be generalized. This reduction can
furthermore be done effectively by a polynomial time algorithm.

\smallskip\noindent {\bf Lower bound on space usage.}  We provide two lower bounds
on space addressing different aspects of our strategies. One property
of our strategies is that the smaller the space used is, the smaller
the probabilities of actions employed are. The reciprocal of the
smallest non-zero probability is the {\em patience} of a
strategy. This is a parameter of interest for strategies. We show that
the patience of our strategies is close to optimal. In particular, we
show that the first $f(T)$ memory states must use probabilities close
to $1/T^{f(T)}$, where $s(T)=\log f(T)$ is the space usage. We can
almost match this bound by our strategies.

\smallskip\noindent{\bf Finite-memory deterministic-update Markov strategies are no good.}
Beside the lower bound on patience we investigate the possibility of
using a good strategy for Player~1 which would use only a constant
number of states but where the actions could also depend on the round
number. This is what we call a {\em finite-memory Markov strategy}. We
show that such a strategy which also updates its memory state
deterministically cannot exist. This answers a question posed by Abraham Neyman.

\begin{theorem}
  For all $\eps<\frac{1}{2}$, there exists no finite-memory
  deterministic-update $\eps$-optimal Markov strategy for Player~1 in
  the Big Match.
\end{theorem}

\subsection{Our techniques}
%\smallskip\noindent {\bf Our techniques.}
The previously given strategies for Player~1 in the Big Match
\cite{AMS:BlackwellFerguson1968,AS:Kohlberg1974} use space
$\Theta(\log T)$ as they maintain the count of the number of different
actions taken by the other player.  There are two principal ways how
one could try to decrease the number of states for such randomized
strategies: either to use approximate counters
\cite{CACM:Morris78a,BIT:Flajolet85}, or to sub-sample the stream of
actions of the other player and use a good strategy on the sparse
sample. In this paper we use the latter approach.

\smallskip\noindent{\bf Overview over our strategy for the Big Match.}
Our strategies for Player~1 proceed by observing the actions of
Player~2 and collecting statistics on the payoff. Based on these
statistics Player~1 adjusts his actions. The statistics is collected
at random sample points and Player~1 plays according to a ``safe''
strategy on the points not sampled and plays according to a good (but
space-inefficient) strategy on the sample points. If the space of
Player~1 is at least $\log \log T$ then Player~1 is able to collect
sufficient statistics to accurately estimate properties of the actions
of Player~2. Namely, substantial dips in the average reward given to
Player~1 can be detected with high probability and Player~1 can react
accordingly. Thus that during infinite play, the average reward
will not be able drop for extended periods of time, and this will
guarantee that $\liminf$ evaluation of the average rewards is close to
the value of the game.

\smallskip\noindent{\bf The bottle-neck in the $\liminf$ case.}
However, if our space is considerably less than $\log \log T$ we do
not know how to accurately estimate these properties of the actions of
Player~2. Thus, long stretches of actions of Player~2 giving low
average rewards might go undetected as long as they are accompanied by
stretches of high average rewards.  Thus one could design a strategy
for Player~2 that has low $\liminf$ value of the average rewards, but
has large $\limsup$ value. Against such a strategy, our
space-efficient strategy for Player~1 is unlikely to stop. So during
infinite play, while our strategy guarantees that the $\limsup$
evaluation of the average rewards is close to the value of the game,
it performs poorly under $\liminf$ evaluation. It is not clear whether
this is an intrinsic property of all very small space strategies for
Player~1 or whether one could design a very small space strategy
achieving that the $\liminf$ evaluation of the average rewards is
close to the value of the game. We leave this as an interesting open
question.

\smallskip\noindent{\bf Generalizing to repeated games with absorbing states.}
Our extension to general repeated games with absorbing states follow
closely the work of Kohlberg~\cite{AS:Kohlberg1974}. He showed that
all such games have a value and constructed $\eps$-optimal strategies
for them, building on the work of Blackwell and
Ferguson~\cite{AMS:BlackwellFerguson1968}. His construction is in two
steps: The question of value and of $\eps$-optimal strategies are
solved for a special case of repeated games with absorbing states,
generalized Big Match games, that are sufficiently similar to the Big
Match game that one of the strategies given by Blackwell and
Ferguson \cite{AMS:BlackwellFerguson1968} can be extended to this more
general class of games. Having done this, Kohlberg shows how to reduce
general repeated games with absorbing states to generalized Big Match
games.

In a similar way we can extend our small-space strategies for the Big
Match to the larger class of generalized Big Match games. These can
then directly be used for Kohlberg's reduction. This reduction is
however only given as an existence statement. We show how the
reduction can be made explicit and computed by a polynomial time
algorithm. This is done using linear programming formulations and
fundamental root bounds of univariate polynomials.  This also provides
explicit bounds on the bitsize of the reduced generalized Big Match
games. We also give a simple polynomial time algorithm for
approximating the value of any repeated game with absorbing states
based on bisection and linear programming.

\section{Definitions}
\label{sec:definitions}
\paragraph{Probability distributions.} 
A \emph{probability distribution} over a finite set $S$, is a map
$d:S\rightarrow [0,1]$, such that $\sum_{s\in S}d(s)=1$. Let
$\Dist(S)$ denote the set of all probability distributions over $S$.

\paragraph{Repeated games with absorbing states.} 
The games we consider are special cases of two player, zero-sum
concurrent mean-payoff games in which all states except at most one are
\emph{absorbing}, i.e.\ never left if entered (note also that an
absorbing state can be assumed to have just a single action for each
player). We restrict our definitions to this special case, introduced
by Kohlberg~\cite{AS:Kohlberg1974} as repeated games with absorbing
states. Such a game $G$ is given by sets of actions $A_1$ and $A_2$
for each player together with maps $\payoff : A_1\times A_2 \rightarrow
\RR$ (the stage payoffs) and $\stopprob: A_1 \times A_2 \rightarrow [0,1]$
(the absorption probabilities).

The game $G$ is played in rounds. In every round $T=1,2,3,\dots$, each
player $k\in\{1,2\}$ independently picks an action $a_k^T \in
A_k$. Player~1 then receives the stage payoff $\payoff(a_1^T,a_2^T)$
from Player~2. Then, with probability $\stopprob(a_1^T,a_2^T)$ the game
stops and all payoffs of future rounds are fixed to be
$\stopprob(a_1^T,a_2^T)$ (we may think of this as the game proceeding to an
absorbing state where the (unique) stage payoff for future rounds is
$\payoff(a_1^T,a_2^T)$). Otherwise, the game just proceeds to the next
round.

The sequence $(a_1^1,a_2^1),(a_1^2,a_2^2),(a_1^3,a_2^3),\dots$ of
actions taken by the two players is called a \emph{play}. A finite
play occurs when the game stops after the last pair of
actions. Otherwise the play is infinite.  To a given play $P$ we
associate an infinite sequence of rewards $(r_T)_{T\geq 1}$ received
by Player~1. If
$P=(a_1^1,a_2^1),(a_1^2,a_2^2),\dots,(a_1^\ell,a_2^\ell)$ is a finite
play of length $\ell$ we let $r_T = \payoff(a_1^T,a_2^T)$ for $1\leq T
\leq \ell$, and $r_T = \payoff(a_1^\ell,a_2^\ell)$ for $T>\ell$. In this
case we say that the game stops with \emph{outcome} $r_\ell$.

Otherwise, if $P=(a_1^1,a_2^1),(a_1^2,a_2^2),\dots$ is infinite we
simply let $r_T = \payoff(a_1^T,a_2^T)$ for all $T\geq 1$.

To evaluate the sequence of the rewards we consider both the $\liminf$
and $\limsup$ value of the average reward $\frac{1}{T} \sum_{t=1}^T
r_t$. We thus define the limit-infimum payoff to Player~1 of the play as
\[
\uinf(P) = \liminf_{n \rightarrow \infty} \frac{1}{n} \sum_{T=1}^n r_T \enspace ,
\]
and similarly the limit-supremum payoff to Player~1 of the play as
\[
\usup(P) = \limsup_{n \rightarrow \infty} \frac{1}{n} \sum_{T=1}^n r_T \enspace .
\]

\paragraph{Strategies.} 
A \emph{strategy} for Player~$k$ is a function $\sigma_k : (A_1 \times
A_2)^* \rightarrow \Dist(A_k)$ describing the probability distribution
of the next chosen action after each finite play. We say that
Player~$k$ \emph{follows} a strategy $\sigma_k$ if for every finite
play $P$ of length $T-1$, at round $T$ Player~$k$ picks the next
action according to $\sigma_k(P)$. We say that a strategy $\sigma_k$
is \emph{pure} if for every finite play $P$ the distribution
$\sigma_k(P)$ assigns probability 1 to one of the actions of $A_k$
(i.e.\ the next action is uniquely determined).  Also, we say that a
strategy $\sigma_k$ is a \emph{Markov strategy} if for every $T$ and
every play $P$ of length $T-1$, the distribution $\sigma_k(P)$ does
not depend on the particular actions during the first $T-1$ rounds but
is just a function of $T$. Thus Markov strategy $\sigma_k$ can be
viewed as a map $\ZZ_+ \rightarrow \Dist(A_k)$ or simply a sequence of
distributions over $A_k$.

A \emph{strategy profile} $\sigma$ is a pair of strategies
$(\sigma_1,\sigma_2)$, one for each player. A strategy profile
$\sigma$ defines a probability measure on plays in the natural way. We
define the expected limit-infimum payoff to Player~1 of the strategy
profile $\sigma=(\sigma_1,\sigma_2)$ as $\uinf(\sigma) =
\uinf(\sigma_1,\sigma_2) = \Exp_{P\sim (\sigma_1,\sigma_2)}[\uinf(P)]$ and similarly the expected
limit-supremum payoff to Player~1 of the strategy profile $\sigma$ as
$\usup(\sigma) = \usup(\sigma_1,\sigma_2) = \Exp_{P\sim (\sigma_1,\sigma_2)}[\usup(P)]$.

\paragraph{Values and near-optimal strategies.} We define the \emph{lower values} of $G$ by
$\lowvinf = \sup_{\sigma_1}\inf_{\sigma_2}\uinf(\sigma_1,\sigma_2)$
and $\lowvsup=\sup_{\sigma_1}\inf_{\sigma_2}\usup(\sigma_1,\sigma_2)$,
and we define the \emph{upper values} of $G$ by
$\uppvinf=\inf_{\sigma_2}\sup_{\sigma_1}\uinf(\sigma_1,\sigma_2)$ and
$\uppvsup=\inf_{\sigma_2}\sup_{\sigma_1}\usup(\sigma_1,\sigma_2)$. Clearly
$\lowvinf \leq \lowvsup \leq \uppvsup$ and $\lowvinf \leq \uppvinf
\leq \uppvsup$. Kohlberg showed that all these values coincide and we call
this common number $\val(G)$ the value $v$ of $G$.
\begin{theorem}[Kohlberg, Theorem 2.1]
$\lowvinf = \uppvsup$.
\end{theorem}
Note that this also shows that for the purpose of defining
the value of $G$ the choice of the limit of the
average rewards does not matter. But a given strategy $\sigma_1$ for
Player~1 could be close to guaranteeing the value with respect to
$\limsup$ evaluation of the average rewards, while being far from
doing so with respect to the more restrictive $\liminf$ evaluation. We
shall hence distinguish between these different guarantees.

Let $\eps>0$ and let $\sigma_1$ be a strategy for Player~1. We say
that $\sigma_1$ is \emph{$\eps$-supremum-optimal}, if
\[
\val(G)-\eps\leq \inf_{\sigma_2}\usup(\sigma_1,\sigma_2)
\]
and that $\sigma_1$ is \emph{$\eps$-optimal}, if 
\[
\val(G)-\eps\leq \inf_{\sigma_2}\uinf(\sigma_1,\sigma_2) \enspace .
\]
\begin{observation}
\label{REM:pure-strategies}
  Clearly it is sufficient to take the infimum over just pure
  strategies $\sigma_2$ for Player~2, and hence when showing that a
  particular strategy $\sigma_1$ is $\eps$-supremum-optimal or
  $\eps$-optimal we may restrict our attention to pure strategies
  $\sigma_2$ for Player~2.
\end{observation}

One can naturally make similar definitions for Player~2, where the
roles of $\liminf$ and $\limsup$ would then be interchanged, but we
shall restrict ourselves here to the perspective of Player~1.

If the strategy $\sigma_1$ is $0$-supremum-optimal ($0$-optimal) we
simply say that $\sigma_1$ is supremum-optimal (optimal). The Big
Match gives an example where Player~1 does not have a supremum-optimal
strategy~\cite{AMS:BlackwellFerguson1968}.

\paragraph{Memory and memory-based strategies.}
A \emph{memory configuration} or \emph{state} is simply a natural number. We will often think of memory configurations as
representing discrete objects such as tuples of integers. In such a case we will
always have a specific encoding of these objects in mind.

Let $\calM \subseteq \NN$ be a set of memory states. A
\emph{memory-based strategy} $\sigma_1$ for Player~1 consists of a
\emph{starting state} $m_s\in \calM$ and two maps, the \emph{action
  map} $\sigma_1^a:\calM\rightarrow \Dist(A_1)$ and the \emph{update
  map} $\sigma_1^u:A_1\times A_2\times\calM\rightarrow \Dist(\calM)$.
We say that Player~1 \emph{follows} the memory-based strategy
$\sigma_1$ if in every round $T$ when the game did not stop yet, he
picks his next move $a_1^T$ at random according to $\sigma_1^a(m_T)$,
where the sequence $m_1,m_2,\dots$ is given by letting $m_1=m_s$ and
for $T=1,2,3,\dots$ choosing $m_{T+1}$ at random according to
$\sigma_k^u(a_1^T,a_2^T,m_T)$, where $a_2^T$ is the action chosen by
Player~2 at round $T$. 

The strategies we construct in this paper have the property that their
action maps do not depend on the action $a_1^T$ of Player~1. In these
cases we simplify notation and write just $\sigma_1^u(a_2^T,m^T)$.

Since each finite play can be encoded by a binary string, and thus a natural number, we can view
any strategy $\sigma_k$ for Player~$k$ as a memory-based strategy.
One can find similarly defined types of strategies in the literature,
but typically, the function corresponding to the update function is
deterministic.%, see e.g. \nb{Insert reference}.

\paragraph{Memory sequences and space usage of memory-based strategies.}
Let $\sigma_1$ be a memory-based strategy for Player~$1$ on memory
states $\calM$ and $\sigma_2$ be a strategy for Player~$2$.  Assume
that Player~$1$ follows $\sigma_1$ and Player~$2$ follows
$\sigma_2$. The strategy profile $(\sigma_1,\sigma_2)$ defines a
probability measure on (finite and infinite) sequences over $\calM$ in
the natural way. For a (finite) sequence $M\in \calM^*$, let
$\omega_1(M)$ denote the probability that Player~1 follows this
sequence of memory states during the first $|M|$ rounds of the game,
while the game does not stop before round $|M|$.

Fix a non-decreasing function $f:\NN\rightarrow\NN$ and a probability $p$.
The strategy $\sigma_1$ {\em uses $\log f(T)$ space with probability at 
least $p$ against $\sigma_2$}, if for all $T$, the probability 
$\Pr_{(\sigma_1,\sigma_2)}[\forall i\leq T:M_i\leq  f(T)]\geq p$ (i.e., with probability at least $p$, the current memory has stayed below that of $f(T)$ 
before round $T$, for all $T$). If $\sigma_1$ uses $\log f(T)$ 
space with probability at least $p$ against every strategy 
$\sigma_2'$, then we say that $\sigma_1$ {\em uses $\log f(T)$ 
space with probability at least $p$}.

\paragraph{The Big Match.}
The Big Match, introduced by Gillette~\cite{AMS:Gillette1957} is a
simply defined repeated game with absorbing states, where each player
has only two actions.  In each round Player~1 has the choice to stop
the game (action $\rig$), or continue with the next round (action
$\lef$). Player~2 has the choice to declare the round safe (action
$\lef$) or unsafe (action $\rig$). If play continues in a round
declared safe, or if play stops in a round declared unsafe,
Player~2 must give Player~1 a reward $1$. In the other two cases no
reward is given.

More formally, let the action sets be $A_1=A_2=\{\lef,\rig\}$. The
rewards are given by $\payoff(a_1,a_2)=1$ if $a_1 = a_2$ and
$\payoff(a_1,a_2)=0$ if $a_1 \neq a_2$. The stopping probabilities are
given by $\stopprob(\rig,a_2)=1$ and $\stopprob(\lef,a_2)=0$.

We can illustrate this game succinctly in a matrix form as shown in
Figure~\ref{fig:big-match2}, where rows are indexed by the actions of
Player~1, columns are indexed by the actions of Player~2, entries give
the rewards, and a star on the reward means that the game stops with
probability~1 (See \cref{sec:matrix-form-of-games} for a general
definition of the matrix form of a repeated game with absorbing
states).

\begin{figure}
\centering
\renewcommand{\arraystretch}{1.3}
$
\begin{array}{ r|r|r| }
\multicolumn{1}{r}{}
 &  \multicolumn{1}{c}{\lef}
 & \multicolumn{1}{c}{\rig} \\
\cline{2-3}
\lef & \nonabsorb{1} & \nonabsorb{0} \\
\cline{2-3}
\rig & \absorb{0} & \absorb{1} \\
\cline{2-3}
\end{array}
$
\renewcommand{\arraystretch}{1}
\caption{The Big Match in a matrix form.
\label{fig:big-match2}}
\end{figure}

% \begin{figure}
% \centering
% \renewcommand{\arraystretch}{1.3}
% $\begin{array}{l|l|l|}
% &\text{\lef} & \text{\rig}\\
% \hline
% \text{\lef} &1 & 0 \\
% \hline
% \text{\rig} &\absorb{0} & \absorb{1} \\\hline
% \end{array}
% $
% \renewcommand{\arraystretch}{1}
% \vspace{0.1cm}
% \caption{The matrix illustrates the repeating game with absorbing states, the Big Match.\label{fig:big-match2}}
% \end{figure}

\paragraph{Density of pure Markov Strategies in the Big Match.}
When constructing strategies for Player~1 in the Big Match, not only
is it sufficient to consider only pure strategies for Player~2 as noted
in \cref{REM:pure-strategies}, but we may restrict our consideration
to pure Markov strategies, since Player~2 only ever observes the
action $\lef$ of Player~1. An important property of a pure Markov
strategy $\sigma$ for Player~2 in the Big Match is the \emph{density}
of $\lef$ actions of a prefix of $\sigma$.

Denote by $\sigma^T \in \{\lef,\rig\}^T$, the length $T$ prefix of
$\sigma$. The \emph{density of $\lef$} in $\sigma^T$, denoted
$\dens(\sigma^T)$, is defined by
\[
\dens(\sigma^T)=\frac{\abs{\{i\mid (\sigma^T)_i=\lef\}}}{T} \enspace .
\]
Further, for $T'<T$ we define
\[
\dens(\sigma,T',T)=\frac{\abs{\{i\geq T'\mid (\sigma^T)_i=\lef\}}}{T-T'+1} \enspace .
\]

\begin{observation}
  Suppose Player~2 follows a pure Markov strategy $\sigma$. Then for
  any play $P$ and $T<\abs{P}$ we have
\[
\dens(\sigma^T) = \frac{1}{T} \sum_{T'=1}^T r_T \enspace ,
\]
where $r_T$ is the reward given to Player~1 in round $T$.  In
particular, when $P$ is infinite, we have
\[
\uinf(P) = \liminf_{T\rightarrow\infty}\dens(\sigma^T) \enspace ,
\]
and
\[
\usup(P) = \limsup_{T\rightarrow\infty}\dens(\sigma^T) \enspace .
\]
\end{observation}

\section{Small space \texorpdfstring{$\eps$}{epsilon}-supremum-optimal
  strategies in the Big Match}
\label{sec:eps-sup}

For given $\eps>0$, let $\xi = \eps^2$.  For any
non-decreasing and \emph{unbounded} function $f : \ZZ_+
\rightarrow \ZZ_+$, we will now give an $\eps$-supremum optimal strategy
$\sigma_1^*$ for Player~1 in the Big Match that for all $\delta>0$
with probability $1-\delta$ uses $O(\log f(T))$ space.  Let $\ov{f}$
be a strictly increasing unbounded function from $\ZZ_+$ to
$\RR_+$, such that $\ov{f}(x)\leq f(x)$ for all $x\in \ZZ_+$, and let $F$
be the inverse of $\ov{f}$. For simplicity, and without loss of
generality, we assume that $F(1)=1$ and $F(T+1)\geq 2\cdot F(T)$. Note
that in particular $\ov{f}(2\cdot T)\leq 2\cdot \ov{f}(T)$.

\paragraph{Intuitive description of the strategy and proof.}
The main idea for building the strategy is to partition the rounds of the game into epochs, such that epoch
$i$ has expected length $F(i)$. The $i$'th epoch is further split 
into $i$ sub-epochs. In each sub-epoch $j$ of the $i$-th epoch we
sample $i^2$ rounds uniformly at random. In every round not sampled we
simply stay in the same memory state and play $\lef$ with probability
1. We view the $i^2$ samples as a stream of actions chosen by
Player~2. We then follow a particular $\xi$-optimal base strategy
$\sigma_1^{i,\xi}$ for the Big Match on the samples of sub-epoch $j$. This
strategy $\sigma_1^{i,\xi}$ is a suitably modified version of a
strategy by Blackwell and Ferguson~\cite{AMS:BlackwellFerguson1968}
and Kohlberg~\cite{AS:Kohlberg1974}.

More precisely, if $\sigma_1^{i,\xi}$ stops in its $k$-th round when
run on the samples of sub-epoch $j$, the strategy $\sigma_1^*$ stops
on the $k$-th sample in sub-epoch $j$. This will ensure that if
$\sigma_1^*$ stops with probability at least $\sqrt{\xi}$, the outcome
is at least $\frac{1}{2}-\xi$.

Also, for any $0<\delta<\frac{1}{2}$ and for sufficiently large $i$,
depending on $\delta$, if the samples have density of $\lef$ at most
$\frac{1}{2} - \delta$ then $\sigma_1^{i,\xi}$ stops on the samples
with a positive probability depending only on $\xi$, namely $\xi^4$.
For $f(T) = \Theta(\log T)$, the division into sub-epochs ensures that 
if $\liminf_{T\rightarrow\infty}\dens(\sigma^T) <
\frac{1}{2}$ then infinitely many sub-epochs have density of $\lef$ smaller than $1/2$, 
and thus the play stops with probability 1 in one of such epochs. This is not necessarily true
for $f(T)$ smaller than $\log T$.

\paragraph{The base strategy.} The important inner part of our
strategy is a $\xi$-optimal strategy $\sigma_1^{i,\xi}$ parametrized
by a non-negative integer $i$. These strategies are similar to
$\xi$-optimal strategies given by Blackwell and
Ferguson~\cite{AMS:BlackwellFerguson1968} and
Kohlberg~\cite{AS:Kohlberg1974} (in fact, setting $i=0$ and replacing
$\xi^4$ by $\xi^2$ below one obtains the strategy used by Kohlberg).

The strategy $\sigma_1^{i,\xi}$ uses deterministic updates of memory,
and uses integers as memory states (we think of the memory as an
integer counter). The memory update function is given by
\[
\sigma_1^{i,u}(a,j) = \begin{cases}
j+1 & \text{if } a=\lef\\
j-1 & \text{if } a=\rig
\end{cases}
\]
and the action function is given by
\[
\sigma_1^{i,a}(j)(\rig) = \begin{cases}
\xi^4(1-\xi)^{i+j} & \text{if } i+j>0\\
\xi^4 & \text{if } i+j\leq 0
\end{cases}
\]

\paragraph{The complete strategy} We are now ready to define
$\sigma_1^*$. The memory states of this strategy are 5-tuples
$(i,j,k,\ell,b) \in \ZZ_+ \times \ZZ_+ \times \ZZ \times \NN \times
\{0,1\}$. Here $i$ denotes the current epoch and $j$ denotes the
current sub-epoch of epoch $i$. The number of samples already made in
the current sub-epoch is $k$. The memory state of the inner strategy is
stored as $\ell$. Finally $b$ is 1 if and only if the strategy will sample
to the inner strategy in the next step.

The memory update function $\sigma_1^{*,u}$ is as follows. Let
$(i,j,k,\ell,b)$ be the current memory state and let $a$ be the action
of Player~2 in the current step. We then describe the distribution of
the next memory state $(i',j',k',\ell',b')$.

\begin{itemize}
\item The current step is not sampled if $b=0$. In that case we keep
  $i'=i$, $j'=j$, $k'=k$, and $\ell'=\ell$.
\item The current epoch is ending if $j=i$, $k=i^2-1$, and $b=1$. In that case
$i'=i+1$, $j'=1$, $k'=0$, and $\ell'=0$.
\item A sub-epoch is ending within the current epoch if $j<i$, $k=i^2-1$, and
$b=1$. In that case $i'=i$, $j'=j+1$, $k'=0$, and $\ell'=0$.
\item We sample within a sub-epoch if $k<i^2-1$ and $b=1$. In that case
  $i'=i$, $j'=j$, $k'=k+1$, and $\ell'=\sigma_1^{i,u}(a,\ell)$.
\end{itemize}
Finally, in every case, we make a probabilistic choice whether to
sample in the next step by letting $b'=1$ with probability
$\frac{(i')^3}{F(i')}$.

The action function $\sigma_1^{*,a}$ is given by
 \[
\sigma_1^{*,a}((i,j,k,\ell,b))(a)=\begin{cases}
\sigma_1^{i,\xi}(j)(a) &\text{if  $b=1$} \\
1  &\text{if $b=0$ and $a=\lef$} \\
0 & \text{otherwise} 
\end{cases} \enspace .
\]

In other words, if the current step is sampled, Player~1 follows
the current base strategy, and otherwise always plays $\lef$.

The starting memory state is $m_s=(1,1,1,0,0)$. The states that can be
reached in sub-epoch $j$ of epoch $i$ are of the form $(i,j,k,\ell,b)$
where $0 \leq k < i^2$ and $-i^2 < \ell < i^2$. Thus at most $4i^4$
states can be reached.
The states are mapped to the natural numbers as follows: The memory $(1,1,1,0,0)$ is mapped to 0 and for each epoch $i$, all states in epoch $i$ are mapped to the numbers (in an arbitrary order) following the numbers mapped to by epoch $i-1$.

\paragraph{Proof preliminaries}
It will be useful to consider the strategy modified to never
stop. Thus denote by $\widetilde{\sigma}_1$ the strategy for Player~1,
where $\widetilde{\sigma}^{u}=\sigma_1^{*,u}$ and
$\widetilde{\sigma}_1^{a}(m)(\lef)=1$ for all memory states $m$.

We next define random variables indicating the locations of the sample
steps.  Fix some strategy $\sigma_2$ for Player~2. Let
$\calM^{\tilde{\sigma}_1,\sigma_2}_{m_s}$ be the memory sequence
assigned to Player~1 when Player~1 follows $\tilde{\sigma}_1$ and
Player~2 follows $\sigma_2$.  For positive integers $i,j,k$ let
$t(i,j,k)$ be the random variable indicating the round in which we
sample the $k$'th time in sub-epoch $j$ of epoch $i$. For simplicity
of notation we let $t(i,0,i^2)$ denote $t(i-1,i-1,(i-1)^2)$, and we
let $t(i,j)=t(i,j,0)$ denote $t(i,j-1,i^2)$.

% In this section, we will focus on showing \cref{thm:sup_eps}.

% \begin{theorem}\label{thm:sup_eps}
%   The strategy $\sigma_1$ uses $\log f(T)$ space with probability
%   $1-\delta$ for all $\delta>0$ and is
%   $\sqrt{\eps}$-supremum-optimal.
% \end{theorem}

% To do so we will use three main lemmas,
% \cref{lem:spaceusage,l-winning,l-stopping-sup}. \Cref{lem:spaceusage}
% focuses on space usage and \cref{l-winning,l-stopping-sup} focuses
% on $\sqrt{\eps}$-supremum-optimality.

% \begin{lemma}
% \label{lem:spaceusage}
% The space usage of $\sigma_1$ is at most $O(\log f(T))$ with probability $1-\gamma$ for all $\gamma>0$.
% \end{lemma}

% \begin{lemma}\label{l-winning}
%   Let $\eps\in(0,1)$. Let $\sigma$ be an pure Markov strategy for
%   Player~2. If the probability that we stop is at least
%   $\sqrt{\eps}$, then we have
%   that $$\Pr\nolimits^{\sigma_1,\sigma}[\mbox{ Player~1 wins } |
%   \mbox{ play stops } ] \geq \frac{1}{2} - \sqrt{\eps}.$$
% \end{lemma}

% \begin{lemma}\label{l-stopping-sup}
%   Let $\eps \in(0,1)$. Let $\sigma$ be an arbitrary pure Markov
%   strategy for Player~2.  If $\limsup_{T\rightarrow \infty}
%   \dens(\sigma^T)< 1/2$ then the play stops with probability 1.
% \end{lemma}

\subsection{Space usage of the strategy}
We will here consider the space usage of $\sigma_1^*$. First we will
argue that with high probability, for all large enough $i$ and any $j$, the
length of sub-epoch $j$ of epoch $i$, $t(i,j,i^2)-t(i,j-1,i^2)$, is close to $F(i)$.

\begin{lemma}\label{l-goodbalance}
  For any $\gamma,\delta\in (0,1/4)$, there is a constant $M$ such
  that with probability at least $1-\gamma$, for all $i\geq M$ and all
  $j\in\{1,\dots,i\}$, we have that 
\[
t(i,j,i^2)-t(i,j-1,i^2) \in  [(1-\delta) F(i)/i,(1+\delta) F(i)/i] \enspace .
\]
\end{lemma}

\begin{proof}
  The expected number of times we sample during $(1-\delta)F(i)/i$
  steps of the $i$'th epoch is $(1-\delta)i^2$.  If
  $t(i,j,i^2)-t(i,j-1,i^2) < (1-\delta) F(i)/i$ then we sampled at
  least $i^2$ times during these $(1-\delta)F(i)/i$ steps of epoch
  $i$.  This means that the actual number of samples is larger than
  its expectation by a factor $\delta/(1-\delta)$.  Thus by the
  multiplicative Chernoff bound, \cref{THM:chernoff}, we see that,
\[
\Pr[t(i,j,i^2)-t(i,j-1,i^2) < (1-\delta) F(i)/i] < \exp\left(-c(1-\delta)i^2\right) \enspace , 
\]
where $c=(\frac{\delta}{1-\delta})^2/(2+\frac{\delta}{1-\delta})$.
Similarly, if $t(i,j,i^2)-t(i,j-1,i^2) > (1+\delta) F(i)/i$ then we
sampled less than $i^2$ times during $(1+\delta)F(i)/i$ steps which is
less than the expected by a factor $\delta/(1+\delta)$ of its
expectation.  Again, by the multiplicative Chernoff bound,
\[
\Pr[t(i,j,i^2)-t(i,j-1,i^2) > (1+\delta) F(i)/i] < \exp\left({-\frac{\delta^2}{2}(1+\delta)i^2}\right) \enspace .
\]
Thus for any $M$, we can bound from above the probability of any of
the differences for $i\geq M$ being outside of the required range by:
\[
\sum_{i=M}^{\infty} i \cdot \left( \exp\left(-c (1-\delta)i^2\right) + \exp\left(- \frac{\delta^2}{2}(1+\delta)i^2\right) \right) \enspace .
\]
This sum is convergent so for sufficiently large $M$ it can be bounded
by $\gamma$. The lemma follows.
\end{proof}

Now we bound the space usage of the strategy $\sigma^*_1$.
\begin{lemma}\label{lem:spaceusage}
  For all constants $\gamma>0$, with probability at least $1-\gamma$, the
  space usage of $\sigma^*_1$ is $O(\log f(T))$.
\end{lemma}
\begin{proof}
  Recall that there are at most $4\cdot i^4$ distinct possible memory
  states reachable during the $j$'th sub-epoch of epoch $i$, for all
  $i,j$. Thus, since there are $i$ sub-epochs in epoch $i$ there are
  at most $4\cdot i^5$ distinct possible memory states reachable
  during the $i$'th epoch. It is then clear that there are at most
  $\sum_{r=0}^i 4 \cdot r^5\leq 4\cdot i^6$ distinct possible
  memory states that can have been reached before the end of epoch
  $i$, for all $i$. Because memory states in earlier epochs are mapped to smaller numbers than latter epochs, we have that the strategy has not been in any state above that of $4\cdot i^6$ before the end of epoch $i$.

  Fix some $\gamma>0$. By \cref{l-goodbalance}, with probability at
  least $1-\gamma$, there is a $M$ such that for all $i\geq M$, the
  number $t(i,i,i^2)-t(i,1,i^2)$ is greater than
  $\frac{F(i)}{2}$. Thus also $t(i,i,i^2)\geq \frac{F(i)}{2}$.
  Consider any round $T$, for $T\geq t(M,M,M^2)$. Let $i$ be the epoch
  containing round $T$.  By the preceding we know that, before time
  step $t(i+1,i+1,(i+1)^2)$ (which is greater than $T$), the strategy
  have only been in memory states below $4\cdot (i+1)^6$. We 
  also have that $T\geq t(i,i,i^2)\geq \frac{F(i)}{2}$ and hence
  $2\ov{f}(T) \geq \ov{f}(2T)\geq i$, where the first inequality is by
  our assumption on $F$. Thus the strategy can only have been in states below that of $4\cdot
  (2\cdot \ov{f}(T)+1)^6$  before time step $T$. This is
  true for all sufficiently large $T$ and thus the strategy uses at
  most $O(\log f(T))$ space with probability $1-\gamma$.
\end{proof}

\subsection{Play stopping implies good outcome}

We first establish some properties of the base strategy
$\sigma_1^{i,\xi}$. The proof of these uses ideas similar to proofs by
Blackwell and Ferguson~\cite{AMS:BlackwellFerguson1968} and
Kohlberg~\cite{AS:Kohlberg1974}, where they showed
$\eps$-optimality of their strategies.

\begin{lemma}\label{l-base}
  Let $T,i\geq 1$ be integers and $0<\xi<1$ be a real number. Let
  $\sigma \in \{\lef,\rig\}^T$ be a arbitrary prefix of a pure Markov
  strategy for Player~2.  Consider the first $T$ rounds where the
  players play the Big Match following $\sigma_1^{i,\xi}$ and $\sigma$
  respectively. Let $\pwin$ be the probability that Player~1 stops the
  game (i.e.\ plays $\rgt$) and wins. Let $\ploss$ be the probability
  that Player~1 stops the game and loses. Then we have:
\begin{enumerate}
\item \[ \ploss \leq (1-\xi)^i \xi^{3} +
  (1-\xi)^{-1}\pwin \enspace . \]
\item For any $0<\delta \leq \frac{1}{2}$ and any $T>i/(2 \delta)$, if
  $\dens(\sigma) \leq \frac{1}{2} - \delta$ then
 \[ \pwin +  \ploss \geq \xi^4\enspace . \]
\end{enumerate}
\end{lemma}

\begin{proof}
%   Define $d_\ell = \abs{\{\ell'< \ell\mid \sigma_{\ell'}=\lft\}| -
%     |\{\ell' < \ell \mid \sigma_{\ell'}=\rgt\}}$. Note that $d_\ell$
%   is precisely the value of the counter used by $\sigma_1^{i,\xi}$ as
%   memory in step $\ell$. For integer $d$, define
%  \[
%  K_d = \{ \ell \in \{1,\dots,T\}\mid (d_\ell = d \;\&\; \sigma_\ell =
%  \lft) \text{ or } (d_\ell = d+1 \;\&\; \sigma_\ell = \rgt) \} \enspace
%  .
% \]
%   Intuitively, $K_d$ is the set of times at which the counter goes
%   from $d$ to $d+1$ or from $d+1$ to $d$. 
%\nb{Changes by Rasmus inserted}
  Define $d_\ell = \abs{\{\ell'< \ell\mid \sigma_{\ell'}=\lft\}| -
    |\{\ell' < \ell \mid \sigma_{\ell'}=\rgt\}}$. Note that $d_\ell$
  is precisely the value of the counter used by $\sigma_1^{i,\xi}$ as
  memory in step $\ell$. For integer $d$, define
 \[
 K_d = \{ \ell \in \{1,\dots,T\}\mid (d_\ell = d \;\&\; \sigma_\ell =
 \lft) \text{ or } (d_\ell = d+1 \;\&\; \sigma_\ell = \rgt) \} \enspace
 .
\]
  There is an illustration of how $d_\ell$ could evolve through the steps in \cref{fig:counter_movement}.
The set $K_d$ is then the times the counter moves between the pair of rows $d$ and $d+1$.
   Observe that the counter is alternately moving up and down in each $K_d$ (for instance, in the gray row, the arrows are wider and first moves up, then down and then up again).

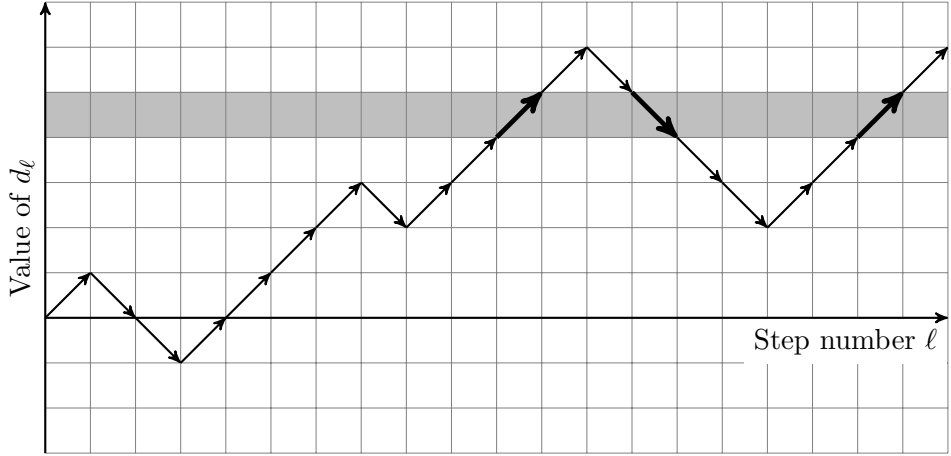
\begin{figure}
\begin{tikzpicture}[thick,scale=0.6]
\coordinate (o) at (0,0);
\coordinate (tr) at (20,7);
\coordinate (bl) at (0,-3);

\fill[light-gray] (0,4) rectangle (20,5);

\draw[gray,very thin] (bl) grid (tr);
\node[anchor=north east,fill=white] at (tr |- o) {Step number $\ell$};
\draw[thick,->,>=stealth'] (o) -- (tr |- o);
\draw[thick,->,>=stealth'] (bl) -- (bl |- tr) node[midway,anchor=south,rotate=90] {Value of $d_\ell$};

\xdef\lx{0}
\xdef\ly{0}
\foreach \x/\y/\v in {1/1/1, 2/0/1, 3/-1/1, 4/0/1, 5/1/1, 6/2/1, 7/3/1, 8/2/1, 9/3/1, 10/4/1, 11/5/1.5, 12/6/1, 13/5/1, 14/4/1.5, 15/3/1, 16/2/1, 17/3/1, 18/4/1, 19/5/1.5, 20/6/1}{\draw[line width=\v*\pgflinewidth, ->,>=stealth'] (\lx,\ly) -- (\x,\y); 
\xdef\lx{\x}
\xdef\ly{\y}
}

\end{tikzpicture}
\caption{Possible movement of the counter $\sigma_1^{i,\xi}$ uses as memory through the steps.\label{fig:counter_movement}}
\end{figure}  
  
  Notice that $K_d$
  partitions $\{1,\dots, T\}$. Let $p_{\mathrm{loss},d}$ be the
  probability that the game stops at some step $\ell\in K_d$ where
  $\sigma_\ell = \lft$, and let $p_{\mathrm{win},d}$ be the
  probability that the game stops at some step $\ell\in K_d$ where
  $\sigma_\ell = \rgt$. We see that $\ploss$ is the sum of
  $p_{\mathrm{loss},d}$, and that $\pwin$ is the sum of
  $p_{\mathrm{win},d}$.  Let $K_d = \{k_1 < k_2 < \cdots <k_m\}$, for
  some $m$. Observe that for any $j$, $\sigma_{k_j} = \lft$ if and
  only if $\sigma_{k_{j+1}} = \rgt$, so
  $\sigma_{k_1},\sigma_{k_2},\dots,\sigma_{k_m}$ is an alternating
  sequence (as mentioned in relation to the illustration) starting with $\lft$ when $d\geq 0$ and starting with
  $\rgt$ otherwise. For any $d\geq -i$, the probability that the game
  stops in round $k_j \in K_d$, conditioned on the event that it did
  not stop before round $k_j$, is $\xi^4 (1-\xi)^{i+d_{k_j}}$ so, it
  is $\xi^4 (1-\xi)^{i+d}$ when $\sigma_{k_j} = \lft$, and it is
  $\xi^4 (1-\xi)^{i+d+1}$ when $\sigma_{k_j} = \rgt$.

%\nb{K exchanged even and odd below. Isn't that correct?}

Hence, for $d\geq 0$, $p_{\mathrm{loss},d} \leq \xi^4 (1-\xi)^{i+d} +
(1-\xi)^{-1} p_{\mathrm{win},d}$. This is because we stop at $k_1$
with probability at most $\xi^4 (1-\xi)^{i+d}$ (in which case Player~1
loses) and then for each even $j$, the probability of stopping at step
$k_j$ (and winning) is at least $(1-\xi)$-times the probability of
stopping at step $k_{j+1}$ (and losing).  (Indeed, the probability of
stopping at step $k_{j+1}$ might be even substantially smaller as the
probability of stopping between $k_j$ and $k_{j+1}$ might be
non-zero.)

For $d\in \{-i,-i+1,\dots,-1\}$, $p_{\mathrm{loss},d} \leq
(1-\xi)^{-1} p_{\mathrm{win},d}$ as for each \emph{odd} $j$, the
probability of stopping at step $k_j$ (and winning) is at least
$(1-\xi)$-times the probability of stopping at step $k_{j+1}$ (and
losing).

Finally, for $d < -i$, $p_{\mathrm{loss},d} \leq p_{\mathrm{win},d}$,
as for each odd $j$, the probability of stopping at step $k_j$ (and
winning) is at least the probability of stopping at step $k_{j+1}$
(and losing).  (The probability of stopping at any such $k_j$
conditioned on not stopping sooner is $\xi^4$ in this case.)

Hence, 
\begin{eqnarray*}
  \ploss = \sum_d p_{\mathrm{loss},d} &\le& \sum_{d\geq 0} \xi^4 (1-\xi)^{i+d} + (1-\xi)^{-1} \sum_d p_{\mathrm{win},d} \\
  &\le& \xi^{3} (1-\xi)^{i} + (1-\xi)^{-1} \pwin.
\end{eqnarray*}

For the second part, if the density of $\sigma_1,\dots,\sigma_T$ is at
most $\frac{1}{2} - \delta$ then it must contain at most $T/2 - \delta
T < T/2 - i/2$ occurrences of the letter $\lft$. Hence, it contains
more than $t/2+i/2$ occurrences of the letter $\rgt$. This implies
that when the game reaches round $T$, we have that the memory state
$j$ (recalling that the memory states are integers and in any round
corresponds to the difference between the number of times Player~2 has
played $\rig$ minus the time he played $\lef$ up till now) is such
that $j \leq -i$ and hence Player~1 plays $\rgt$ at step $t$ with
probability $\xi^4$ if the play did not stop, yet.
\end{proof}

We can now prove the main statement of this subsection, that if the
probability of stopping is not too small, Player~1 wins with
probability close to 1/2 if play the stops.

\begin{lemma}\label{l-winning}
  Let $\xi\in(0,1)$. Let $\sigma$ be an pure Markov strategy for
  Player~2. If the probability that play stops is at least
  $\sqrt{\xi}$, then we have
  that \[\Pr\nolimits^{\sigma^*_1,\sigma}[\text{ Player~1 wins } \mid
  \text{ play stops } ] \geq \frac{1}{2} - \sqrt{\xi} \enspace .\]
\end{lemma}
\begin{proof}
  We will in this proof continue playing in a state $v$, even if
  Player~1 plays $\rig$. Note that $\sigma^*_1$ can still generate a
  choice in this case.  Let $A_{i,j}$ be set of plays in which,
  between round $1$ and round $t(i,j,1)-1$, Player~1 does not play
  $\rgt$.  Let $W_{i,j}$ be the set of plays, in which, between round
  $t(i,j,1)$ and round $t(i,j,i^2)$, Player~1 plays $\rgt$ and when he
  plays $\rgt$ for the first time between these rounds, Player~2 plays
  $\rgt$ as well.  Similarly, let $L_{i,j}$ be the set of plays, in
  which, between round $t(i,j,1)$ and round $t(i,j,i^2)$, Player~1
  plays $\rgt$ and when he does so for the first time Player~2 plays
  $\lft$.  Let $S$ be the set of plays, in which Player~1 plays $\rig$
  in some round.  Let $W$ be the set of plays, in which Player~1 plays
  $\rig$ in some round and the first time he does so Player~2 also
  plays $\rig$.  Let $L$ be the set of plays, in which Player~1 plays
  $\rig$ in some round and the first time he does so Player~2 plays
  $\lef$. We see that $S=W\cup L$.  Clearly,
  $\Pr\nolimits^{\sigma^*_1,\sigma}[W] = \sum_{i,j}
  \Pr\nolimits^{\sigma^*_1,\sigma}[W_{i,j} \;\&\; A_{i,j}]$ and
  $\Pr\nolimits^{\sigma^*_1,\sigma}[L] = \sum_{i,j}
  \Pr\nolimits^{\sigma^*_1,\sigma}[L_{i,j} \;\&\; A_{i,j}]$.

  Fix a possible value of all $t(i,j,k)$'s and denote by $Y$ the
  event that these particular values actually occurs. Fix $i$ and $j$.
  Conditioned on $Y$ and $A_{i,j}$, between time $t(i,j,1)$ and
  $t(i,j,i^2)$ Player~1 plays $\sigma_1^{i,\xi}$ against a fixed
  strategy $\sigma_{t(i,j,1)},\sigma_{t(i,j,2)},\dots,
  \sigma_{t(i,j,i^2)}$ for Player~2. By \cref{l-base}, the probability
  of losing such a game for Player~1 is at most $\xi^3 (1-\xi)^{i}$
  plus the probability of winning in this game divided by
  $(1-\xi)$. Hence,
\begin{eqnarray*}
  \Pr[L_{i,j}\mid Y,A_{i,j}] \leq \xi^3 (1-\xi)^{i} + (1-\xi)^{-1} \Pr[ W_{i,j}\mid Y,A_{i,j}]. 
\end{eqnarray*}
Since the above inequality is true conditioned on arbitrary values of
$t(i,j_1,j_2)$'s, it is true also without the conditioning:
\begin{eqnarray*}
\Pr[L_{i,j}\mid A_{i,j}] \leq \xi^3 (1-\xi)^{i} + (1-\xi)^{-1} \Pr[ W_{i,j}\mid A_{i,j}]. 
\end{eqnarray*}
Thus,
\begin{eqnarray*}
  \Pr\nolimits^{\sigma^*_1,\sigma}[L] &=& \sum_{i=1}^\infty \sum_{j=1}^i \Pr[L_{i,j} \;\&\; A_{i,j}] \\ 
  &=& \sum_{i=1}^\infty \sum_{j=1}^i \Pr[L_{i,j} \mid A_{i,j}] \cdot \Pr[A_{i,j}] \\ 
  &\le& \sum_{i=1}^\infty \sum_{j=1}^i \left( \xi^3 (1-\xi)^{i} + (1-\xi)^{-1} \Pr\nolimits^{\sigma^*_1,\sigma}[W_{i,j}\mid A_{i,j}] \right) \cdot \Pr[A_{i,j}]\\
  &\le& \sum_{i=1}^\infty \sum_{j=1}^i \xi^3 (1-\xi)^{i} + (1-\xi)^{-1} \sum_{i=1}^\infty \sum_{j=1}^i \Pr\nolimits^{\sigma^*_1,\sigma}[W_{i,j}\;\&\; A_{i,j}]\\
    &=& \sum_{i=1}^\infty i\cdot \xi^3 (1-\xi)^{i} + (1-\xi)^{-1} \sum_{i=1}^\infty \sum_{j=1}^i \Pr\nolimits^{\sigma^*_1,\sigma}[W_{i,j}\;\&\; A_{i,j}]\\
  &\le& \xi  + (1-\xi)^{-1} \Pr\nolimits^{\sigma^*_1,\sigma}[W]\enspace .
\end{eqnarray*}
Hence, $(1-\xi)\Pr\nolimits^{\sigma^*_1,\sigma}[L] \leq \xi - \xi^2 +
\Pr\nolimits^{\sigma^*_1,\sigma}[W]$, and so
$\Pr\nolimits^{\sigma^*_1,\sigma}[L] -
\Pr\nolimits^{\sigma^*_1,\sigma}[W] \leq 2\xi$.  By our assumption we
also have that
$\Pr\nolimits^{\sigma^*_1,\sigma}[L]+\Pr\nolimits^{\sigma^*_1,\sigma}[W]=\Pr\nolimits^{\sigma^*_1,\sigma}[S]\geq
\sqrt{\xi}$.

We want to find the minimum probability that we might win, conditioned
on us stopping \[\Pr\nolimits^{\sigma^*_1,\sigma}[W\mid
S]=\frac{\Pr\nolimits^{\sigma^*_1,\sigma}[W]}{\Pr\nolimits^{\sigma^*_1,\sigma}[S]}=\frac{\Pr\nolimits^{\sigma^*_1,\sigma}[W]}{\Pr\nolimits^{\sigma^*_1,\sigma}[W]+\Pr\nolimits^{\sigma^*_1,\sigma}[L]}\]

We see that it is greater than the solution of: 

\[
\begin{array}{lll}
\min &\frac{x}{x+y}\\
\text{s.t. } & x+y &\geq \sqrt{\xi}\\
&y-x &\leq 2\xi
\end{array}
\]

Solving the above, we see that $\Pr\nolimits^{\sigma^*_1,\sigma}[W\mid
S]\geq \frac{1}{2}-\sqrt{\xi}$ and the lemma follows.
\end{proof}

\subsection{Low density means play stops}

In this subsection we will prove that play will stop with
probability~1 if the density of prefixes of the pure Markov strategy
used by Player~2 is not infinitely often at least 1/2. First, we will
show that low density implies that some sequence of sub-epochs also
have low density.

% We will be using the following
% instance of a bound by Hoeffding, \cref{THM:Hoeffding}.
% \begin{proposition}
%   Let $N \geq n \geq 1$ be integers.  Let $c_1,c_2,\dots,c_N$ be items
%   with labels from $\{-\Delta_{\max},\dots,\Delta_{\max}\}$. Let $S_n$ be the sum
%   of the labels of a randomly chosen subset of $c_1,\dots,c_N$ of size
%   $n$. For any $t>0$:
% \[
% \Pr[\abs{ S_n - {\Exp[S_n]}} \geq t ] \leq 2 \exp\left(-\frac{t^2}{2n(\Delta_{\max})^2}\right).
% \]
% \label{PROP:hoeffding}
% \end{proposition}
% We will in the next proof only need the case where $\Delta_{\max}=1$.

\begin{lemma}\label{l-subdivion-sup}
  Let $\delta \in(0,1)$. Let $\sigma$ be an arbitrary pure Markov
  strategy for Player~2.  Let $a_{i,j}$ be some numbers. Consider the
  event $Y$ where $t(i,j)=a_{i,j}$ for all $i,j$.  If
  $\limsup_{T\rightarrow \infty} \dens(\sigma^T)\leq 1/2-\delta$, then
  conditioned on $Y$, there is an infinite sequence of sub-epochs and
  epochs $(i_n,j_n)_n$ such that
  $\dens(\sigma,a_{i_n,j_n}+1,a_{i_n,(j_n+1)})\leq 1/2-\delta/4$.
\end{lemma}
\begin{proof}
  Let $M$ be such that for every $T'\geq M$ we have that
  $\dens(\sigma^{T'})\leq 1/2-\delta/2$.  Let $(T_n)_n$ be a sequence
  such that $T_1\geq M$ and for all $n\geq 1$ we have that
  $T_{n+1}\cdot \delta/4\geq T_{n}$ and $T_n=a_{i,j}$ for some
  $i,j$. Let $(i_n',j_n')_n$ be the sequence such that
  $T_n=a_{i_n',j_n'}$. This means that even if
  $\dens(\sigma^{T_{n}})=0$, the density $\dens(\sigma,T_{n}+1,T_{n+1})$
  is at most $1/2-\delta/4$, because $\dens(\sigma^{T_{n+1}})\leq
  1/2-\delta/2$. But, we then get that there exists some sub-epoch
  $j_n$ in epoch $i_n$, such that $j_{n}' \leq j_n\leq j_{n+1}'$ and
  such that $i_n'\leq i_n\leq i_{n+1}'$ for which the density of that
  sub-epoch $\dens(\sigma,a_{i_n,j_n}+1,a_{i_n,(j_{n}+1)})$ is at most
  $1/2-\delta/4$, because not all sub-epochs can have density below
  that of the average sub-epoch. But then $(i_n,j_n)_n$ satisfies the
  lemma statement.
\end{proof}

We are now ready to prove the main statement of this subsection.
\begin{lemma}
\label{l-stopping-sup}
Let $\sigma$ be an arbitrary pure Markov strategy for Player~2. If
\[
\limsup_{T\rightarrow \infty} \dens(\sigma^T)< 1/2 \enspace ,
\]
then when played against $\sigma_1^*$ the play stops with
probability~1.
\end{lemma}
\begin{proof}
  Let $\delta>0$ be such that $\limsup_{T\rightarrow \infty}
  \dens(\sigma^T)\leq 1/2-\delta$.  Consider arbitrary numbers
  $a_{i,j}$ and the event $Y$ stating that $t(i,j)=a_{i,j}$ for all
  $i,j$.  Let $(i_n,j_n)_n$ be the sequence of sub-epochs and epochs
  shown to exists by \cref{l-subdivion-sup} with probability~1. That
  is, for each $(i_n,j_n)$ we have that sub-epoch $j_n$ of epoch $i_n$
  has density at most $1/2-\delta/4$.  We see that, conditioned on $Y$
  that each sample are sampled uniformly at random in each sub-epoch
  $j$ of each epoch $i$, except for the last sample.  Now consider
  some fixed $n$.  By Hoeffding's inequality, \cref{THM:Hoeffding}
  (setting $a_i=0$ and $b_i=1$, and letting $c_i$ be $i$'th payoff for
  Player~1), the probability that in sub-epoch $j_n$ of epoch $i_n$
  that among our $i_n^2-1$ first samples we have $\frac{\delta}{8}
  \cdot ((i_n)^2-1)$ additional $\lft$ on top of the expectation
  (which is at most $(1/2-\delta/4)\cdot ((i_n)^2-1)$) is bounded as
\[
\begin{split}
\Pr\left[\dens(\sigma_{t(i_n,j_n,1)},\sigma_{t(i_n,j_n,2)},\dots,\sigma_{t(i_n,j_n,(i_n)^2-1)}) \geq \frac{1}{2} - \frac{\delta}{8}\right] \\ \leq  2 \exp\left(-\frac{2(\frac{\delta}{8}((i_n)^2-1))^2}{(i_n)^2-1}\right) = 2\exp\left(-\frac{\delta^2}{32}((i_n)^2-1)\right)
 \enspace .
\end{split}
\]
For sufficiently large $i_n$, this is less than $\frac{1}{2}$. If on
the other hand the number of $\lef$'s we sample is less than
$(\frac{1}{2}-\frac{\delta}{8})((i_n)^2-1)$, we see that
$(i_n)^2-1>i_n/(16\delta)$ for large enough $i_n$ and in that case we
have, by \cref{l-base}, that we stop with probability at least
$\xi^4$ in sub-epoch $j_n$ of epoch $i_n$.  Thus, for each of the
infinitely many $n$'s for which $i_n$ is sufficiently high, we have a
probability of at least $\frac{\xi^4}{2}$ of stopping. Thus play
must stop with probability~1.

The argument was conditioned on some fixed assignment of endpoints of
sup-epochs and epochs, but since there is such a assignment with
probability~1 (since they are finite with probability~1), we conclude
that the proof works without the condition.
\end{proof}

\subsection{Proof of main result}

 \begin{theorem}\label{thm:sup_eps}
   The strategy $\sigma^*_1$ is $\sqrt{\xi}$-supremum-optimal, and for
   all $\delta>0$, with probability at least $1-\delta$ it uses
   space $O(\log f(T))$.
 \end{theorem}
\begin{proof}%[Proof of \cref{thm:sup_eps}]
  The space usage follows from \cref{lem:spaceusage}.  Let $s$ be the
  probability that the play stops. We now consider three cases, either
  (i)~$s=1$; or (ii)~$\sqrt{\xi}<s<1$; or (iii)~$s\leq
  \sqrt{\xi}$. In case (i), if $s=1$, Player~1 wins with
  probability $\frac{1}{2}-\sqrt{\xi}$, by \cref{l-winning}.  In
  case (ii), if $\sqrt{\xi}<s<1$, then, by \cref{l-winning},
  conditioned on the play stopping, Player~1 wins with probability
  $W_s=\frac{1}{2}-\sqrt{\xi}$ and, since $s<1$, by
  \cref{l-stopping-sup}
  $\widetilde{W}_s=\limsup_{T\rightarrow\infty}\dens(\sigma^T)\geq
  \frac{1}{2}$. Thus, Player~1 wins with probability\[s\cdot
  W_s+(1-s)\cdot \widetilde{W}_s\geq s\cdot
  (\frac{1}{2}-\sqrt{\xi})+(1-s)\frac{1}{2}\geq
  \frac{1}{2}-\sqrt{\xi}\] In case (iii), if $s\leq
  \sqrt{\xi}$, then by \cref{l-stopping-sup}
  $\widetilde{W}_s=\limsup_{T\rightarrow\infty}\dens(\sigma^T)\geq
  \frac{1}{2}$. The winning probability is then at least \[s\cdot 0+(1-s)\cdot
  \frac{1}{2}\geq \frac{1-\sqrt{\xi}}{2}\geq
  \frac{1}{2}-\sqrt{\xi}\]
\end{proof}

\section{An \texorpdfstring{$\eps$}{epsilon}-optimal strategy that uses \texorpdfstring{$\log \log T$}{log log T} space for the Big Match}

\label{sec:eps-inf}

For given $\eps>0$, let $\xi=\eps^2$. In this section we give a
$\eps$-optimal strategy $\sigma_1^*$ for Player~1 in the Big Match
that for all $\delta>0$ with probability $1-\delta$ uses $O(\log\log
T)$ space. The strategy is simply an instantiation of the strategy
$\sigma_1^*$ from \cref{sec:eps-sup}, setting $f(T)=\lceil{\log
  T}\rceil$. We can then let $\ov{f}=\log T$ and $F(T)=2^T$.

The claim about the space usage of $\sigma_1^*$ is thus already
established in \cref{sec:eps-sup}. To obtain the stronger property of
$\eps$-optimality rather than just $\eps$-supremum-optimality, we just
need to establish a $\liminf$ version of \cref{l-stopping-sup}.

First we show a technical lemma. For a pure Markov strategy $\sigma$
and a sequence of integers $I=\{i_1,i_2,\dots,i_m\}$, let $\sigma_I$
be the sequence, $\sigma_{i_1},\sigma_{i_2},\dots,\sigma_{i_m}$. Note
that $\sigma^k=\sigma_{\{1,\dots,k\}}$.

\begin{lemma}\label{l-subdivision}
  Let $\sigma$ be a pure Markov strategy for Player~2, $\delta<1/4$ be
  a positive real, and $M$ be a positive integer.  Let $\liminf_{T
    \rightarrow \infty} \dens(\sigma^T) \leq \frac{1}{2} - \delta.$
  Let $\ell_1,\ell_2,\dots$ be such that for all $i \geq M$, we have
  that $\ell_i \in [(1-\delta)\cdot (2^{i+1}-1),(1+\delta)\cdot
  (2^{i+1}-1)]$.  Then there exists a sequence $k_2,k_3,\dots$ such
  that for infinitely many $i>M$, we have that $\ell_{i-1}+\delta 2^{i-2} \leq
  k_i \leq \ell_i$ and $\dens(\sigma_{\{\ell_{i-1}+1,\dots,k_i\}}) \leq \frac{1}{2} -
  \frac{\delta}{4}.$
\end{lemma}

\begin{proof}
  Let $\ell_i$ be as required. If there are infinitely many $i$ such
  that $\dens(\sigma_{\{\ell_{i-1}+1,\dots,\ell_i\}}) \leq \frac{1}{2}
  - \frac{\delta}{4}$ then set $k_i = \ell_{i+1}$ and the lemma
  follows by observing $(k_i - \ell_{i-1}) \geq (1-\delta)(2^{i+1}-1)-
  (1+\delta)(2^{i}-1) = (1-3\delta)2^i \geq \delta 2^{i-2}$, for $i>
  M$.  So assume that only for finitely many $i$,
  $\dens(\sigma_{\{\ell_{i-1}+1,\dots,\ell_i\}}) \leq \frac{1}{2} -
  \frac{\delta}{4}$.  Thus the following claim can be applied for
  arbitrary large $i_0$.

\begin{claim}
  Let $i_0\geq M$ be given. If for every $i\geq i_0$,
  $\dens(\sigma_{\{\ell_{i-1}+1,\dots,\ell_i\}}) > \frac{1}{2} -
  \frac{\delta}{4}$ then there exist $j>i_0$ and $k$ such that
  $\ell_{j-1}+\delta 2^{j-2} \leq k \leq \ell_j$ and
  $\dens(\sigma_{\{\ell_{j-1}+1,\dots,k\}}) \leq \frac{1}{2} -
  \delta$.
\end{claim}

We can use the claim to find $k_2,k_3,\dots$ inductively. Start with
large enough $i_0\geq M$ and set $k_i = \ell_i$ for all $i\leq i_0$.
Then provided that we already inductively determined
$k_2,k_3,\dots,k_{i_0}$, we apply the above claim to obtain $j$ and
$k$, and we set $k_j=k$ and $k_i=\ell_i$, for all $i=i_0+1,\dots,j-1$.

So it suffices to prove the claim. For any $d\geq 1$, $\ell_{i_0}
\cdot 2^{d-1} \leq \ell_{i_0+d}$ and 
\[
\dens(\sigma^{\ell_{i_0+d}}) \geq \frac{(\frac{1}{2} -
  \frac{\delta}{4})(\ell_{i_0+d} - \ell_{i_0})}{\ell_{i_0+d}} \enspace .
\]
Furthermore, if $d \geq 1+\log(4/\delta)$ then $\ell_{i_0} \leq
\frac{\delta}{4} \ell_{i_0+d}$ and
\[
\dens(\sigma^{\ell_{i_0+d}}) \geq \left(\frac{1}{2} - \frac{\delta}{4}\right) - \frac{\delta}{4} = \frac{1}{2} - \frac{\delta}{2} \enspace .
\]
Since $\liminf_{k \rightarrow \infty} \dens(\sigma^k) \leq \frac{1}{2}
- \delta$, there must be $k$ and $d\geq 1+\log(4/\delta)$ such that
$\ell_{i_0+d-1} \leq k \leq \ell_{i_0+d}$ and $\dens(\sigma^k) \leq
\frac{1}{2} - \delta$. Set $j=i_0+d$. Also
\[
 \dens(\sigma^k) = \frac{ \dens(\sigma^{\ell_{j-1}}) \ell_{j-1} + \dens(\sigma_{\{\ell_{j-1}+1,\dots,k\}}) (k-\ell_{j-1}) }{\ell_{j-1} + (k-\ell_{j-1})} \enspace ,
\]
which means
\[
\begin{split}
  \left( \dens(\sigma^k) -
    \dens(\sigma_{\{\ell_{j-1}+1,\dots,k\}})\right) (k-\ell_{j-1})
  &= \left( \dens(\sigma^{j-1}) - \dens(\sigma^k) \right) \ell_{j-1} \\
  \geq \left[\left(\frac{1}{2} - \frac{\delta}{2}\right) -
    \left(\frac{1}{2} - \delta\right) \right] \ell_{j-1} &=
  \frac{\delta}{2} \ell_{j-1} \enspace .
\end{split}
\]
Thus $\dens(\sigma_{\{\ell_{j-1}+1,\dots,k\}}) \leq \dens(\sigma^k)$
which in turn is less than $\frac{1}{2} - \delta$.  Furthermore,
$k-\ell_{j-1} \geq \frac{\delta}{2} \ell_{j-1} \geq \frac{\delta}{2}
(1-\delta) ( 2^{j} - 1 ) \geq \frac{\delta}{4} 2^{j}$, provided that
$j\geq 2$.  Hence, $k$ and $j$ have the desired properties.
\end{proof}

We are now ready to prove the $\liminf$ version of \cref{l-stopping-sup}.  
\begin{lemma}
\label{l-stopping-inf}
Let $\sigma$ be an arbitrary pure Markov strategy
for Player~2. If
\[
\liminf_{t\rightarrow \infty} \dens(\sigma_1,\dots,\sigma_t)< 1/2 \enspace ,
\]
then when played against $\sigma_1^*$ the play stops with
probability~1.
\end{lemma}
\begin{proof}
Let $0<\delta<\frac{1}{4}$ be such that \[
\liminf_{t\rightarrow \infty} \dens(\sigma_1,\dots,\sigma_t)\leq 1/2-\delta
\]
  Pick arbitrary $\gamma\in (0,1)$. We will show that with probability
  at least $1-\gamma$ the game stops, and this implies the
  statement. Let $M$ be given by \cref{l-goodbalance} applied for
  $\gamma$ and $\delta/2$.  Then we have that with probability at
  least $1-\gamma$, for all $i\geq M$ and $j \in\{1,\dots,i\}$,
\[
t(i,j,i^2)-t(i,j-1,i^2) \in [(1-\frac{\delta}{2})
  2^i/i,(1+\frac{\delta}{2}) 2^i/i] \enspace .
\]

  Pick $t_{i,j} \in \NN$, for $i=1,2,\dots$ and $j\in\{1,\dots,i\}$, so
  that $t_{i,j-1} < t_{i,j}$ where $t_{i,0}$ stands for
  $t_{i-1,i-1}$. Let $t_{i,j}-t_{i,j-1} \in [(1-\frac{\delta}{2})
  2^i/i,(1+\frac{\delta}{2}) 2^i/i]$, for all $i\geq M$ and
  $j\in\{1,\dots,i\}$.  Pick $M'$ so that $\frac{\delta}{2}(2^{M'}-1)
  \geq \max\{t_{M,0}, (1-\frac{\delta}{2}) 2^{M}\}$. Define $\ell_i =
  t_{i,i}$ for all $i\geq 1$.  Then for all $i\geq M'$, $\ell_i \in
  [(1-\delta)\cdot (2^{i+1}-1),(1+\delta)\cdot (2^{i+1}-1)]$ as
\begin{eqnarray*}
  \ell_i = t_{i,i} &=& t_{M,0} + \sum_{M \leq i' \leq i, 1\leq j\leq i'} t_{i',j}-t_{i',j-1} \\
  &\le& \frac{\delta}{2} (2^{M'} - 1) +  \sum_{M \leq i' \leq i} i' \cdot (1+\frac{\delta}{2}) 2^{i'}/{i'} \\
  &\le& \frac{\delta}{2} (2^{M'} - 1) + (1 + \frac{\delta}{2})\cdot  ( 2^{i+1}-1) \\
  &\le& (1 + \delta)\cdot  ( 2^{i+1}-1),
\end{eqnarray*}
and similarly for the lower bound: $ \ell_i \geq \sum_{i', j}
t_{i',j}-t_{i',j-1} \geq (1-\frac{\delta}{2}) (2^{i+1} - 2^{M}) \geq
(1 - \delta)\cdot ( 2^{i+1}-1).$ Thus \cref{l-subdivision} is
applicable on $\ell_i$ with $M$ set to $M'$, and we obtain a sequence
$k_2,k_3,\dots$ such that $\dens(\sigma_{\ell_{i-1}+1,\dots,k_i}) \leq
\frac{1}{2} - \frac{\delta}{4}$ and $k_i - \ell_{i-1} \geq \delta
2^{i-2}$ for infinitely many $i$.  Pick any of the infinitely many $i
\geq \max\{M', 32(1+\delta)/\delta\}$ for which $k_i - \ell_{i-1} \geq
\delta 2^{i-2}$ and $\dens(\sigma_{\ell_{i-1}+1,\dots,k_i}) \leq
\frac{1}{2} - \frac{\delta}{4}$. Since $\delta 2^{i-3} \geq
(1+\delta)2^i/i$, there is some $j\in\{1,\dots,i\}$ such that
$\ell_{i-1} + \delta 2^{i-3} \leq k_i - (1+\delta)2^i/i \leq t_{i,j}
\leq k_i$. Fix such $j$. Since $k_i \leq t_{i,j} + (1+\delta)2^i/i$,
we have
\begin{eqnarray*}
  \dens(\sigma_{\ell_{i-1}+1,\dots,t_{i,j}}) &=& \frac{ \dens(\sigma_{\ell_{i-1}+1,\dots,k_i}) (k_i-\ell_{i-1}) }{t_{i,j} + \ell_{i-1}} \\
  &\le& \frac{ \dens(\sigma_{\ell_{i-1}+1,\dots,k_i}) ( (1+\delta)2^i/i + t_{i,j} -\ell_{i-1}) }{t_{i,j} + \ell_{i-1}} \\
  &\le& \left( \frac{1}{2} - \frac{\delta}{4} \right) \cdot \left(1+\frac{8(1+\delta)}{i} \right)\\
  &\le& \frac{1}{2} - \frac{\delta}{4} + \frac{4(1+\delta)}{i} \leq \frac{1}{2} - \frac{\delta}{8}.
\end{eqnarray*}
Hence, $\dens(\sigma_{\ell_{i-1}+1,\dots,t_{i,j}}) \leq \frac{1}{2} -
\frac{\delta}{8}$. So for some $j' \in \{1,\dots,j\}$,
$\dens(\sigma_{t_{i,j'-1}+1,\dots,t_{i,j'}}) \leq \frac{1}{2} -
\frac{\delta}{8}$. We can state the following claim.
\begin{claim}
  For $i$ large enough, conditioned on $t(a,b,a^2) = t_{a,b}$, for all
  $a\geq M$ and all $b$, and conditioned on that the game did not stop
  before the time $t_{i,j'-1}+1$, the game stops during times
  $t_{i,j'-1}+1,\dots,t_{i,j'}$ with probability at least $\xi^4/2$.
\end{claim}
Conditioned on $t(a,b,a^2) = t_{a,b}$, for all $a,b$, the claim
implies that the game stops with probability 1. Note that the
condition is true for some valid choice of $t_{a,b}$ with probability
$1-\gamma$.  This is because the claim can be invoked for infinitely
many $i$'s and for each such $i$ we will have $\xi^4 /2$ chance
of stopping.

It remains to prove the claim. Assume $t(i,j'-1,i^2) = t_{i,j'-1}$ and
$t(i,j',i^2) = t_{i,j'}$. Clearly, \[\dens(\sigma_{t_{i,j'-1}+1,
  \dots, t_{i,j'}-1}) \leq \dens(\sigma_{t_{i,j'-1}+1, \dots,
  t_{i,j'}}) \cdot (2^{i-1}/(2^{i-1}-1)) \leq
\frac{1}{2}-\frac{\delta}{16}\enspace ,\] for $i$ large enough.  So if
we sample $i^2-1$ times from $\sigma_{t_{i,j'-1}+1}, \dots,
\sigma_{t_{i,j'}-1}$ we expect at most $(\frac{1}{2} -
\frac{\delta}{16})(i^2-1)$ of the letters to be $\lft$. By Hoeffding's
inequality, \cref{THM:Hoeffding} (again setting $a_i=0$ and $b_i=1$,
and letting $c_i$ be $i$'th payoff for Player~1), the probability that
we get at least $\frac{\delta}{32}(i^2-1)$ additional $\lft$ items on
top of the expectation is given by
\[
\begin{split}
\Pr\left[\dens(\sigma_{t(i,j',1)},\sigma_{t(i,j',2)},\dots,\sigma_{t(i,j',i^2-1)})
  \geq \frac{1}{2} - \frac{\delta}{32}\right] \\ \leq 2 \exp\left(-\frac{2(\frac{\delta}{32}(i^2-1))^2}{i^2-1}\right) = 
 2 \exp\left( -\frac{\delta^2}{512}(i^2-1)\right) \enspace .
\end{split}
\]
The probability is taken over the possible choices of
$t(i,j',1)<t(i,j',2)<\cdots<t(i,j',i^2-1)$ assuming $t(i,j'-1,i^2) =
t_{i,j'-1}$ and $t(i,j',i^2) = t_{i,j'}$.  For $i$ sufficiently large,
$2e^{-\delta^2(i^2-1)/512}\leq 1/2$. Whenever
\[\dens(\sigma_{t(i,j',1)},\sigma_{t(i,j',2)},\dots,\sigma_{t(i,j',i^2-1)}) \leq \frac{1}{2} - \frac{\delta}{32}\] we have at least $\xi^4$ chance of stopping
by \cref{l-base}, as Player~1 plays $\sigma_1^{i+j',\xi}$
against \[\sigma_{t(i,j',1)},\sigma_{t(i,j',2)},\dots,\sigma_{t(i,j',i^2-1)}\]
and $i^2-1 > i/\delta \geq (i+j')/2\delta$ for sufficiently large $i$.
Hence, the game stops with probability at least $(1-1/2) \cdot \xi^4 =
\xi^4/2$. The claim, and thus the lemma, follows.
\end{proof}

We can now conclude with the main result of this section.
\begin{theorem}\label{thm:inf_eps}
  The strategy $\sigma_1^*$ is $\sqrt{\xi}$-optimal, and for all
  $\delta>0$, with probability at least $1-\delta$ it uses space
  $O(\log \log T)$.
\end{theorem}
\begin{proof}
  This is proved just like \cref{thm:sup_eps}, except that
  \cref{l-stopping-inf} is used in place of \cref{l-stopping-sup}.
\end{proof}
We can also improve the strategy and get the following corollary.

 \begin{corollary}\label{cor:tm}
 For any natural number $k$, there is a strategy which is  $2^{-k}$-optimal, has patience 2 and can be implemented on a Turing machine, using at most 1 random bit and amortized constant time\footnote{i.e. for all $T$, let $c(T)$ be the computation used for the first $T$ rounds, then $\frac{c(T)}{T}$ is some constant} per round and with probability at least $1-\delta$ does it use
  tape space $O(\log \log (T)+\log k)$ upto round $T$.  
 \end{corollary}
\begin{proof}
We will only sketch the necessary ingredients for the proof.
We will modify the strategy  $\sigma_1^{*}$ as defined in \cref{sec:eps-inf} to use less patience.
First, it is easy to see that the probability that two sample points
are ever within a polynomial in $i$ distance of each other is small,
in any sufficiently high epoch $i$, and thus we can ignore this
event.
The idea is as follows: Each sub-epoch of epoch $i$ is split into blocks of  length $B=i^{O(1)}\cdot k$ (where we simply keep track of the current location within the current block, but not the current block number). 
 The purpose of each block is to provide all the randomness needed for the next block. Hence, it needs to (1)~decide if there is a sample point in the next block; and (2)~if so where; and (3)~if we should stop on that sample point. Observe that we are ignoring the possibility that there are more than 1 sample point in a block.
 It is easy to see that the number of random bits needed in total to generate all those events is less than $i^{O(1)}\cdot k$ and thus it can be done with at most 1 random bit per round, implying the patience bound.
 To get the tape space bound, we utilize the three following simple ideas: 
 \begin{enumerate}
 \item[\bf Idea 1:] To get a probability like $\prod a_j$, for some sequence of probabilities $a_j$ of length $\ell$, one can test if sub-events that happens with probability $a_i$ for all $i$ all happens. This requires only space for a counter counting up to $\ell$ and space for the event that uses the most space (by reusing the space for the event). 
 \item[\bf Idea 2:] To get an probability like $2^{-x}$ (or similarly $1-2^{-x}$) one can simply flip $x$ coins and if all comes up tails, then the event happens. This requires only space for a counter counting up to $x$.
 \item[\bf Idea 3:] To get an probability like $\frac{y}{2^x}$, for any natural number $x$ and $y$, one can simply use $x$ random bits. This uses $x$ many bits of space.
 \end{enumerate}
Direct application of these three ideas suffice to get all three events in $O(\log i+\log k)$ many bits. Observe that we can do this easily on a two tape Turing machine (one is used for events following idea 3 and one for the constant number of counters) and we only need amortized constant time in each round (to increment and/or reset some subset of the counters on the first tape and perhaps to add one more random bit or reset the second tape). 
\end{proof}

\section{Lower bound on patience}

\newcommand{\pat}{{\mathrm{pat}}}

When considering a strategy of a player one may want to look at how small or large the probabilities 
occurring in that strategy are. The parameter of interest is the {\em patience} of the strategy which is the reciprocal
of the smallest non-zero  probability occurring in the strategy. Patience is closely related to the expected
length of finite plays as small probability events will not occur if the play is too short so they will
have little influence on the overall outcome \cite{AS:Kohlberg1974,HKM-LICS}. Care has to be taken how to define
patience for strategies with infinitely many possible events.
One thing to note of our space efficient strategies is that the patience of the states
in which we are with high probability during the first $T$ steps is approximately $T$, for rounds $T$ close to the end of an epoch. 
In this section  we show  that this is essentially necessary. So if the space used
by the strategy with high probability is $\log f(T)$, then the first $f(T)$ states must have
patience about $T$. Thus the smaller the space the strategy uses the larger the patience the states must have.

The main theorem of this section states that if the patience of the $f(T)$ states in which Player 1 is with
high probability is less than about $T^{1/f(T)}$ then the strategy is bad for Player 1. It is easy to observe
that events with probability substantially less than $1/T$ are unlikely to occur during the first $T$ steps
of the game so one should expect the patience of a good strategy for Player 1 to be in the range
between roughly  $T^{1/f(T)}$ and $T$.

One may wonder whether the exponent $1/f(T)$ in $T^{1/f(T)}$ is
necessary. It turns out that it is, so our lower bound is close to
optimal using a technique like in Corollary~\ref{cor:tm}.

We use the following definitions to deal with the fact that our
strategies use infinitely many transitions so their overall patience
is infinite. 

For a memory based strategy $\sigma_1$ of Player~1 in a repeated
zero-sum game with absorbing states, the \emph{patience} of a set of
memory states $M$ is defined as:
\[
\pat(M)=\max
\left\{\frac{1}{\sigma_1^a(m)(a_1)},\frac{1}{\sigma_1^u(a_1,a_2,m)(m')},\;m,m'\in
  M, a_1\in A_1,a_2\in A_2\right\} \enspace .
\]
The patience of the other player is defined similarly.

\begin{theorem}
  Let $\delta,\eps>0$ be reals and $f:\NN\rightarrow \NN$ be an
  unbounded non-decreasing function such that $f(T) \le \frac{1}{4}
  \log_{1/\eps} T$ for all large enough $T$.  If a strategy $\sigma_1$
  of Player~1 uses space $\log f(T)$ before time
  $T$ with probability at least $1-\delta$, and the patience of the
  set of lexicographically first $f(T)$ memory states is at most $T^{1/(2f(T))}$ for all $T$ large enough,
  then there is a strategy $\sigma_2$ of Player~2 such that
  $u(\sup,(\sigma_1,\sigma_2)) \le \delta+2\eps$.
\end{theorem}

\begin{proof}
  Assume that $\eps<1/2$ and pick an integer $k$ sufficiently large.
  For $i>0$, define $\ell_i = \eps^{-i}$ and $T_i = \sum_{j=1}^{i}
  \ell_j$.  The strategy $\sigma_2$ of Player~2 proceeds in phases,
  each phase $i$ is of length $\ell_i$.  In the first $k$ phases,
  Player~2 plays $\lef$ with probability $1-\eps$ and $\rig$ with
  probability $\eps$. In each phase $i>k$, Player~2 plays $\rig$ for
  the first $(1-\eps)\ell_i$ steps, and afterwards he plays $\lef$
  with probability $1-\eps$ and $\rig$ with probability
  $\eps$. Notice, if the game does not stop by Player~1 playing $\rig$
  at some point then the expected $\limsup$ payoff to Player~1 is at
  most $\eps$.

  Our goal is to show that if the game stops then the expected payoff
  to Player~1 is at most $\delta+2\eps$.  If the game stops during the
  last $\eps \ell_i$ steps of a phase $i$, then the expected payoff to
  Player~1 is $\eps$ as the probability of Player~2 playing $\rig$ at
  that time is $\eps$. If the game stops during the first
  $(1-\eps)\ell_i$ steps of a phase $i>k$, then the payoff of Player~1
  is 1.  Our goal is to argue that the overall probability that
  Player~1 stops during the first $(1-\eps)\ell_i$ steps of some phase
  $i>k$ is small.

  For any $t>0$, denote by $M(t)$ the set of the lexicographically
  first $f(t)$ memory states (i.e. those mapped to a number below $f(t)$.  Let $C$ be the
  event that for all steps $t$, Player~1 is in one of the states in
  $M(t)$.  For $t<t'$, let $S(t,t')$ be the event that Player~1 plays
  $\rig$ in one of the steps $[t,t')$. Let $A_i(t)$ be the event that
  at time $t$, Player~1 is in one of the states in $M(T_i)$ and there
  is some memory state in $M(T_i)$ that can be reached from the state
  current at time $t$ and in which there is a non-zero probability of
  playing $\rig$ (i.e., stopping).

  The probability that $C$ does not occur is at most $\delta$ so for
  the rest of the proof we will assume that $C$ occurs.  Let $k$ be
  large enough, and $i>k$.  It is clear that if
  $S(T_{i-1},T_{i-1}+(1-\eps)\ell_i)$ occurs then $A_i(T_{i-1})$ must
  have occurred as well so:
\[
\Pr[S(T_{i-1},T_{i-1}+(1-\eps)\ell_i)] \le \Pr[A_i(T_{i-1})] \enspace
.
\]
Furthermore, if $A_i(T_{i-1})$ occurs then $A_i(t)$ occurs for all
$t<T_{i-1}$. For $t\in [T_{i-1}-\eps \ell_{i-1},T_{i-1}-f(T_i))$, if
$A_i(t)$ occurs then within the next $f(T_i)$ steps the strategy of
Player~1 might reach a state in which Player~1 chooses the stopping
action $\rig$ with non-zero probability. Because of the patience of
$M(T_i)$ and the fact that Player~2 plays each of his possible actions
with probability at least $\eps$ during that time steps we have for
any $t\in [T_{i-1}-\eps \ell_{i-1},T_{i-1}-f(T_i))$,
\[
\Pr[S(t,t+f(T_i)) \;|\;A_i(t)] \ge
\left(\frac{\eps}{\pat(M(T_i))}\right)^{f(T_i)} \enspace .
\]
Hence, the probability that $A_i(T_{i-1})$ occurs and the game did not
stop yet is at most:
\[
\left(1- \left(\frac{\eps}{\pat(M(T_i))}\right)^{f(T_i)}\right)^{\eps
  \ell_{i-1}/f(T_i)} \le
{\mathrm{e}}^{-\left(\frac{\eps}{\pat(M(T_i))}\right)^{f(T_i)} \cdot
  \eps \ell_{i-1}/f(T_i)}.
\]
For sufficiently large $T$, $f(T) \le \frac{1}{4} \log_{1/\eps}
T$. Furthermore, $T_i \le 2 \ell_i$ and $\eps T_i - 1 \le T_{i-1}$.
Since $i$ is sufficiently large we have:
\begin{eqnarray*}
      \left(\frac{\eps}{\pat(M(T_i))}\right)^{f(T_i)} \cdot \eps \cdot \frac{\ell_{i-1}}{f(T_i)} &\ge&
      \left(\frac{\eps}{T_i^{\frac{1}{2 f(T_i)}}}\right)^{f(T_i)} \cdot \eps \cdot \frac{T_{i-1}}{2 f(T_i) } \cr
&\ge& \eps^{f(T_i)} \cdot \eps^2 \cdot \frac{T_{i}^{1/2}}{2 f(T_i)} -1 \cr
&\ge& \eps^2 \cdot \frac{T_{i}^{1/4}}{2 f(T_i)} -1 \ge T_i^{1/5}.
\end{eqnarray*}
Since $T_i^{1/5} \gg i$ for $i$ large enough, we get 
\[
\Pr[S(T_{i-1},T_{i-1}+(1-\eps)\ell_i)] \le \eps \cdot {\mathrm{e}}^{-i}.
\]
We set $k$ to be large enough so that the above analysis would work
for $i>k$.  Thus except for probability at most $\delta + \eps$,
Player~1 stops in a step when Player~2 plays $\rig$ with probability
only $\eps$.  Thus the expected payoff to Player~1 is at most $\delta
+ 2\eps$.
\end{proof}

%%%%%%%%%%%%%%%%%%%%%%%%%%%%%%%%%%%%%%%%%%%%%%%%%%%%%%%%%%%%%%%%%%%%%%%%%%%%%%%%

\section{No finite-memory \texorpdfstring{$\eps$}{epsilon}-optimal
  deterministic-update Markov strategy exists}
A {\em memory-based Markov strategy} is an extension of a memory-based
strategy that may also depend on the round number.  More precisely,
for Player~1, the action map $\sigma_1^a$ for a memory-based Markov
strategy $\sigma_1$ is a map from $\ZZ_{+}\times \calM$ to
$\Dist(A_1)$ and the update map $\sigma_1^u$ for memory-based Markov
strategies is a map from $\ZZ_{+}\times A_1\times A_2\times\calM$ to
$\Dist(\calM)$.  We say that Player~1 \emph{follows} the memory-based
Markov strategy $\sigma_1$ if in every round $T$ when the game did not
stop yet, he picks his next move $a_1^T$ at random according to
$\sigma_1^a(T,m_T)$, where the sequence $m_1,m_2,\dots$ is given by
letting $m_1=m_s$ and for $T=1,2,3,\dots$ choosing $m_{T+1}$ at random
according to $\sigma_k^u(a_1^T,a_2^T,m_T,T)$, where $a_2^T$ is the
action chosen by Player~2 at round $T$. The definition of memory-based
Markov strategies is similar for Player~2.  Note that memory-based
Markov strategies are more general than memory-based strategies.

A memory-based (resp. Markov) strategy $\sigma_1$ for Player~1 has
{\em deterministic-update}, if for all $a_1\in A_1$, all $a_2\in A_2$
and all $m\in \calM$ (resp. all $T\in Z_{+}$) the distribution
$\sigma_1^u(a_1,a_2,m)$ (resp. $\sigma_1^u(T,a_1,a_2,m)$) is
deterministic.

In this section we will argue that no finite-memory $\eps$-optimal
deterministic-update Markov strategy for Player~1 in the Big Match
exists, for $\eps<\frac{1}{2}$.  Let $n$ be some positive integer. Let
$\sigma_1$ be some Markov strategy using at most $n$ memory for
Player~1. We will show that for all $\delta>0$, there exists a
strategy $\sigma_2$ for Player~2 that ensures that
$u(\inf,(\sigma_1,\sigma_2))<\delta$. This shows that $\sigma_1$ can
only ensure value~0.

\paragraph*{Construction of $\sigma_2$ and the
  sequence of strategies $\sigma_2^k$.} We will now describe the
construction of $\sigma_2$. The strategy $\sigma_2$ will be the final
strategy in a finite sequence of strategies $(\sigma_2^k)_{k}$. Each
of the strategies $\sigma_2^k$ (and thus also $\sigma_2$) is a
deterministic memory-based Markov strategy and will use the same set
of memory states $\calM$ of size $n$ and update map $\sigma_1^u$ as
$\sigma_1$. Observe that the action map $\sigma_2^a$ for such a
strategy can be thought of as a $(\infty,n)$-matrix $A$ over
$\{\lef,\rig\}$, where $A_{T,m}=\sigma_2^a(T,m)$.
\begin{itemize}
\item Let $\sigma_2^{k,a}$ be the action map for $\sigma_2$.
\item Let $S_{\rig}^k=\{(T,m)\mid \sigma_2^{k,a}(T,m)=\rig\}$ (i.e. the pairs under which $\sigma_2^k$ plays $\rig$).
\item For all $T$, let $\calM^T=\{(T',m)\mid T'\leq T\}$ (i.e. the memory states before round $T$).
\item For all $T$, let 
 $S_{\rig}^{k,T}=S_\rig^k\cap \calM^T$ (i.e. the pairs under which $\sigma_2^k$ plays $\rig$ before round $T$).
\end{itemize}

\paragraph*{Properties of strategies in the sequence $\sigma_2^k$.}
Besides ensuring that the last strategy $\sigma_2$ in the sequence is
such that $u(\inf,(\sigma_1,\sigma_2))<\delta$, our construction of
$\sigma_2^k$ will ensure the following properties:
\begin{enumerate}
\item {\bfseries Property~1.} The probability to stop (using union
  bound) while Player~2 plays $\rig$ is at most \[\sum_{(T,m)\in
    S_{\rig}^{k}}\sigma_1^a(T,m)(\rig)\leq (1-2^{-k})\delta\enspace
  .\]\label{pro:low stop}
\item {\bfseries Property~2.} The infimum limit, for $T$ going to
  infinite, of the fraction of all pairs before round $T$ under which
  $\sigma_2^k$ plays $\rig$ is at
  least \[\liminf_{t\rightarrow\infty}\frac{|S_\rig^{k,T}|}{n\cdot
    T}\geq \frac{\delta\cdot k}{n}\enspace .\]\label{pro:fraction
    pairs}
\end{enumerate}

\paragraph*{The sequence has finite length.} Observe that Property~2
ensures that the strategy $\sigma_2^{k}$ cannot exists, for $k>
\frac{n}{\delta}$, implying that the sequence has finite length. This
is because $\sigma_2^k$, for such $k$, otherwise would require that
there is some $T$, such that the number of pairs such that
$\sigma_2^k$ plays $\rig$ before round $T$ is strictly more than the
number of pairs before round $T$.

\paragraph*{The strategy $\sigma_2^0$.} The action map
$\sigma_2^{0,a}$ is such that $\sigma_2^{0,a}(T,m)=\lef$ for all $T\in
\NN$ and $m\in \calM$. The strategy has the wanted properties (because
$S_{\rig}^0$ is the empty set).

\paragraph*{Informal inductive construction of the sequence of
  strategies $(\sigma_2^k)_k$.}
Consider playing $\sigma_1$ against $\sigma_2^k$ and let the play be
$P$.  If the outcome is above $\delta$ (that is, $\sigma_2^k$ does not
satisfy the properties of $\sigma_2$ and we therefore need to
construct $\sigma_2^{k+1}$) we know two things: 1) The probability
that $P$ stops is not~1, since otherwise the outcome is at most
$(1-2^{-k})\delta$, by Property~1. 2) Player~2 plays $\lef$ above a
$\delta$ fraction of the time (in lim-inf).  In that case, since the
probability to stop is below~1, there is a round $M$, such that if the
play reaches that round, the probability to stop in the rest of the
play is below $2^{-(k+1)}\delta$. Let the sequence of memory states
the players uses in the play be $(m_T^k)_T$.  The strategy
$\sigma_2^{k+1}$ is then like the strategy $\sigma_2^k$, except for
playing $\rig$ then in memory $m_T^k$ in round $T$. By choice of $M$,
we get Property~1. Also, since the fraction of the time player~2
played $\lef$ in $P$ was above $\delta$, then $\sigma_2^{k+1}$
contains $\frac{\delta}{n}$ more $\rig$'s than $\sigma_2^k$, which is
Property~2.  In \cref{fig:seq_sigma2}, we have an illustration of the
action map of a possible sequence of strategies $(\sigma_2^k)_k$ for
$n=4$ and some strategy $\sigma_1$.  The line illustrates the memory
sequence $(m_T^k)_T$ (for instance, the sequence $(m_T^1)_T$ is
4132114442333313131...) where the probability to stop somewhere on the
solid line is below $2^{-(k+1)}\delta$.
  
\def\vx{18}

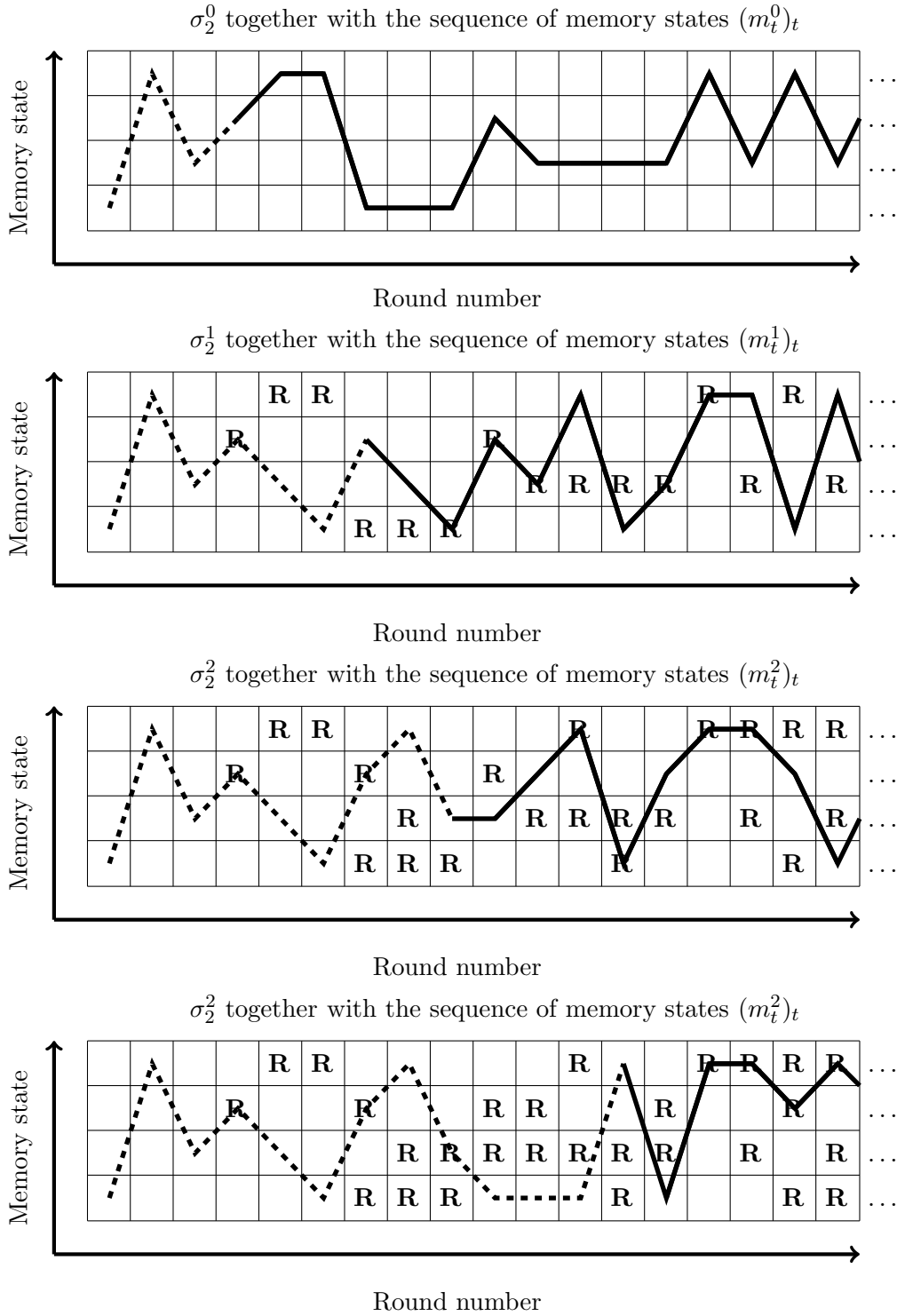
\begin{figure}
\begin{tikzpicture}[node distance=3cm]

  \matrix (v1) [label=above:$\sigma_2^0$ together with the sequence of
  memory states $(m_t^0)_t$,minimum height=1.5em,minimum width=1.5em,
  anchor=center,text depth=.5ex,text height=2ex,text width=1em, matrix
  of math nodes,nodes in empty cells, left delimiter={.},right
  delimiter={.}]  {
    &&&&&&&&&&&&&&&&&&\dots\\
    &&&&&&&&&&&&&&&&&&\dots\\
    &&&&&&&&&&&&&&&&&&\dots\\
    &&&&&&&&&&&&&&&&&&\dots\\
  }; \foreach \y in {1,...,4} { \draw[black] (v1-\y-1.north west) --
    (v1-\y-\vx.north east); } \draw[black] (v1-4-1.south west) --
  (v1-4-\vx.south east); \foreach \x in {1,...,\vx} { \draw[black]
    (v1-1-\x.north west) -- (v1-4-\x.south west); } \draw[black]
  (v1-1-\vx.north east) -- (v1-4-\vx.south east); \draw[->,ultra
  thick] ($(v1-4-1.south west) + (-0.5,-0.5)$) -- ($(v1-1-1.north
  west)+(-0.5,0)$); \draw[->,ultra thick] ($(v1-4-1.south west) +
  (-0.5,-0.5)$) -- ($(v1-4-\vx.south east)+(0,-0.5)$);

\path let \p1 = ($(v1-4-1.south west) + (-0.5,-0.5)$), 
\p2=($(v1-1-1.north west)+(-0.5,0)$) in node[anchor=center,rotate=90] (state) at ($(\p1)!0.5!(\p2)+(-0.5,0)$) {Memory state};

\path let \p1 = ($(v1-4-1.south west) + (-0.5,-0.5)$), 
\p2=($(v1-4-\vx.south east)+(0,-0.5)$) in node (round) at ($(\p1)!0.5!(\p2)+(0,-0.5)$) {Round number};

\draw[dashed,black,line width=2pt] (v1-4-1.center) -- (v1-1-2.center) -- (v1-3-3.center) -- (v1-2-4.center);
\xdef\llx{4}
\xdef\lly{2}
\xdef\lx{5}
\xdef\ly{1}
\foreach \x/\y in {6/1,7/4,8/4,9/4,10/2,11/3,12/3,13/3,14/3,15/1,16/3,17/1,18/3} {
\draw[black,line width=2pt] (v1-\lly-\llx.center) -- (v1-\ly-\lx.center)-- (v1-\y-\x.center);
\xdef\llx{\lx}
\xdef\lly{\ly}
\xdef\lx{\x}
\xdef\ly{\y}
}
\draw[black,line width=2pt] (v1-\lly-\llx.center) -- (v1-\ly-\lx.center)-- (v1-2-18.east);

\end{tikzpicture}
\begin{tikzpicture}[node distance=3cm]

  \matrix (v1) [label=above:$\sigma_2^1$ together with the sequence of
  memory states $(m_t^1)_t$,minimum height=1.5em,minimum width=1.5em,
  anchor=center,text depth=.5ex,text height=2ex,text width=1em, matrix
  of math nodes,nodes in empty cells, left delimiter={.},right
  delimiter={.}]  {
    &&&&\rig&\rig&&&&&&&&&\rig&&\rig&&\dots\\
    &&&\rig&&&&&&\rig&&&&&&&&&\dots\\
    &&&&&&&&&&\rig&\rig&\rig&\rig&&\rig&&\rig&\dots\\
    &&&&&&\rig&\rig&\rig&&&&&&&&&&\dots\\}; \foreach \y in {1,...,4} {
    \draw[black] (v1-\y-1.north west) -- (v1-\y-\vx.north east); }
  \draw[black] (v1-4-1.south west) -- (v1-4-\vx.south east); \foreach
  \x in {1,...,\vx} { \draw[black] (v1-1-\x.north west) --
    (v1-4-\x.south west); } \draw[black] (v1-1-\vx.north east) --
  (v1-4-\vx.south east); \draw[->,ultra thick] ($(v1-4-1.south west) +
  (-0.5,-0.5)$) -- ($(v1-1-1.north west)+(-0.5,0)$); \draw[->,ultra
  thick] ($(v1-4-1.south west) + (-0.5,-0.5)$) -- ($(v1-4-\vx.south
  east)+(0,-0.5)$);

  \path let \p1 = ($(v1-4-1.south west) + (-0.5,-0.5)$),
  \p2=($(v1-1-1.north west)+(-0.5,0)$) in
  node[anchor=center,rotate=90] (state) at
  ($(\p1)!0.5!(\p2)+(-0.5,0)$) {Memory state};

  \path let \p1 = ($(v1-4-1.south west) + (-0.5,-0.5)$),
  \p2=($(v1-4-\vx.south east)+(0,-0.5)$) in node (round) at
  ($(\p1)!0.5!(\p2)+(0,-0.7)$) {Round number};

  \draw[dashed,black,line width=2pt] (v1-4-1.center) --
  (v1-1-2.center) -- (v1-3-3.center) -- (v1-2-4.center)--
  (v1-3-5.center)-- (v1-4-6.center)-- (v1-2-7.center); \xdef\llx{7}
  \xdef\lly{2} \xdef\lx{8} \xdef\ly{3} \foreach \x/\y in
  {9/4,10/2,11/3,12/1,13/4,14/3,15/1,16/1,17/4,18/1} {
    \draw[black,line width=2pt] (v1-\lly-\llx.center) --
    (v1-\ly-\lx.center)-- (v1-\y-\x.center); \xdef\llx{\lx}
    \xdef\lly{\ly} \xdef\lx{\x} \xdef\ly{\y} } \draw[black,line
  width=2pt] (v1-\lly-\llx.center) -- (v1-\ly-\lx.center)--
  (v1-3-18.north east);

\end{tikzpicture}

\begin{tikzpicture}[node distance=3cm]

  \matrix (v1) [label=above:$\sigma_2^2$ together with the sequence of
  memory states $(m_t^2)_t$,minimum height=1.5em,minimum width=1.5em,
  anchor=center,text depth=.5ex,text height=2ex,text width=1em, matrix
  of math nodes,nodes in empty cells, left delimiter={.},right
  delimiter={.}]  {
    &&&&\rig&\rig&&&&&&\rig&&&\rig&\rig&\rig&\rig&\dots\\
    &&&\rig&&&\rig&&&\rig&&&&&&&&&\dots\\
    &&&&&&&\rig&&&\rig&\rig&\rig&\rig&&\rig&&\rig&\dots\\
    &&&&&&\rig&\rig&\rig&&&&\rig&&&&\rig&&\dots\\
  }; \foreach \y in {1,...,4} { \draw[black] (v1-\y-1.north west) --
    (v1-\y-\vx.north east); } \draw[black] (v1-4-1.south west) --
  (v1-4-\vx.south east); \foreach \x in {1,...,\vx} { \draw[black]
    (v1-1-\x.north west) -- (v1-4-\x.south west); } \draw[black]
  (v1-1-\vx.north east) -- (v1-4-\vx.south east); \draw[->,ultra
  thick] ($(v1-4-1.south west) + (-0.5,-0.5)$) -- ($(v1-1-1.north
  west)+(-0.5,0)$); \draw[->,ultra thick] ($(v1-4-1.south west) +
  (-0.5,-0.5)$) -- ($(v1-4-\vx.south east)+(0,-0.5)$);

  \path let \p1 = ($(v1-4-1.south west) + (-0.5,-0.5)$),
  \p2=($(v1-1-1.north west)+(-0.5,0)$) in
  node[anchor=center,rotate=90] (state) at
  ($(\p1)!0.5!(\p2)+(-0.5,0)$) {Memory state};

  \path let \p1 = ($(v1-4-1.south west) + (-0.5,-0.5)$),
  \p2=($(v1-4-\vx.south east)+(0,-0.5)$) in node (round) at
  ($(\p1)!0.5!(\p2)+(0,-0.7)$) {Round number};

  \draw[dashed,black,line width=2pt] (v1-4-1.center) --
  (v1-1-2.center) -- (v1-3-3.center) -- (v1-2-4.center)--
  (v1-3-5.center)-- (v1-4-6.center)-- (v1-2-7.center)--
  (v1-1-8.center)-- (v1-3-9.center); \xdef\llx{9} \xdef\lly{3}
  \xdef\lx{10} \xdef\ly{3} \foreach \x/\y in
  {11/2,12/1,13/4,14/2,15/1,16/1,17/2,18/4} { \draw[black,line
    width=2pt] (v1-\lly-\llx.center) -- (v1-\ly-\lx.center)--
    (v1-\y-\x.center); \xdef\llx{\lx} \xdef\lly{\ly} \xdef\lx{\x}
    \xdef\ly{\y} } \draw[black,line width=2pt] (v1-\lly-\llx.center)
  -- (v1-\ly-\lx.center)-- (v1-3-18.east);

\end{tikzpicture}

\begin{tikzpicture}[node distance=3cm]

  \matrix (v1) [label=above:$\sigma_2^2$ together with the sequence of
  memory states $(m_t^2)_t$,minimum height=1.5em,minimum width=1.5em,
  anchor=center,text depth=.5ex,text height=2ex,text width=1em, matrix
  of math nodes,nodes in empty cells, left delimiter={.},right
  delimiter={.}]  {
    &&&&\rig&\rig&&&&&&\rig&&&\rig&\rig&\rig&\rig&\dots\\
    &&&\rig&&&\rig&&&\rig&\rig&&&\rig&&&\rig&&\dots\\
    &&&&&&&\rig&\rig&\rig&\rig&\rig&\rig&\rig&&\rig&&\rig&\dots\\
    &&&&&&\rig&\rig&\rig&&&&\rig&&&&\rig&\rig&\dots\\
  }; \foreach \y in {1,...,4} { \draw[black] (v1-\y-1.north west) --
    (v1-\y-\vx.north east); } \draw[black] (v1-4-1.south west) --
  (v1-4-\vx.south east); \foreach \x in {1,...,\vx} { \draw[black]
    (v1-1-\x.north west) -- (v1-4-\x.south west); } \draw[black]
  (v1-1-\vx.north east) -- (v1-4-\vx.south east); \draw[->,ultra
  thick] ($(v1-4-1.south west) + (-0.5,-0.5)$) -- ($(v1-1-1.north
  west)+(-0.5,0)$); \draw[->,ultra thick] ($(v1-4-1.south west) +
  (-0.5,-0.5)$) -- ($(v1-4-\vx.south east)+(0,-0.5)$);

  \path let \p1 = ($(v1-4-1.south west) + (-0.5,-0.5)$),
  \p2=($(v1-1-1.north west)+(-0.5,0)$) in
  node[anchor=center,rotate=90] (state) at
  ($(\p1)!0.5!(\p2)+(-0.5,0)$) {Memory state};

  \path let \p1 = ($(v1-4-1.south west) + (-0.5,-0.5)$),
  \p2=($(v1-4-\vx.south east)+(0,-0.5)$) in node (round) at
  ($(\p1)!0.5!(\p2)+(0,-0.7)$) {Round number};

\draw[dashed,black,line width=2pt] (v1-4-1.center) -- (v1-1-2.center) -- (v1-3-3.center) -- (v1-2-4.center)-- (v1-3-5.center)-- (v1-4-6.center)-- (v1-2-7.center)-- (v1-1-8.center)-- (v1-3-9.center)-- (v1-4-10.center)-- (v1-4-11.center)-- (v1-4-12.center)-- (v1-1-13.center);
\xdef\llx{13}
\xdef\lly{1}
\xdef\lx{14}
\xdef\ly{4}
\foreach \x/\y in {15/1,16/1,17/2,18/1} {
\draw[black,line width=2pt] (v1-\lly-\llx.center) -- (v1-\ly-\lx.center)-- (v1-\y-\x.center);
\xdef\llx{\lx}
\xdef\lly{\ly}
\xdef\lx{\x}
\xdef\ly{\y}
}
\draw[black,line width=2pt] (v1-\lly-\llx.center) -- (v1-\ly-\lx.center)-- (v1-2-18.north east);

\end{tikzpicture}

\caption{Posible sequence of strategies $(\sigma_2^k)_k$ and
  corresponding sequence of memory states
  $(m_t^k)_t$.\label{fig:seq_sigma2}}
\end{figure}

\paragraph*{Formal inductive construction of $\sigma_2^{k+1}$.}
We will now show that, given $\sigma_2^{k}$, we either have that
$u(\inf,(\sigma_1,\sigma_2^k))<\delta$ or we can construct
$\sigma_2^{k+1}$ with the wanted properties. In the first case
$\sigma_2^k$ has the properties we wanted from $\sigma_2$ and can then
stop the sequence. We now consider the case where
$u(\inf,(\sigma_1,\sigma_2^k))\geq \delta$.
\begin{itemize}
\item Let $p$ be the probability that the play stops, when the players
  follows $(\sigma_1,\sigma_2^k)$.
\item For all $T$, let $p_T$ be the probability that the play stops in
  round $T$, conditioned on it not stopping before round $T$.
\item Let $(m_T^k)_T$ be the sequence of memory states associated with
  $\sigma_1$ (or equally $\sigma_2^k$), then played against $\sigma_2^k$. Note
  that given $\sigma_1$ and $\sigma_2^k$, the sequence is
  deterministic, because the choices of Player~2 are deterministic and
  if the game has not stopped, Player~1 has played $\lef$ at all
  earlier times and he updates its memory state deterministically.
\item For all $T$, let $S_{\tail}^{k,T}=\{(T',m_{T'}^k)\mid T'\geq T\}$
  (i.e. the pairs in the memory state sequence after round $T$).
\item Let $v$ be the value of the game if it does not stop, i.e. \[
  v=\liminf_{T\rightarrow\infty}\frac{|(S_{\tail}^{k,0}\setminus
    S_{\rig}^{k})\cap \calM^T|}{T} \enspace .
\]
Note that the set $(S_{\tail}^{k,0}\setminus S_{\rig}^{k})\cap \calM^T$
is the set of pairs, in the memory state sequence in which Player~2 plays
$\lef$ before round $T$.
\end{itemize}

Using the above definitions, we see that \[
u(\inf,(\sigma_1,\sigma_2^k)) \leq p (1-2^{-k})\delta + (1-p)v
\enspace .
\]
Thus, since $u(\inf,(\sigma_1,\sigma_2^k))\geq \delta$ we must have
that $p<1$ and $\delta\leq v$.

We will now use the following mathematical lemma.

\begin{lemma}\label{lem:tail has small pr}
  Let $x<1$ be some real number. Let $(x_T)_T$ be an infinite
  sequence, in which $x_T\in [0,1]$ and $1-\prod_T (1-x_T)=x$, then
  $\sum_T x_T\leq -\log (1-x)<\infty$.
\end{lemma}
\begin{proof}
  We have that all $x_T<1$, since otherwise the product would be 0,
  and hence $x=1$.  Also, we get that \begin{align*} \prod_T
    (1-x_T)&=1-x \Rightarrow \sum_T \log (1-x_T)=\log (1-x)\\ \Rightarrow
    -\sum_T \log (1-x_T)&=-\log (1-x)
    \Rightarrow \sum_T x_T\leq -\log
    (1-x) \enspace .
\end{align*}
The last inequality comes from that $f(y)=-\log (1-y)\geq y$ for $y\in
[0,1)$.  This follows from $f(0)=0$ and $f'(y)=\frac{1}{1-y}\geq 1$,
for $y\in [0,1)$.
\end{proof}

We have that $1-\prod_T(1-p_T)=p$ and by \cref{lem:tail has small
  pr} we then get that $\sum_T p_T$ is finite. Hence, there exists
some $M$ such that $\sum_{T=M}^\infty p_T\leq \frac{\delta}{2^{k+1}}$.

We can now define $\sigma_2^{k+1,a}$. Let 
\[\sigma_2^{k+1,a}(T,m)=\begin{cases} \rig & \text{if }(T,m)\in S_{\tail}^{k,M}\\
\sigma_2^{k,a}(T,m) &\text{otherwise}
\end{cases}\]

\paragraph*{The strategy $\sigma_2^{k+1}$ satisfies the wanted
  properties.} We will now show that $\sigma_2^{k+1}$ satisfies the
wanted properties.
\begin{enumerate}
\item 
That Property~\ref{pro:low stop} is satisfied comes from that
\begin{align*}
\sum_{(T,m)\in S_{\rig}^{k+1}}\sigma_1^a(T,m)(\rig)&\leq
\sum_{(T,m)\in S_{\rig}^{k}}\sigma_1^a(T,m)(\rig)+\sum_{(T,m)\in S_{\tail}^{k,M}}\sigma_1^a(T,m)(\rig)\\&\leq (1-2^{-k})\delta+2^{-k-1}\delta
=(1-2^{-k-1})\delta \enspace .
\end{align*}
\[\]
\item That Property~\ref{pro:fraction pairs} is satisfied can be seen
  as follows. We have that
\begin{align*}
  \liminf_{T\rightarrow\infty}\frac{|S_\rig^{k+1,T}|}{n\cdot
    T}&=\liminf_{T\rightarrow\infty}\frac{|S_\rig^{k,T}|+|(S_{\tail}^{k,M}\setminus
    S_{\rig}^{k})\cap \calM^T|}{n\cdot T} \\ &\geq
  \liminf_{T\rightarrow\infty}\frac{|S_\rig^{k,T}|}{n\cdot
    T}+\liminf_{T\rightarrow\infty}\frac{|(S_{\tail}^{k,M}\setminus
    S_{\rig}^{k})\cap \calM^T|}{n\cdot T} \enspace .
 \end{align*}
 Since the properties are satisfied for $\sigma_2^k$, we get that
 $\liminf_{T\rightarrow\infty}\frac{|S_\rig^{k,T}|}{n\cdot T}\geq
 \frac{\delta k}{n}$.  We thus just need to argue that \[
 \liminf_{T\rightarrow\infty}\frac{|(S_{\tail}^{k,M}\setminus
   S_{\rig}^{k})\cap \calM^T|}{n'\cdot T} \geq \frac{\delta}{n} \] and we are
 done.  That statement can be seen as follows
\begin{align*}
  \delta&\leq v \Rightarrow \delta\leq
  \liminf_{T\rightarrow\infty}\frac{|(S_{\tail}^{k,0}\setminus
    S_{\rig}^{k})\cap \calM^T|}{T} \Rightarrow
  \frac{\delta}{n}\leq \liminf_{T\rightarrow\infty}\frac{|(S_{\tail}^{k,0}\setminus S_{\rig}^{k})\cap \calM^T|}{n'\cdot T} \Rightarrow \\
  \frac{\delta}{n}&\leq
  \liminf_{T\rightarrow\infty}\frac{|(S_{\tail}^{k,M}\setminus
    S_{\rig}^{k})\cap \calM^T|}{n\cdot T} \enspace ,
\end{align*}
where the last inequality comes from that $S_{\tail}^{k,0}$ consists of the
same pairs as $S_{\tail}^{k,M}$ and then $M$ more.
\end{enumerate}

The above leads to the following theorem.

\begin{theorem}\label{thm:no-finite-markov-str}
  For all $\eps<\frac{1}{2}$, there exists no finite-memory
  deterministic-update $\eps$-optimal Markov strategy for Player~1
  in the Big Match.
\end{theorem}

%%%%%%%%%%%%%%%%%%%%%%%%%%%%%%%%%%%%%%%%%%%%%%%%%%%%%%%%%%%%%%%%%%%%%%%%%%%%%%%%

\section{Generalized Big Match Games}
\label{sec:generalized-big-match}
In order to generalize our results to arbitrary repeated games with
absorbing states, we follow the approach of Kohlberg and consider
first the subset of such games where Player~1, like in the Big Match,
has just the choice whether to declare that the game should stop or
continue. But unlike the Big Match, in case Player~1 declares the game
should stop, the game will only stop with some non-zero probability
(that depends on the action of Player~2). Furthermore, Player~2 can
have any number of actions and the rewards can be arbitrary. We call
such games \emph{generalized} Big Match games.

More formally, a generalized Big Match game $G$ is specified as
follows. Let $A_1 = \{\lef,\rig\}$ and let $A_2$ be any finite
set. The rewards $\payoff(a_1,a_2)$ are arbitrary, but the stop
probabilities $\stopprob(a_1,a_2)$ must satisfy that
$\stopprob(\lef,a_2)=0$ and $\stopprob(\rig,a_2)>0$ for all $a_2 \in
A_2$.

Our strategies for generalized Big Match games will follow the same
template as those given in \cref{sec:eps-sup,sec:eps-inf}. The change
required is a modification of the base strategies
$\sigma_1^{i,\xi}$. The proof that these new base strategies
$\tau_1^{i,\xi}$ have the desired properties follow those for
$\sigma_1^{i,\xi}$, but uses also additional ideas similar to those of
Kohlberg\cite{AS:Kohlberg1974}.

Given $G$ we define the \emph{derived matrix game} $\derivedgame{G}$ by
\[
\derivedgame{G}_{a_1,a_2} = \begin{cases}
 \payoff(a_1,a_2) &\text{if }a_1 = \lef\\
 \stopprob(a_1,a_2)\cdot \payoff(a_1,a_2) &\text{if } a_1 = \rig
\end{cases} \enspace.
\]

\begin{assumption}
  \label{matrixgameassumption}In the remainder of this section we make
  the following assumptions about the given generalized Big Match game
  $G$:
\begin{itemize}
\item The entries of $\derivedgame{G}$ are integer.
\item The value $\derivedgame{G}$ is 0.
\item Player~1 does not have a pure optimal strategy in $\derivedgame{G}$.
\end{itemize}
\end{assumption}
We observe that the last requirement means that in the matrix game
$\derivedgame{G}$ Player~1 has a unique optimal strategy and it plays
each action with non-zero probability.

Define $\omega=\min_{a_2} \stopprob(\rig,a_2)$ to be the minimum
non-zero stop probability of $G$ and $K=\max_{a_1,a_2}
\abs{\derivedgame{G}_{a_1,a_2}}$ be the maximum magnitude of an entry of
$\derivedgame{G}$.

\begin{remark}
As noted in \cref{REM:pure-strategies}, it is sufficient to consider
only pure strategies for Player~2. Unlike for the Big Match, here play
may continue even when Player~1 chooses the action $\rig$, and hence
it is not sufficient to consider only pure Markov strategies for
Player~2. We shall however (as done also by Kohlberg) give only the
proof for this special case and just note that the proof for general
pure strategies of Player~2 is done along the same lines
\end{remark}

\paragraph{Generalized density of pure Markov strategies}
In order to generalize our strategies given in
\cref{sec:eps-sup,sec:eps-inf}, we must first generalize the notion of
density of a pure Markov strategy $\sigma$ for Player~2. We then
define
\[
\gdens(\sigma^T)=\frac{\sum_{i=1}^T \payoff(\lef,\sigma_i)}{T} \enspace ,
\]
and further, for $T'< T$ we define
\[
\gdens(\sigma,T',T)=\frac{\sum_{i=T'}^{T} \payoff(\lef,\sigma_i)}{T-T'} \enspace .
\]

We can make a similar correspondence between the generalized density
of a play and the average of the rewards received by Player~1. The
main difference here is that the play may possibly continue whenever
Player~1 plays the action $\rig$. However, the event that Player~1
plays action $\rig$ an infinite number of times happens with
probability 0, since each time Player~1 does play action $\rig$, the
game stops with probability at least $\omega>0$ by definition.

\begin{observation}
\label{OBS:FiniteLotteries}
  Suppose Player~2 follows a pure Markov strategy $\sigma$. Consider a
  play $P$ in which Player~1 plays only the action $\lef$. Then for
  any $T<\abs{P}$ we have
\[
\gdens(\sigma^T) = \frac{1}{T} \sum_{T'=1}^T r_T \enspace ,
\]
where $r_T$ is the reward given to Player~1 in round $T$. Consider now
an infinite play $P$ in which Player~1 plays action $\rig$ only a
finite number of times. Then we have
\[
\uinf(P) = \liminf_{T\rightarrow\infty}\gdens(\sigma^T) \enspace ,
\]
and
\[
\usup(P) = \limsup_{T\rightarrow\infty}\gdens(\sigma^T) \enspace .
\]
\end{observation}

We need the following simple statement, which can be viewed as a
quantified version of \cite[Lemma~2.5]{AS:Kohlberg1974}.

\begin{lemma}
\label{l-gen-kohlberg}
Let $\sigma$ be a pure Markov strategy for Player~2, and let $j$ be
any integer. If $T \cdot \gdens(\sigma^T) \leq -j \cdot 2K$ then
\[
\sum_{i=1}^T \stopprob(\rig,\sigma^T_i) \cdot \payoff(\rig,\sigma^T_i) \geq j \enspace .
\]
\end{lemma}
\begin{proof}
  Let $\sigma_1$ be the (unique) optimal strategy in
  $\derivedgame{G}$. By \cref{matrixgameassumption} we have that
  $\sigma_1(\lef)>0$ and that the value of $\derivedgame{G}$
  is~0. This immediately implies that $\sigma_1(\lef)>1/2K$, since
  $p=\sigma_1(\lef)$ must satisfy an equation of the form
  \[
  p \cdot \derivedgame{G}_{\lef,a_2} + (1-p) \cdot
  \derivedgame{G}_{\rig,a_2} = 0 \enspace ,
  \]
  for some $a_2 \in A_2$, where the entries
  $\derivedgame{G}_{\lef,a_2}$ and $\derivedgame{G}_{\rig,a_2}$ are
  non-zero and integers of magnitude at most $K$.

  Now, by optimality of $\sigma_1$ we have
\[
\sigma_1(\lef)\cdot \payoff(\lef,a_2)+\sigma_1(\rig)\cdot
\stopprob(\rig,a_2)\cdot \payoff(\rig,a_2)\geq 0\enspace ,
\]
for any $a_2 \in A_2$, and thus also
\[
\sum_{i=1}^T \sigma_1(\lef)\cdot
\payoff(\lef,\sigma^T_i)+\sigma_1(\rig)\cdot
\stopprob(\rig,\sigma^T_i)\cdot \payoff(\rig,\sigma^T_i))\geq 0
\enspace .
\label{eqt:matrix2}
\]
By assumption $T \cdot \gdens(\sigma^T) = \sum_{i=1}^T
\payoff(\lef,\sigma^T_i) \leq -j \cdot 2K$. Hence
\[
\sum_{i=1}^T \stopprob(\rig,\sigma^T_i) \cdot \payoff(\rig,\sigma^T_i)
\geq - \frac{\sigma_1(\lef)}{\sigma_1(\rig)} \sum _{i=1}^T
\payoff(\lef,\sigma^T_i) \geq \frac{1}{2K} \cdot j \cdot 2K = j
\enspace .
\]
\end{proof}

\paragraph{The Big Match as a generalized Big Match game}
The Big Match as defined in \cref{sec:definitions} does not
immediately fit our definition of generalized Big Match games, since
the value of the derived matrix game is $1/2$ rather than $0$. To
achieve this we may simply replace the rewards of value~$0$ with
value~$-1$. Doing this we see that
\[
\begin{split}
\gdens(\sigma^T) &=\frac{ \abs{\{i\mid (\sigma^T)_i=\lef\}} - \abs{\{i\mid (\sigma^T)_i=\rig\}}}{T}\\ &= \frac{ 2\abs{\{i\mid (\sigma^T)_i=\lef\}} - T}{T} = 
 2\dens(\sigma^T)-1
\end{split}
\]
for any $\sigma^T$.

\begin{figure}
\centering
\renewcommand{\arraystretch}{1.3}
$
\begin{array}{ r|r|r| }
\multicolumn{1}{r}{}
 &  \multicolumn{1}{c}{\lef}
 & \multicolumn{1}{c}{\rig} \\
\cline{2-3}
\lef & \nonabsorb{1} & \nonabsorb{-1} \\
\cline{2-3}
\rig & \absorb{-1} & \absorb{1} \\
\cline{2-3}
\end{array}
$
\renewcommand{\arraystretch}{1}
\caption{The Big Match as a generalized Big Match game.
\label{fig:big-match-zeroval}}
\end{figure}

\subsection{Small space
  \texorpdfstring{$\eps$}{epsilon}-supremum-optimal strategies in
  generalized Big Match games}
\label{sec:eps-sup-gen}

For given $\eps$, let $\xi = \eps^2/(4K^4)$. Thus
$\eps=2K^2\sqrt{\xi}$. The strategy $\sigma_1^*$ for Player~1 is
obtained from the strategy of \cref{sec:eps-sup}, by exchanging the
base strategy $\sigma_1^{i,\xi}$ with the new base strategy
$\tau_1^{i,\xi}$ defined next.

\paragraph{The base strategy.}
Similar to $\sigma_1^{i,\xi}$, the strategy $\tau_1^{i,\xi}$ uses
deterministic updates of memory, and uses integers as memory states
The memory update function is changed to
\[
\tau_1^{i,\xi,u}(a,j) = j - \derivedgame{G}_{\rig,a} = j - \stopprob(\rig,a)\payoff(\rig,a) \enspace ,
\]
whereas the action function is unchanged as
\[
\tau_1^{i,\xi,a}(j)(\rig) = \begin{cases}
\xi^4(1-\xi)^{i+j} & \text{if } i+j>0\\
\xi^4 & \text{if } i+j\leq 0
\end{cases} \enspace .
\]
Note that when the Big Match is redefined as a generalized Big Match
game, we have that $\tau_1^{i,\xi}=\sigma_1^{i,\xi}$.

We will next generalize the statements of \cref{sec:eps-sup} in the
following paragraphs.

\subsubsection{Space usage of the strategy}
We will here consider the space usage of $\sigma_1^*$. Like for the strategy $\sigma_1^*$ for the Big Match we are interested in the length of the epochs and can generalize \cref{l-goodbalance} to generalized Big Match. 

\begin{lemma}\label{l-goodbalance2}
  For any $\gamma,\delta\in (0,1/4)$, there is a constant $M$ such
  that with probability at least $1-\gamma$, for all $i\geq M$ and all
  $j\in\{1,\dots,i\}$, we have that 
\[
t(i,j,i^2)-t(i,j-1,i^2) \in  [(1-\delta) F(i)/i,(1+\delta) F(i)/i] \enspace .
\]
\end{lemma}
\begin{proof}
The proof is precisely the same as for \cref{l-goodbalance}, since it only concerns itself with the length of epochs and not with the base strategy, and only the base strategy has changed as compared to section \cref{sec:eps-sup}.
\end{proof}

We next give a generalization of \cref{lem:spaceusage}.
 
\begin{lemma}\label{lem:spaceusage2}
  For all constants $\gamma>0$, with probability at least $1-\gamma$, the
  space usage of $\sigma^*_1$ is $O(\log f(T)+\log K)$.
\end{lemma}
\begin{proof}
Observe that the base strategy $\tau_1^{i,\xi}$ can reach a factor of $2K$ more memory states upto round $T$ than $\sigma_1^{i,\xi}$, since $\sigma_1^{i,\xi}$ changes its counter by $\pm 1$ in each round while $\tau_1^{i,\xi}$ changes its counter by some number in $\{-K,\dots,K\}$.
Thus, $\sigma_1^*$ uses at most a factor $2K$ more memory states then using $\tau_1^{i,\xi}$ as the base strategy instead of $\sigma_1^{i,\xi}$ in round $T$ for any $T$. 
The statement then follows from a similar proof as the one for \cref{lem:spaceusage}.
\end{proof}

\subsubsection{Play stopping implies good outcome}

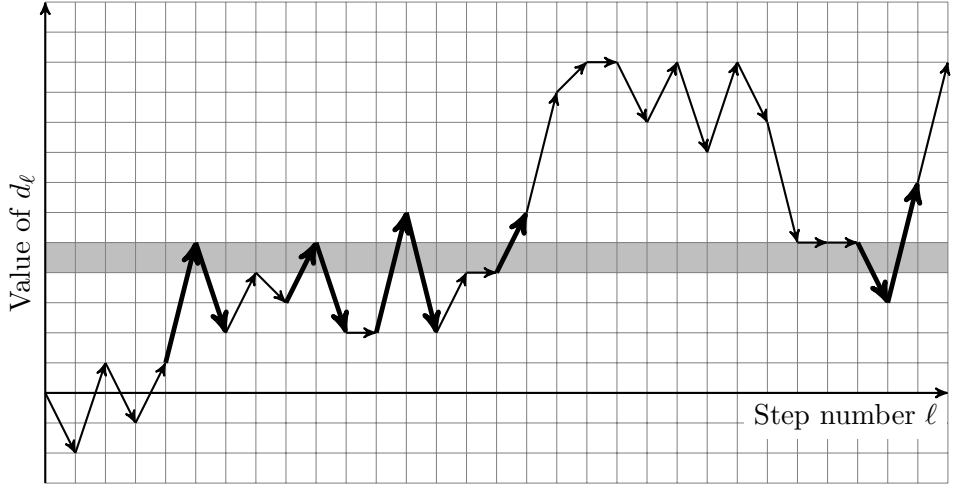
\begin{figure}
\begin{tikzpicture}[thick,scale=0.4]
\coordinate (o) at (0,0);
\coordinate (tr) at (30,13);
\coordinate (bl) at (0,-3);

\fill[light-gray] (0,4) rectangle (30,5);

\draw[gray,very thin] (bl) grid (tr);
\node[anchor=north east,fill=white] at (tr |- o) {Step number $\ell$};
\draw[thick,->,>=stealth'] (o) -- (tr |- o);
\draw[thick,->,>=stealth'] (bl) -- (bl |- tr) node[midway,anchor=south,rotate=90] {Value of $d_\ell$};

\xdef\lx{0}
\xdef\ly{0}
\foreach \x/\y/\v in {1/-2/1,2/1/1,3/-1/1,4/1/1,5/5/1.5,6/2/1.5,7/4/1,8/3/1,9/5/1.5,10/2/1.5,11/2/1,12/6/1.5,13/2/1.5,14/4/1,15/4/1,16/6/1.5,17/10/1,18/11/1,19/11/1,20/9/1,
21/11/1,22/8/1,23/11/1,24/9/1,25/5/1,26/5/1,27/5/1,28/3/1.5,29/7/1.5,30/11/1}{\draw[line width=\v*\pgflinewidth, ->,>=stealth'] (\lx,\ly) -- (\x,\y); 
\xdef\lx{\x}
\xdef\ly{\y}
}

\end{tikzpicture}
\caption{Possible movement of the counter $\tau_1^{i,\eps}$ uses as memory through the rounds.\label{fig:counter_movement2}}
\end{figure}

\begin{lemma}
  \label{l-base2} 
  Let $T,i\geq 1$ be integers and $0<\xi<1$ be a real number. Let
  $\sigma \in (A_2)^T$ be an arbitrary prefix of a pure Markov
  strategy for Player~2.  Consider the first $T$ rounds where the
  players play $G$ following $\tau_1^{i,\eps}$ and $\sigma$
  respectively.

  Let $S \in \{1,\dots,T\} \cup \{\infty\}$ be a random variable in
  case the game stops in the first $T$ rounds denotes that round, and
  is $\infty$ otherwise. Let $U$ be the random variable that denotes
  the outcome in case the game stops in the first $T$ rounds, and is 0
  otherwise. Then
\begin{enumerate}
\item 
  \[
  -\Exp[U \mid U<0]\Pr[U<0] \leq \xi^3(1-\xi)^{i-K+1} +
  (1-\xi)^{1-2K}\Exp[U \mid U>0]\Pr[U>0] \enspace .
  \]
\item If $\gdens(\sigma^T) \leq -i \cdot 2K/T$ then
  \[
  \Pr[S<\infty] \geq \xi^4 \cdot \omega \enspace .
  \]
\end{enumerate}
\end{lemma}
\begin{proof}
Define 
\[
d_\ell = - \sum_{\ell'<\ell} \derivedgame{G}_{\rig,\sigma_{\ell'}} \enspace .
\]
and note that $d_\ell$ is the value of the counter used by
$\tau_1^{i,\xi}$ as memory in step $\ell$. 
There is an illustration of how $d_\ell$ could evolve through the steps in \cref{fig:counter_movement2}.
Let $I = \{\ell \in
\{1,\dots,T\} \mid \derivedgame{G}_{\rig,\sigma_{\ell}} < 0\}$ and $D
= \{\ell \in \{1,\dots,T\} \mid \derivedgame{G}_{\rig,\sigma_{\ell}}
>0\}$ be the sets of times where the counter is incremented and
decremented, respectively. For integer $d$, define sets $K_d$ by
\[
\begin{split}
K_d = \{ \ell \in \{1,\dots,T\} \mid & (\derivedgame{G}_{\rig,\sigma_\ell}<0 \;\&\; d_\ell \leq d < d_\ell-\derivedgame{G}_{\rig,\sigma_\ell}) \text{ or }\\
&(\derivedgame{G}_{\rig,\sigma_\ell}>0 \;\&\; d_\ell -\derivedgame{G}_{\rig,\sigma_\ell} \leq d < d_\ell)\}\enspace .
\end{split}
\]
Intuitively, $K_d$ is now the set of times where the counter is either
incremented to pass trough the value $d$ and end above, or is
decreased from above $d$ passing through the value $d$. 
For instance, the gray row of \cref{fig:counter_movement2} corresponds to a set $K_d$, which consists of the edges that starts below or on the bottom of the gray row and ends over or on the top of the gray row. Each such edge are wider, in the figure, then the remaining edges. Note that horizontal edges are not in $K_d$.  
Notice each
$\ell \in \{1,\dots,T\}$ belongs to exactly
$\abs{\derivedgame{G}_{\rig,\sigma_\ell}}$ many of the sets
$K_d$. Also if $K_d = \{k_1 < k_2 < \dots < k_m\}$, for some $m$, then
$k_j \in I$ if and only if $k_{j+1} \in k_{j+1} \in D$. Thus $K_d$ is
an sequence of elements alternately from $I$ and $D$, and starts
with an element of $I$ when $d\geq 0$ and starts with an element of
$D$ otherwise.

Finally let, $A_\ell \in \{\lef,\rig\}$ be the random variable
indicating the action of Player~1 at time step $\ell$, and define
$p_\ell = \Pr[S\geq \ell \;\&\; A_\ell=\rig]$. We now have
\[
\begin{split}
\Exp[U] 
& = \sum_{\ell=1}^T \Pr[S=\ell] \cdot \payoff(\rig,\sigma_\ell)
 = \sum_{\ell=1}^T p_\ell \cdot \stopprob(\rig,\sigma_\ell) \payoff(\rig,\sigma_\ell)\\
& = \sum_{\ell=1}^T p_\ell \cdot \derivedgame{G}_{\rig,\sigma_\ell}
 = \sum_{\ell=1}^T \sum_{d : \ell \in K_d} \sgn(\derivedgame{G}_{\rig,\sigma_\ell}) \cdot p_\ell\\
& = \sum_{d} \sum_{\ell\in K_d} \sgn(\derivedgame{G}_{\rig,\sigma_\ell}) \cdot p_\ell \enspace .
\end{split}
\]
Similarly, 
\[
\Exp[U \mid U<0]\Pr[U<0] = - \sum_{d} \sum_{\ell\in K_d \cap I}
p_\ell \enspace ,
\] 
and
\[
\Exp[U \mid U>0]\Pr[U>0] = \sum_{d} \sum_{\ell\in K_d \cap D}
p_\ell \enspace .
\]
For integer $d$, define $E_{\mathrm{loss},d} = \sum_{\ell\in K_d \cap
  I} p_\ell$ and $E_{\mathrm{win},d} = \sum_{\ell\in K_d \cap D}
p_\ell$. Recall that by definition of $\tau^{i,\xi}$ we have
$\Pr[A_\ell = \rig \mid S \geq \ell] =
\xi^4(1-\xi)^{\max(0,i+d_\ell)}$.

Consider now $d \geq 0$. Then $p_{k_1} \leq \Pr[A_\ell =
\rig \mid S \geq k_1] \leq \xi^4(1-\xi)^{i+d-K+1}$, since $d_{k_1} >
d-K$. For even $j$ we have
\[
\begin{split}
p_{k_j} &= \Pr[A_{k_j}=\rig | S \geq k_j]\Pr[S \geq k_j] \\
 &\geq (1-\xi)^{2K-1}\Pr[A_{k_j}=\rig | S \geq k_{j+1}] \Pr[S \geq k_{j+1}] = (1-\xi)^{2K-1}p_{k_{j+1}} \enspace ,
\end{split}
\]
since $d_{k_j} \leq d+K < d_{k_{j+1}} + 2K$. It follows that
\[
E_{\mathrm{loss},d} \leq \xi^4(1-\xi)^{i+d-K+1} + (1-\xi)^{1-2K}
E_{\mathrm{win},d} \enspace .
\]

For $d < 0$, $E_{\mathrm{loss},d} \leq (1-\xi)^{1-2K}
E_{\mathrm{win},d}$, since for \emph{odd} $j$, $p_{k_j} \geq
(1-\xi)^{2K-1}p_{k_{j+1}}$ as above. For $d<-i$ we can give better
estimates, like in the case of \cref{l-base}, but it is not needed.

Taking the summation over $d$ then gives
\[
\begin{split}
-\Exp[U \mid U <0]\Pr[U<0] &= \sum_d E_{\mathrm{loss},d} \\ &\leq \sum_{d \geq 0} \xi^4(1-\xi)^{i+d-K+1} + (1-\xi)^{1-2K}\sum_d E_{\mathrm{win},d}\\
&= \xi^3(1-\xi)^{i-K+1} + (1-\xi)^{1-2K}\Exp[U \mid U>0]\Pr[U>0] \enspace .
\end{split}
\]

For the second part, if $\gdens(\sigma^T) \leq -i \cdot 2K/T$ then
\cref{l-gen-kohlberg} gives $\sum_{\ell=1}^T
\derivedgame{G}_{\rig,\sigma_\ell} \geq i$, and hence $d_T \leq
-i$. This implies that if the game reaches round $T$, Player~1 plays
$\rig$ with probability $\xi^4$, which means the game stops with
probability at least $\xi^4 \cdot \omega$.
\end{proof}

\begin{lemma}
\label{l-winning2}
Let $0<\xi<0$. Let $\sigma$ be a pure Markov strategy for
Player~2. Let $S$ be the event that the game stops. Let $U$ be the
random variable that denotes the outcome in case the game stops, and
is 0 otherwise. If $\Pr[S] \geq K\sqrt{\xi}$ then $\Exp[U \mid S] \geq
-2K\sqrt{\xi}$.
\end{lemma}
\begin{proof}
  The proof follows along that of \cref{l-winning2}. Let $A_{i,j}$ be
  the set of plays in which between round 1 and round $t(i,j,1)-1$,
  the game does not stop. Let $U_{i,j}$ be the random variable that
  denotes the outcome in case the game stops between round $t(i,j,1)$
  and round $t(i,j,i^2)$, and is 0 otherwise.  Note that $\Exp[U] =
  \sum_{i,j} \Exp[U_{i,j} \mid A_{i,j}] \Pr[A_{i,j}]$.

  Fix a possible value of all $t(i,j,k)$'s and denote by $Y$ the event
  that these particular values actually occur. Fix $i$ and
  $j$. Conditioned on $Y$ and $A_{i,j}$, between time $t(i,j,1)$ and
  $t(i,j,i^2)$ Player~1 plays $\tau_1^{i,\xi}$ against a fixed
  strategy $\sigma_{t(i,j,1)},\sigma_{t(i,j,2)},\dots,
  \sigma_{t(i,j,i^2)}$ for Player~2. By \cref{l-base2} we thus have
\begin{multline*}
  -\Exp[U_{i,j} \mid U_{i,j}<0, Y, A_{i,j}]\Pr[U_{i,j}<0 \mid Y,
  A_{i,j}] \\ \leq \xi^3(1-\xi)^{i-K+1} + (1-\xi)^{1-2K}\Exp[U_{i,j} \mid
  U_{i,j}>0,Y,A_{i,j}]\Pr[U_{i,j}>0 \mid Y,A_{i,j}] \enspace .
\end{multline*}
Since the above inequality is true conditioned on arbitrary values
  of $t(i,j_1,j_2)$'s, it is true also without the conditioning:
\begin{multline*}
  -\Exp[U_{i,j} \mid U_{i,j}<0, A_{i,j}]\Pr[U_{i,j}<0 \mid A_{i,j}]\\
  \leq \xi^3(1-\xi)^{i-K+1} + (1-\xi)^{1-2K}\Exp[U_{i,j} \mid
  U_{i,j}>0,A_{i,j}]\Pr[U_{i,j}>0 \mid A_{i,j}] \enspace .
\end{multline*}
Thus,
\[
\begin{split}
&-\Exp[U \mid U<0]\Pr[U<0] \\
&= -\sum_{i=1}^\infty \sum_{j=1}^i  \Exp[U_{i,j} \mid U_{i,j}<0 , A_{i,j}]\Pr[U_{i,j}<0 \mid A_{i,j}]\Pr[A_{i,j}]\\
&\begin{split}\leq  \sum_{i=1}^\infty \sum_{j=1}^i \Big(&\xi^3(1-\xi)^{i-K+1} + \\ &(1-\xi)^{1-2K}\Exp[U_{i,j} \mid
  U_{i,j}>0,A_{i,j}]\Pr[U_{i,j}>0 \mid A_{i,j}]\Big)\Pr[A_{i,j}]
\end{split}
\\
&\leq \xi + (1-\xi)^{1-2K}\sum_{i=1}^\infty \sum_{j=1}^i \Exp[U_{i,j} \mid
  U_{i,j}>0,A_{i,j}]\Pr[U_{i,j}>0 \mid A_{i,j}]\Pr[A_{i,j}]\\
&= \xi + (1-\xi)^{1-2K} \Exp[U \mid U>0]\Pr[U>0]
\end{split}
\]
Hence
\[
-(1-\xi)^{2K-1}\Exp[U \mid U<0]\Pr[U<0] \leq \xi(1-\xi)^{2K-1} + \Exp[U\mid U>0]\Pr[U>0] \enspace ,
\]
and so
\[
\begin{split}
-\Exp[U] &= -\Exp[U \mid U<0]\Pr[U<0] - \Exp[U \mid U>0]\Pr[U>0] \\
&\leq -(1-(1-\xi)^{2K-1})\Exp[U \mid U<0]\Pr[U<0] + \xi(1-\xi)^{2K-1}\\
&\leq K(1-(1-\xi)^{2K-1}) + \xi(1-\xi)^{2K-1} \\
&\leq K(2K-1)\xi + \xi \leq \enspace 2K^2 \cdot \xi,
\end{split}
\]
where we use the equation $1-(1-x)^k \leq kx$, which is valid for
positive integer $k$ and $0\leq x \leq 1$. Hence the expected outcome
conditioned on the game stopping can be estimated by
\[
\Exp[U \mid S] = \frac{\Exp[U]}{\Pr[S]} \geq \frac{-2K^2\xi}{K\sqrt{\xi}} = -2K\sqrt{\xi} \enspace ,
\]
using the above estimate on $\Exp[U]$ and the assumption $\Pr[S] \geq K\sqrt{\xi}$.
\end{proof}

\subsubsection{Low density means play stops}

\begin{lemma}\label{l-subdivion-sup2}
  Let $\eps,\delta \in(0,1)$. Let $\sigma$ be an arbitrary pure Markov
  strategy for Player~2.  Let $a_{i,j}$ be some numbers. Consider the
  event $Y$ where $t(i,j)=a_{i,j}$ for all $i,j$.  If
  $\limsup_{T\rightarrow \infty} \gdens(\sigma^T)\leq -\delta$, then
  conditioned on $Y$, there is an infinite sequence of sub-epochs and
  epochs $(i_n,j_n)_n$ such that
  $\gdens(\sigma,a_{i_n,j_n},a_{i_n,j_n+1})\leq -\delta/4$.
\end{lemma}
\begin{proof}
  The proof follows that of \cref{l-subdivion-sup} with small
  changes. Thus, let $M$ be such that for every $T' \geq M$ we have
  $\gdens(\sigma^{T'})\leq -\delta/2$.  Let $(T_n)_n$ be a sequence
  such that $T_1\geq M$ and for all $n\geq 1$ we have that
  $T_{n+1}\cdot \delta/4\geq K \cdot T_{n}$ and $T_n=a_{i,j}$ for some
  $i,j$. Let $(i_n',j_n')_n$ be the sequence such that
  $T_n=a_{i_n',j_n'}$. This means that even if
  $\gdens(\sigma^{T_{n}})=-K$, the generalized density
  $\gdens(\sigma,T_{n}+1,T_{n+1})$ is at most $-\delta/4$, because
  $\gdens(\sigma^{T_{n+1}})\leq -\delta/2$. But, we then get that
  there exists some sub-epoch $j_n$ in epoch $i_n$, such that $j_{n}'
  \leq j_n\leq j_{n+1}'$ and such that $i_n'\leq i_n\leq i_{n+1}'$ for
  which the generalized density of that sub-epoch
  $\dens(\sigma,a_{i_n,j_n}+1,a_{i_n,(j_{n}+1)})$ is at most
  $1/2-\delta/4$, because not all sub-epochs can have generalized
  density below that of the average sub-epoch. But then $(i_n,j_n)_n$
  satisfies the lemma statement.
\end{proof}

\begin{lemma}\label{l-stopping-sup2}
  Let $\sigma$ be an arbitrary pure Markov strategy for Player~2.  If
\[
\limsup_{T\rightarrow \infty} \gdens(\sigma^T)< 0
\]
then when played against $\sigma_1^*$ the play stops with
probability~1.
\end{lemma}
\begin{proof}
  The proof follows that of \cref{l-stopping-sup} with small changes.
  Let $\delta>0$ be such that $\limsup_{T\rightarrow \infty}
  \gdens(\sigma^T)\leq -\delta$. Consider arbitrary numbers $a_{i,j}$
  and the event $Y$ stating that $t(i,j)=a_{i,j}$ for all $i,j$.  Let
  $(i_n,j_n)_n$ be the sequence of sub-epochs and epochs shown to
  exists by \cref{l-subdivion-sup2} with probability~1. That is, for
  each $(i_n,j_n)$ we have that sub-epoch $j_n$ of epoch $i_n$ has
  generalized density at most $-\delta/4$.  We see that, conditioned
  on $Y$ that each sample are sampled uniformly at random in each
  sub-epoch $j$ of each epoch $i$, except for the last sample.

  Now consider some fixed $n$. By Hoeffding's inequality,
  \cref{THM:Hoeffding} (setting $a_i=-K$ and $b_i=K$, and letting
  $c_i=\payoff(\lef,\sigma_{t(i_n,j_n,i)})$), the probability that the
  generalized density of the subsequence given by the first
  $(i_n)^2-1$ samples is more that $-\delta/8$ is bounded by
\[
2\exp\left(-\frac{2(\frac{\delta}{8}((i_n)^2-1))^2}{((i_n)^2-1)(2K)^2}\right)
= 2\exp\left( -\frac{\delta^2}{128\cdot K^2}((i_n)^2-1)\right) \enspace .
\]
For sufficiently large $i_n$, this is less than $\frac{1}{2}$. If on
the other hand the generalized density of the subsequence is more than
$-\delta/8$, we see that $(i_n)^2-1 \geq i_n \cdot 2K \cdot 8/\delta$
for large enough $i_n$, and in that case we have, by \cref{l-base2},
that the game stops with probability at least $\xi^4 \cdot \omega$ in
sub-epoch $j_n$ of epoch $i_n$. Thus, for each of the infinitely many
$n$'s for which $i_n$ is sufficiently high, we have a probability of
at least $\frac{\xi^4\cdot \omega}{2}$ of stopping. Thus play must
stop with probability~1.

The argument was conditioned on some fixed assignment of endpoints of
sup-epochs and epochs, but since there is such a assignment with
probability~1 (since they are finite with probability~1), we conclude
that the proof works without the condition.
\end{proof}

\subsubsection{Proof of main result}
 \begin{theorem}\label{thm:sup_eps2}
   The strategy $\sigma^*_1$ is $2K^2\sqrt{\xi}$-supremum-optimal, and for all
   $\delta>0$, with probability at least $1-\delta$ does it use space
   $O(\log f(T) + \log K)$.
 \end{theorem}
\begin{proof}%[Proof of \cref{thm:sup_eps}]
  The space usage follows from \cref{lem:spaceusage2}. Let $\sigma$ be
  a pure Markov strategy for Player~2. Let $S$ be the event that the
  game stops. Let $U$ be the random variable that denotes the outcome
  in case the game stops, and is equal to $\limsup_{T \rightarrow
    \infty} \gdens(\sigma^T)$ otherwise. By \cref{OBS:FiniteLotteries}
  we have $\usup(\sigma^*_1,\sigma)=\Exp[U]$. Let $s=\Pr[S]$ the the
  probability that the game stops. We now consider three cases, either
  (i)~$s=1$; or (ii)~$K\sqrt{\xi}<s<1$; or (iii)~$s\leq
  K\sqrt{\xi}$. In case (i), by \cref{l-winning2} we have
  $\Exp[U]=\Exp[U \mid S]\geq -2K\sqrt{\xi}$.

  In case~(ii), by \cref{l-winning2} $\Exp[U \mid S] \geq
  -2K\sqrt{\xi}$, since $K\sqrt{\xi}<s$. And by \cref{l-stopping-sup2}
  we have $\Exp[U \mid \overline{S}] \geq 0$, since $s<1$. Thus
\[
\Exp[U] \geq s\cdot(-2K\sqrt{\xi}) + (1-s)\cdot 0 \geq -2K\sqrt{\xi} \enspace .
\]
In case~(iii), again by \cref{l-stopping-sup2} we have $\Exp[U \mid
\overline{S}] \geq 0$, since $s<1$. Thus
\[
\Exp[U] \geq s\cdot(-K) + (1-s)\cdot 0 \geq -K^2\sqrt{\xi} \enspace .
\]
\end{proof}

\subsection{An \texorpdfstring{$\eps$}{epsilon}-optimal strategy that
  uses \texorpdfstring{$\log \log T$}{log log T} space for generalized
  Big Match games}

In this section we give a
$\eps$-optimal strategy $\sigma_1^*$ for Player~1 in any generalized Big Match
game that for all $\delta>0$ with probability $1-\delta$ uses $O(\log\log
T)$ space. Similarly to how \cref{sec:eps-inf} showed that the strategy $\sigma_1^*$, if initialized correctly, from \cref{sec:eps-sup} was $\eps$-optimal for Player~1 in the Big Match, we here show that the strategy $\sigma_1^*$ from \cref{sec:eps-sup-gen}, if initialized correctly, is $\eps$-optimal. Similarly to \cref{sec:eps-inf}, the $\eps$-optimal strategy for generalized Big Match games is simply an instantiation of the strategy
$\sigma_1^*$ from \cref{sec:eps-sup-gen}, setting $f(T)=\lceil{\log
  T}\rceil$. We can then let $\ov{f}=\log T$ and $F(T)=2^T$.
The proofs of the statements in this section is nearly identical to those of \cref{sec:eps-inf}, but there are minor changes and they are thus given here in full.

The claim about the space usage of $\sigma_1^*$ is already
established in \cref{sec:eps-sup-gen}. To obtain the stronger property of
$\eps$-optimality rather than just $\eps$-supremum-optimality, we just
needs to establish a $\liminf$ version of \cref{l-stopping-sup2}, like how we in \ref{sec:eps-inf} gave a $\liminf$ version of \cref{l-stopping-sup}.

First we show a technical lemma similarly to \cref{l-subdivision}. Recall that for a pure Markov strategy $\sigma$
and a sequence of integers $I=\{i_1,i_2,\dots,i_m\}$, we have that $\sigma_I$
is the sequence, $\sigma_{i_1},\sigma_{i_2},\dots,\sigma_{i_m}$. Again, note
that $\sigma^k=\sigma_{\{1,\dots,k\}}$. 

\begin{lemma}\label{l-subdivision2}
  Let $\sigma$ be a pure Markov strategy for Player~2, $\delta<1/4$ be
  a positive real, and $M$ be a positive integer.  Let $\liminf_{T
    \rightarrow \infty} \gdens(\sigma^T) \leq - \delta.$
  Let $\ell_1,\ell_2,\dots$ be such that for all $i \geq M$, we have
  that $\ell_i \in [(1-\delta)\cdot (2^{i+1}-1),(1+\delta)\cdot
  (2^{i+1}-1)]$.  Then there exists a sequence $k_2,k_3,\dots$ such
  that for infinitely many $i> M$, we have that $\ell_{i-1}+\delta/K 2^{i-2} \leq
  k_i \leq \ell_i$ and that $\liminf_{i \rightarrow \infty}
  \gdens(\sigma_{\{\ell_{i-1}+1,\dots,k_i\}}) \leq  -
  \frac{\delta}{4}.$
\end{lemma}

\begin{proof}
The proof follows  \cref{l-subdivision} with small changes.
  Let $\ell_i$ be as required. If there are infinitely many $i$ such
  that $\gdens(\sigma_{\{\ell_{i-1}+1,\dots,\ell_i\}}) \leq
  - \frac{\delta}{4}$ then set $k_i = \ell_{i+1}$ and the lemma
  follows by observing $(k_i - \ell_{i-1}) \geq (1-\delta)(2^{i+1}-1)-
  (1+\delta)(2^{i}-1) = (1-3\delta)2^i \geq \delta/K 2^{i-2}$, for $i>
  M$.  So assume that only for finitely many $i$,
  $\gdens(\sigma_{\{\ell_{i-1}+1,\dots,\ell_i\}}) \leq -
  \frac{\delta}{4}$.  Thus the following claim can be applied for
  arbitrary large $i_0$.

\begin{claim}
  Let $i_0\geq M$ be given. If for every $i\geq i_0$,
  $\gdens(\sigma_{\{\ell_{i-1}+1,\dots,\ell_i\}}) >-
  \frac{\delta}{4}$ then there exist $j>i_0$ and $k$ such that
  $\ell_{j-1}+(\delta/K) 2^{j-2} \leq k \leq \ell_j$ and
  $\gdens(\sigma_{\{\ell_{j-1}+1,\dots,k\}}) \leq -
  \delta$.
\end{claim}

We can use the claim to find $k_2,k_3,\dots$ inductively. Start with
large enough $i_0\geq M$ and set $k_i = \ell_i$ for all $i\leq i_0$.
Then provided that we already inductively determined
$k_2,k_3,\dots,k_{i_0}$, we apply the above claim to obtain $j$ and
$k$, and we set $k_j=k$ and $k_i=\ell_i$, for all $i=i_0+1,\dots,j-1$.

So it suffices to prove the claim. For any $d\geq 1$, $\ell_{i_0}
\cdot 2^{d-1} \leq \ell_{i_0+d}$ and 
\[
\gdens(\sigma^{\ell_{i_0+d}}) \geq \frac{ -
  \frac{\delta}{4}(\ell_{i_0+d} - \ell_{i_0})-K\cdot \ell_{i_0}}{\ell_{i_0+d}} \enspace .
\]
Furthermore, if $d \geq 1+\log(4K/\delta)$ then $\ell_{i_0} \leq
\frac{\delta}{4K} \ell_{i_0+d}$ and
\[
\gdens(\sigma^{\ell_{i_0+d}}) \geq - \frac{\delta}{4} - \frac{\delta}{4} =  - \frac{\delta}{2} \enspace .
\]
Since $\liminf_{k \rightarrow \infty} \gdens(\sigma^k) \leq
- \delta$, there must be $k$ and $d\geq 1+\log(4/\delta)$ such that
$\ell_{i_0+d-1} \leq k \leq \ell_{i_0+d}$ and $\gdens(\sigma^k) \leq
- \delta$. Set $j=i_0+d$. Also
\[
 \gdens(\sigma^k) = \frac{ \gdens(\sigma^{\ell_{j-1}}) \ell_{j-1} + \gdens(\sigma_{\{\ell_{j-1}+1,\dots,k\}}) (k-\ell_{j-1}) }{\ell_{j-1} + (k-\ell_{j-1})} \enspace ,
\]
which means
\[
\begin{split}
  \left( \gdens(\sigma^k) -
    \gdens(\sigma_{\{\ell_{j-1}+1,\dots,k\}})\right) (k-\ell_{j-1})
  &= \left( \gdens(\sigma^{\ell_{j-1}}) - \gdens(\sigma^k) \right) \ell_{j-1} \\
  \geq \left[ - \frac{\delta}{2} 
     - \delta \right] \ell_{j-1} &=
  \frac{\delta}{2} \ell_{j-1} \enspace .
\end{split}
\]
Thus $\gdens(\sigma_{\{\ell_{j-1}+1,\dots,k\}}) \leq \gdens(\sigma^k)$
which in turn is less than $-\delta$.  Furthermore,
$k-\ell_{j-1} \geq \frac{\delta}{2K} \ell_{j-1} \geq \frac{\delta}{2K}
(1-\delta) ( 2^{j} - 1 ) \geq \frac{\delta}{4K} 2^{j}$, provided that
$j\geq 2$.  Hence, $k$ and $j$ have the desired properties.
\end{proof}

We are now ready to prove the $\liminf$ version of \cref{l-stopping-sup2}.  
\begin{lemma}
\label{l-stopping-inf2}
Let $\sigma$ be an arbitrary pure Markov strategy
for Player~2. If
\[
\liminf_{t\rightarrow \infty} \gdens(\sigma_1,\dots,\sigma_t)< 0 \enspace ,
\]
then when played against $\sigma_1^*$ the play stops with
probability~1.
\end{lemma}
\begin{proof}
The proof is similarly to \cref{l-stopping-inf} with minor modifications. 
Let $0<\delta<\frac{1}{4}$ be such that \[
\liminf_{t\rightarrow \infty} \gdens(\sigma_1,\dots,\sigma_t)\leq -\delta
\]
  Pick arbitrary $\gamma\in (0,1)$. We will show that with probability
  at least $1-\gamma$ the game stops, and this implies the
  statement. Let $M$ be given by \cref{l-goodbalance2} applied for
  $\gamma$ and $\delta/2$.  Then we have that with probability at
  least $1-\gamma$, for all $i\geq M$ and $j \in\{1,\dots,i\}$,
\[
t(i,j,i^2)-t(i,j-1,i^2) \in [(1-\frac{\delta}{2})
  2^i/i,(1+\frac{\delta}{2}) 2^i/i] \enspace .
\]

  Pick $t_{i,j} \in \NN$, for $i=1,2,\dots$ and $j\in\{1,\dots,i\}$, so
  that $t_{i,j-1} < t_{i,j}$ where $t_{i,0}$ stands for
  $t_{i-1,i-1}$. Let $t_{i,j}-t_{i,j-1} \in [(1-\frac{\delta}{2})
  2^i/i,(1+\frac{\delta}{2}) 2^i/i]$, for all $i\geq M$ and
  $j\in\{1,\dots,i\}$.  Pick $M'$ so that $\frac{\delta}{2}(2^{M'}-1)
  \geq \max\{t_{M,0}, (1-\frac{\delta}{2}) 2^{M}\}$. Define $\ell_i =
  t_{i,i}$ for all $i\geq 1$.  Then for all $i\geq M'$, $\ell_i \in
  [(1-\delta)\cdot (2^{i+1}-1),(1+\delta)\cdot (2^{i+1}-1)]$ as
\begin{eqnarray*}
  \ell_i = t_{i,i} &=& t_{M,0} + \sum_{M \leq i' \leq i, 1\leq j\leq i'} t_{i',j}-t_{i',j-1} \\
  &\le& \frac{\delta}{2} (2^{M'} - 1) +  \sum_{M \leq i' \leq i} i' \cdot (1+\frac{\delta}{2}) 2^{i'}/{i'} \\
  &\le& \frac{\delta}{2} (2^{M'} - 1) + (1 + \frac{\delta}{2})\cdot  ( 2^{i+1}-1) \\
  &\le& (1 + \delta)\cdot  ( 2^{i+1}-1),
\end{eqnarray*}
and similarly for the lower bound: $ \ell_i \geq \sum_{i', j}
t_{i',j}-t_{i',j-1} \geq (1-\frac{\delta}{2}) (2^{i+1} - 2^{M}) \geq
(1 - \delta)\cdot ( 2^{i+1}-1).$ Thus \cref{l-subdivision2} is
applicable on $\ell_i$ with $M$ set to $M'$, and we obtain a sequence
$k_2,k_3,\dots$ such that $\gdens(\sigma_{\ell_{i-1}+1,\dots,k_i}) \leq
 - \frac{\delta}{4}$ and $k_i - \ell_{i-1} \geq \delta
2^{i-2}/K$ for infinitely many $i$.  Pick any of the infinitely many $i
\geq \max\{M', 32(1+\delta)K/\delta\}$ for which $k_i - \ell_{i-1} \geq
\delta 2^{i-2}/K$ and $\gdens(\sigma_{\ell_{i-1}+1,\dots,k_i}) \leq
- \frac{\delta}{4}$. Since $\delta 2^{i-3}/K \geq
(1+\delta)2^i/i$, there is some $j\in\{1,\dots,i\}$ such that
$\ell_{i-1} + \delta/K 2^{i-3} \leq k_i - (1+\delta)2^i/i \leq t_{i,j}
\leq k_i$. Fix such $j$. Since $k_i \leq t_{i,j} + (1+\delta)2^i/i$,
we have
\begin{eqnarray*}
  \gdens(\sigma_{\ell_{i-1}+1,\dots,t_{i,j}}) &=& \frac{ \gdens(\sigma_{\ell_{i-1}+1,\dots,k_i}) (k_i-\ell_{i-1}) }{t_{i,j} + \ell_{i-1}} \\
  &\le& \frac{ \gdens(\sigma_{\ell_{i-1}+1,\dots,k_i}) ( (1+\delta)2^i/i + t_{i,j} -\ell_{i-1}) }{t_{i,j} + \ell_{i-1}} \\
  &\le& - \frac{\delta}{4}  \cdot \left(1+\frac{8(1+\delta)}{i} \right)\\
  &\le& - \frac{\delta}{4} + \frac{4(1+\delta)}{i} \leq - \frac{\delta}{8}.
\end{eqnarray*}
Hence, $\gdens(\sigma_{\ell_{i-1}+1,\dots,t_{i,j}}) \leq -
\frac{\delta}{8}$. So for some $j' \in \{1,\dots,j\}$,
$\gdens(\sigma_{t_{i,j'-1}+1,\dots,t_{i,j'}}) \leq  -
\frac{\delta}{8}$. We can state the following claim.
\begin{claim}
  For $i$ large enough, conditioned on $t(a,b,a^2) = t_{a,b}$, for all
  $a\geq M$ and all $b$, and conditioned on that the game did not stop
  before the time $t_{i,j'-1}+1$, the game stops during times
  $t_{i,j'-1}+1,\dots,t_{i,j'}$ with probability at least $\xi^4\omega /2$.
\end{claim}
Conditioned on $t(a,b,a^2) = t_{a,b}$, for all $a,b$, the claim
implies that the game stops with probability 1. Note that the
condition is true for some valid choice of $t_{a,b}$ with probability
$1-\gamma$.  This is because the claim can be invoked for infinitely
many $i$'s and for each such $i$ we will have $\xi^4\omega /2$ chance
of stopping.

It remains to prove the claim. Assume $t(i,j'-1,i^2) = t_{i,j'-1}$ and
$t(i,j',i^2) = t_{i,j'}$. Clearly, \[\gdens(\sigma_{t_{i,j'-1}+1,
  \dots, t_{i,j'}-1}) \leq \gdens(\sigma_{t_{i,j'-1}+1, \dots,
  t_{i,j'}}) \cdot (2^{i-1}/(2^{i-1}-1)) \leq
-\frac{\delta}{16}\enspace ,\] for $i$ large enough.  So if
we sample $i^2-1$ times from $\sigma_{t_{i,j'-1}+1}, \dots,
\sigma_{t_{i,j'}-1}$ the generalized density is at most $-
\frac{\delta}{16}$ in expectation. 
 By Hoeffding's inequality,
  \cref{THM:Hoeffding} (setting $a_i=-K$ and $b_i=K$, and letting
  $c_i=\payoff(\lef,\sigma_{t(i_n,j_n,i)})$), the probability that the
  generalized density of the subsequence given by the first
  $i^2-1$ samples is more than $-\delta/32$ (i.e. $\delta/32$ greater than the expectation) is bounded by
\[
2\exp\left(-\frac{2(\delta/32(i^2-1))^2}{(i^2-1)(2K)^2}\right)
= 2\exp\left( -\frac{\delta^2}{2048\cdot K^2}(i^2-1)\right) \enspace .
\]
The probability is taken over the possible choices of
$t(i,j',1)<t(i,j',2)<\cdots<t(i,j',i^2-1)$ assuming $t(i,j'-1,i^2) =
t_{i,j'-1}$ and $t(i,j',i^2) = t_{i,j'}$.  For $i$ sufficiently large,
$2\exp\left( -\frac{\delta^2}{2048\cdot K^2}(i^2-1)\right)\leq 1/2$. Also, whenever
\[\dens(\sigma_{t(i,j',1)},\sigma_{t(i,j',2)},\dots,\sigma_{t(i,j',i^2-1)}) \leq  - \frac{\delta}{32}\] we have at least $\xi^4\cdot \omega$ chance of stopping 
by \cref{l-base2}, as Player~1 plays $\tau_1^{i,\eps}$
against \[\sigma_{t(i,j',1)},\sigma_{t(i,j',2)},\dots,\sigma_{t(i,j',i^2-1)}\]
and $\frac{-\delta}{32}\leq -i \cdot 2K/(i^2-1)$ for sufficiently large $i$.

Hence, the game stops with probability at least $(1-1/2) \cdot \xi^4\cdot\omega =
\xi^4\cdot \omega/2$. The claim, and thus the lemma, follows.
\end{proof}

We can now conclude with the main result of this section.
\begin{theorem}\label{thm:inf_eps2}
   The strategy $\sigma^*_1$ is $2K^2\sqrt{\xi}$-optimal, and for all
   $\delta>0$, with probability at least $1-\delta$ does it use space
   $O(\log \log T + \log K)$.
\end{theorem}
\begin{proof}
  This is proved just like \cref{thm:sup_eps2}, except that
  \cref{l-stopping-inf2} is used in place of \cref{l-stopping-sup2}.
\end{proof}

%%%%%%%%%%%%%%%%%%%%%%%%%%%%%%%%%%%%%%%%%%%%%%%%%%%%%%%%%%%%%%%%%%%%%%%%%%%%%%%%

\section{Reduction of repeated games with absorbing states to generalized Big Match games}

As explained in \cref{sec:generalized-big-match}, for defining
strategies for repeated games with absorbing states, Kohlberg reduced
such games in general form to the special case of generalized Big
Match games. The actual terminology, ``generalized Big-Match games'',
is due to Coulomb~\cite{MOR:Coulomb1999}.

Performing the reduction of Kohlberg requires two things. The first
thing is to determine the value of the repeated game. Kohlberg
showed that the value is the same as the limit of the value of the
associated $n$-stage game as $n$ goes to infinity. The other thing is
finding two optimal strategies in an associated parametrized matrix
game with certain closeness properties. Here Kohlberg appealed just to
semi-continuity of the mapping from the parameter to an optimal
strategy of the matrix game. In this section we show how to make these
two ingredients efficient, namely by describing polynomial time
algorithms for them.

Hansen~et.al.~\cite{STOC:HKLMT11} recently showed the existence of a
polynomial time algorithm for computing the value of any undiscounted
stochastic game with a constant number of non-absorbing states. We
present below a much simpler algorithm for the case of repeated games
with absorbing states based on a characterization of the value of
those given by Kohlberg~\cite{AS:Kohlberg1974}. This algorithm is based
only on bisection together with solving linear programs. 

The algorithm of Hansen~et.al. is in fact similar in spirit, based on
bisection and linear programming as well, but is applied to a
discounted version of the game with a discount factor for which no
explicit expression is readily available.

\paragraph{Additional definitions}
The \emph{bit-size} of an integer $n$ is the smallest non-zero integer
$\tau$ such that $\abs{n} < 2^{\tau}$. Thus $\tau = \lfloor \log_2
\abs{n} \rfloor+1$ for non-zero $n$. For a polynomial $p \in \ZZ[x]$,
we denote by $\norm{p}_\infty$ the maximum magnitude of a coefficient
of $p$.

\subsection{Marginal value of matrix games and value of repeated games}
\label{sec:matrix-form-of-games}

A matrix game is given by a $m \times n$ real matrix $A=(a_{ij})$. The
game is played by the two players simultaneously choosing a pure
strategy, where Player~1 chooses action $i$ among the $m$ rows and
Player~2 chooses action $j$ among the $n$ columns. Hereafter Player~1
receives payoff $a_{ij}$. A strategy of a player is a probability
distribution over the player's actions. Let $\Delta^n$ denote the
strategies of Player~1 and $\Delta^m$ denote the strategies of
Player~2. Given $x \in \Delta^m$ and $y \in \Delta^n$, the expected
payoff to Player~1 when Player~1 uses strategy $x$ and Player~2 uses
strategy $y$ is then $x^\transpose Ay$. As shown by
von~Neumann~\cite{MA:vonNeumann28} every matrix game $A$ has a value
$\val(A)$ in mixed strategies, namely
\[
\val(A) = \max_{x \in \Delta^m} \min _{y \in \Delta^n} x^\transpose A
y = \min_{y \in \Delta^n} \max_{x \in \Delta^m} x^\transpose A y
\enspace .
\]
Let $O_1(A) \subseteq \Delta^m$ and $O_2(A) \subseteq \Delta^n$ denote
the set of optimal strategies for Player~1 and Player~2,
respectively. That is, $O_1(A) = \{x \in \Delta^m \mid \forall y \in
\Delta^n : x^\transpose A y \geq \val(A)\}$ and $O_2(A) = \{y \in
\Delta^n \mid \forall x \in \Delta^m : x^\transpose A y \geq
\val(A)\}$.

Let $B$ be another $m \times n$ real matrix. Mills~\cite{AMS:Mills1956}
showed that the limit
\[
\frac{\partial\val(A)}{\partial B} := \lim_{\alpha \rightarrow 0^+}
\frac{\val(A+\alpha B)-\val(A)}{\alpha}
\]
exists and characterized the limit as the value of the game $B$ when
the strategies of Player~1 and Player~2 are restricted to be optimal
in $A$.
\begin{theorem}[Mills]
\begin{equation}
\margval{A}{B} = \max_{x \in O_1(A)} \min_{y \in O_2(A)} x^\transpose B y 
\label{EQ:Mills}
\end{equation}
\label{THM:Mills}
\end{theorem}
The limit $\margval{A}{B}$ is called the \emph{marginal value} of $A$
with respect to $B$. It is not hard to see that
\cref{EQ:Mills} implies that $\margval{A}{B}$ may be
computed using linear programming. Indeed, we may express that
simultaneously $x \in O_1(A)$ and $y \in O_2(A)$ by linear equalities
and inequalities with auxiliary variable $v$ as\footnote{By $f_n$ we
  mean the vector $(1,\dots,1)^\transpose$ of dimension $n$.}:
\[
\begin{split}
f_n v - A^\transpose x & \leq 0 \\
x & \geq 0 \\
f_m^\transpose x & = 1
\end{split}
\qquad
\begin{split}
f_mv - Ay & \geq 0 \\
y & \geq 0 \\
f_n^\transpose y &= 1
\end{split}
\]
Thus for fixed $x$, the quantity $\min_{y \in O_2(A)} x^\transpose B
y$ may be computed by the linear program
\[
\begin{array}{lrcl}
\min & \multicolumn{1}{l}{x^\transpose B y}\\
\textit{s.t.} 
&A^\transpose x' - f_nv & \geq & 0\\
&f_mv - Ay & \geq & 0 \\
&f_m^\transpose x' & = &1\\
&f_n^\transpose y &=& 1\\
&x', y & \geq & 0 \\
\end{array}
\]
with auxiliary variables $x'$ and $v$. Taking the dual we obtain
\[
\begin{array}{lrcl}
\max & \multicolumn{1}{l}{r+s}\\
\textit{s.t.} 
&Ap + f_mr & \leq & 0\\
&f_ns - A^\transpose q & \leq & x^\transpose B \\
&f_m^\transpose q - f_n^\transpose p& = &0\\
&p,q & \geq & 0 \\
\end{array}
\]
in variables $p,q,r,s$, and then by reintroducing $x \in O_1(A)$ as
variables we obtain the following linear program for computing
$\margval{A}{B}$
\[
\begin{array}{lrcl}
\max & \multicolumn{1}{l}{r+s}\\
\textit{s.t.} 
&Ap + f_mr & \leq & 0\\
&f_ns - A^\transpose q - x^\transpose B & \leq & 0 \\
&f_m^\transpose q - f_n^\transpose p & = &0\\
&f_n v - A^\transpose x &\leq& 0\\
&Ay' - f_m v &\leq& 0\\
&f_m^\transpose x &=& 1\\
&f_n^\transpose y' &=& 1\\
&p,q,x,y' & \geq & 0 \\
\end{array}
\]
Appealing to the existence of polynomial time algorithms for linear programming we get:
\begin{corollary}
  The marginal value $\margval{A}{B}$ can be computed in polynomial
  time in the bit-size of $A$ and $B$.
\label{COR:MarginalValue}
\end{corollary}

In this section it will be useful to introduce an alternative notation
for repeated games with absorbing states, their \emph{matrix
  form}. Consider such a game given by action sets $A_1$ and $A_2$,
the stage payoff function $\payoff : A_1 \times A_2 \rightarrow \RR$,
and the absorption probability function $\stopprob : A_1 \times A_2
\rightarrow \RR$. We shall now assume $A_1 = \{1,2,\dots,m\}$ and
$A_2=\{1,2,\dots,n\}$. We then let $b_{ij} = \payoff(i,j)$ and
$\omega_{ij}=\omega(i,j)$. The game will now be identified by a $m
\times n$ matrix $A=(a_{ij})$, populated by the symbolic entries
$a_{ij}$ defined by letting $a_{ij} = \omega_{ij}\absorb{b_{ij}}$ if
$\omega_{ij}>0$ and $a_{ij}=b_{ij}$ if $\omega_{ij}=0$.

Let $A=(a_{ij})$ be a $m \times n$ repeated game with absorbing
states. The notion of the derived matrix game
$\derivedgame{A}=(\widetilde{a}_{ij})$ obtained from $A$ is generalized from
the definition of \cref{sec:generalized-big-match} to be given by
\[
\widetilde{a}_{ij} = \begin{cases} \omega_{ij}b_{ij} & \text{if } a_{ij}=\omega_j\absorb{b_{ij}}\\ b_{ij} & \text{if } a_{ij} = b_{ij}
\end{cases}
\]

Also, given reals $u$ and $t$, we define an associated $m \times n$
matrix game denoted $A(u,t)$, letting entry $(i,j)$ be
\[
\begin{split}
A(u,t)_{ij} 
& = \omega_{ij}b_{ij} + (1-\omega_{ij})(t b_{ij} + (1-t)u)\\
& = \omega_{ij}b_{ij}+(1-\omega_{ij})u + t((1-\omega_{ij})(b_{ij}-u))
\end{split}
\]

Let $A_1$ be the $m \times n$ matrix with entries
$\omega_{ij}b_{ij}+(1-\omega_{ij})u$, and $A_2$ be the $m \times n$
matrix with entries $(1-\omega_{ij})(b_{ij}-u)$. In other words, we
write $A(u,t) = A_1 + t A_2$.  Then the limit 
\[
\margval{A(u,t)}{t^+} := \lim_{t \rightarrow 0^+} \frac{\val(A(u,t)) - \val(A(u,0))}{t} =
\margval{A_1}{A_2}
\]
exists and may be computed in polynomial time by
\cref{THM:Mills} and
\cref{COR:MarginalValue}. Define the extended real number
$\Delta_A(u)$ by the limit
\[
\Delta_A(u) := \lim_{t \rightarrow 0^+}\frac{\val(A(u,t))-u}{t} \enspace .
\]
Clearly, $\Delta_A(u) = \margval{A(u,t)}{t^+}$ when $\val(A(u,0))=u$,
and otherwise $\Delta_A(u)$ is $\infty$ or $-\infty$ depending on
whether $\val(A(u,0))>u$ or $\val(A(u,0))<u$. Kohlberg showed that the
value of $A$ can be characterized by $\Delta_A(u)$.
\begin{theorem}[Kohlberg]
  Let $A$ be a repeated game with absorbing states.  The value of $A$
  is the unique point $u_0$ for which
\[
u < u_0 \Rightarrow \Delta_A(u) > 0
\]
and
\[
u > u_0 \Rightarrow \Delta_A(u) < 0 \enspace .
\]
\label{THM:ValueCharacterization}
\end{theorem}
Using this characterization and bisection together with
\cref{COR:MarginalValue} yields a very simple algorithm for
approximating the value of a repeated game with absorbing states.
\begin{proposition}
  There is an algorithm that given a repeated game with absorbing
  states $A$ and $\eps>0$ computes the value of $A$ to within an
  additive error $\eps$ in polynomial time in the bit-size of $A$
  and $\log(1/\eps)$.
\label{PROP:ApproxValue}
\end{proposition}

\subsection{Parametrized Matrix Games}

The value of a $m \times n$ matrix game $A$ as well as an optimal
strategy for Player~1 may be computed by the following linear program
in variables $(x,v)$.
\begin{equation}
\label{EQ:MatrixLP}
\begin{array}{lrcl}
\max & \multicolumn{1}{l}{v}\\
\textit{s.t.} &f_nv - A^\transpose x & \leq & 0\\
&x & \geq & 0\\
&f_m^T x & =& 1\\
\end{array}
\end{equation}

A \emph{basic solution} to LP~(\ref{EQ:MatrixLP}) is obtained by
selecting $m+1$ constraints indexed by $B$, that includes the equality
constraint. This gives the $(m+1) \times (m+1)$ matrix $M^A_B$,
consisting of the coefficients of these constraints, appropriately
ordered (we shall assume the equality constraint is ordered last). A
basic solution is determined by $B$ if $M^A_B$ is non-singular, and in
that case it is $(x,v)^\transpose = (M^A_B)^{-1} e_{m+1}$\footnote{By
  $e_n$ we mean the standard $n$th unit vector of appropriate
  dimension.}. By Cramer's rule $x_i=\det((M^A_B)_i)/\det(M^A_B)$ and
$v=\det((M_B^A)_{m+1})/\det(M^A_B)$, where $(M_B^A)_i$ is the matrix
obtained from $M_B^{A}$ by replacing column $i$ with $e_{m+1}$. The
basic solution $(x,v)^\transpose$ is a \emph{basic feasible solution}
(bfs) if also $x \geq 0$ and $f_nv - A^\transpose x \leq 0$.

We consider the setting where each entry of $A$ is a linear function
in a variable $t$. Let $A(t)$ denote this matrix game. When $t_0>0$ is
sufficiently small, then if $B$ defines an optimal basic feasible
solution for $A(t_0)$, then $B$ also defines an optimal basic feasible
solution for any $0<t\leq t_0$. We give an explicit bound for this using
the following fundamental root bound.
\begin{lemma}
~\cite[Chapter~6, equations (4) and (5)]{Yap2000} Let $f \in \ZZ[x]$ be a
  non-zero integer polynomial. Then for any non-zero root $\gamma$ of
  $f$ it holds $(2 \norm{f}_{\infty})^{-1} < \abs{\gamma} < 2
  \norm{f}_{\infty}$.
\label{LEM:Sep-bound}
\end{lemma}
Using this we have the following precise statement.
\begin{proposition}
  Let $A(t)$ be a $m \times n$ matrix game parametrized by $t$, where
  each entry is a linear function in $t$ with integer coefficients of
  bit-size at most $\tau$. Let $t_0 =
  (4((m+1)2^{\tau+1})^{2(m+1)})^{-1}$. If $B$ defines an optimal bfs
  for $A(t_0)$ then $B$ also defines an optimal bfs for $A(t)$ for all
  $0<t\leq t_0$.
\label{PROP:Parametrized-Matrix-Game}
\end{proposition}
\begin{proof}
  Let $P^B_i(t) =\det((M^A_B)_i)$ and $Q^B(t)=\det(M^A_B)$. These are
  polynomials of degree at most $(m+1)$ having coefficients of
  magnitude at most $(m+1)!2^{m+1}(2^\tau)^{m+1} \leq
  ((m+1)2^{\tau+1})^{m+1}$.  By \cref{LEM:Sep-bound} we then have
  that $\sgn(Q^B(t))=\sgn(Q(t_0)$ and
  $\sgn(P^B_i(t))=\sgn(P^B_i(t_0))$ for all $0<t\leq t_0$. This means
  that if $B$ defines a bfs for $t_0$, then $B$ also defines a bfs for
  all $0<t<t_0$. To ensure that the bfs defined by $B$ is optimal we
  shall compare it with any other bfs defined by a different set
  $B'$. We then need to ensure that $\frac{P^B_{m+1}(t)}{Q^B(t)}
  \geq \frac{P^{B'}_{m+1}(t)}{Q^{B'}(t)}$. For this we consider the
  polynomial $H(t) = P^B_{m+1}(t)Q^{B'}(t) - P^{B'}_{m+1}Q^B(t)$. Note
  that $H$ is a polynomial of degree at most $2(m+1)$ having
  coefficients of magnitude at most $2((m+1)2^{\tau+1})^{2(m+1)}$.
  Then by \cref{LEM:Sep-bound} again, also
  $\sgn(H(t))=\sgn(H(t_0))$ for all $0<t<t_0$, which means that if $B$
  defines a bfs that is also optimal for $t_0$ the bfs it defines for
  all $0<t\leq t_0$ is optimal as well.
\end{proof}

\subsection{Reduction to generalized Big Match games}

We give here an effective version of the reduction of Kohlberg of
repeated games with absorbing states to the special case of
generalized big match games~\cite[Lemma~2.8 and
Theorem~2.1]{AS:Kohlberg1974}. We additionally make the (rather
simple) extension to repeated games with generalized absorbing
states. 

We shall need the following lemma.
\begin{lemma}
  Let $P$ and $Q$ be integer polynomials such that $\lim_{t\rightarrow
    0^+} \frac{P(t)}{Q(t)}$ exists. Let $\eta = 1/k$ for a positive
  integer $k$ and suppose $\norm{P}_\infty \leq M$ as well as
  $\norm{Q}_\infty \leq M$. Then
\begin{equation}
\Abs{\frac{P(t)}{Q(t)} - \lim_{t\rightarrow 0^+} \frac{P(t)}{Q(t)}} < \eta
\label{EQ:closetolimit}
\end{equation}
whenever $0<t\leq t_0 = (6kM^2)^{-1}$.
\label{LEM:closetolimit}
\end{lemma}
\begin{proof}
  Note that $\lim_{t\rightarrow 0^+} \frac{P(t)}{Q(t)} = \frac{a}{b}$,
  where $b$ is the non-zero coefficient of $Q$ of lowest degree, and
  $a$ is the coefficient of $P$ of the same degree.  Let
  $H_1(t)=k(bP(t)-aQ(t))-bQ(t)$ and
  $H_2(t)=k(aQ(t)-bP(t))-bQ(t)$. Then \cref{EQ:closetolimit}
  holds if and only if $H_1(t)<0$ and $H_2(t)<0$. Noting that
  $\norm{H_1}_\infty \leq 3kM^2$ as well as $\norm{H_2}_\infty \leq
  3kM^2$ the conclusion follows from \cref{LEM:Sep-bound}.
\end{proof}

Let $A'$ be a $m \times n$ repeated game with absorbing states with
stage payoffs $b'_{ij}$ and stopping probabilities $\omega_{ij}$.  For
an optimal strategy $x$ in $A'(0,0)$, define:
\begin{equation}
\begin{split}
\omega_j &= \sum_{i=1}^m x_i\omega_{ij}\\
b_j &= \begin{cases}
\frac{1}{\omega_j}\sum_{i=1}^m x_i\omega_{ij}b'_{ij} & \text{if } \omega_j > 0\\
0 & \text{if } \omega_j=0
\end{cases}\\
e_j &= \begin{cases}\frac{1}{1-\omega_j}\sum_{i=1}^m x_i(1-\omega_{ij})b'_{ij} & \text{if } \omega_j < 1\\
0 & \text{if } \omega_j=1
\end{cases}
\end{split}
\end{equation}
and similarly for an optimal strategy $x(t)$ in $A'(0,t)$ for some given
$t>0$, define:
\begin{equation}
\begin{split}
\omega(t)_j &= \sum_{i=1}^m x(t)_i\omega_{ij}\\
b(t)_j &= \begin{cases}
\frac{1}{\omega(t)_j}\sum_{i=1}^m x(t)_i\omega_{ij}b'_{ij} & \text{if } \omega(t)_j > 0\\
0 & \text{if } \omega(t)_j=0
\end{cases}\\
e(t)_j &= \begin{cases}\frac{1}{1-\omega(t)_j}\sum_{i=1}^m x(t)_i(1-\omega_{ij})b'_{ij} & \text{if } \omega(t)_j < 1\\
0 & \text{if } \omega(t)_j=1
\end{cases}
\end{split}
\end{equation}

Suppose that the $\omega_{ij}$'s are rational numbers with common
denominator $\beta_1$ and the nominators and $\beta_1$ are of bit-size
$\tau_1$. Similarly suppose that the $b'_{ij}$'s are rational numbers
with common denominator $\beta_2$ and the nominators and $\beta_2$ are
of bit-size $\tau_2$.  By definition
$A'(0,t)_{ij}=\omega_{ij}b'_{ij}+t(1-\omega_{ij})b'_{ij}$. Thus the
entries of $A'(0,t)$ are linear functions in $t$ where the
coefficients are rational numbers with common denominator
$\beta=\beta_1\beta_2$ and the bit-sizes of the nominators and
denominators are at most $\tau=\tau_1+\tau_2$. Multiplying each entry
of $A'(0,t)$ by $\beta$ only scales every bfs, so setting $t_0 =
(4((m+1)2^{\tau+1})^{2(m+1)})^{-1}$, whenever $B$ defines an optimal
bfs in $\beta A'(0,t_0)$ it also defines an optimal bfs for $A'(0,t)$ for all
$0<t\leq t_0$, by \cref{PROP:Parametrized-Matrix-Game}. So
let $B$ define an optimal bfs for $A'(0,t_0)$. Let now $P_i(t)
=\det((M^{\beta A'(0,t)}_B)_i)$ and $Q(t)=\det(M^{\beta A'(0,t)}_B)$,
and define $x_i(t)=P_i(t)/Q(t)$. $P_i(t)$ and $Q(t)$ are polynomials
of degree at most $m+1$ having integer coefficients of magnitude at
most $((m+1)2^{\tau+1})^{m+1}$. Furthermore is $x(t)$ an optimal
strategy in $A'(0,t)$ for all $0<t\leq t_0$. Let $x =
\lim_{t\rightarrow 0^+} x(t)$. Then $x$ is an optimal strategy in
$A'(0,0)$. Each coordinate $x_i$ is the ratio between two coefficients
from $P_i(t)$ and $Q(t)$ and is therefore a rational number with
nominator and denominator of magnitude at most
$((m+1)2^{\tau+1})^{m+1}$.

We have always
\[
\omega(t)_j = \sum_{i=1}^m
\frac{P_i(t)}{Q(t)}\omega_{ij} = \frac{\sum_{i=1}^m
  P_i(t)\omega_{ij}}{Q(t)}
\]
and
\[
1-\omega(t)_j = \sum_{i=1}^m x(t)_i(1-\omega_{ij}) = \frac{\sum_{i=1}^m
  P_i(t)(1-\omega_{ij})}{Q(t)}
\]
Consider now the case of
$\omega(t)_j>0$. Then 
\[
b(t)_j=\frac{Q(t)}{\sum_{i=1}^m
  P_i(t)\omega_{ij}}\sum_{i=1}^m\frac{P_i(t)}{Q(t)}\omega_{ij}b'_{ij}
= \frac{\sum_{i=1}^m P_i(t)\omega_{ij}b'_{ij}}{\sum_{i=1}^m
  P_i(t)\omega_{ij}}
\]
Consider now the case of $\omega(t)_j<1$. Then
\[
e(t)_j = \frac{Q(t)}{\sum_{i=1}^m
  P_i(t)(1-\omega_{ij})}\sum_{i=1}^m\frac{P_i(t)}{Q(t)}(1-\omega_{ij})b'_{ij}
= \frac{\sum_{i=1}^m P_i(t)(1-\omega_{ij})b'_{ij}}{\sum_{i=1}^m
  P_i(t)(1-\omega_{ij})}
\]

Since $x = \lim_{t \rightarrow 0^+} x(t)$ we also have $\lim_{t
  \rightarrow 0^+} \omega_j(t) = \omega_j$. Furthermore, in case that
$\omega_j > 0$, we have $\lim_{t \rightarrow 0^+} b_j(t) = b_j$, and
in case $\omega_j=0$, we have $\lim_{t \rightarrow 0^+} e_j(t) = e_j$.

Let $\eta=1/k$ for a positive integer $k$. Define $t_1 =
(2k((m+1)2^{\tau+1})^{m+2})^{-1}$ and
$t_2=(6k((m+1)2^{\tau+1})^{2(m+1)})^{-1}$. Note that $t_2 \leq
\min(t_0,t_1)$.

Suppose that $\omega_j > 0$. Then $\omega(t)_j > 0$ if $\sum_{i=1}^m
P_i(t) \omega_{ij} > 0$. Since $\omega_j>0$ there exists $i$ such that
$P_i(t)$ is a non-zero polynomial and $\omega_{ij}>0$. Then by
\cref{LEM:Sep-bound} it follows that $\omega(t)_j > 0$ for all
$0<t \leq t_0$.  Also from \cref{LEM:closetolimit}, we have
$\abs{b_j-b(t)_j} < \eta$ whenever $0< t \leq t_1$.  Suppose now that
$\omega_j = 0$. Consider the polynomial $H(t) =
\frac{\beta_1}{\eta}\sum_{i=1}^m P_i(t)\omega_{ij} - \beta_1Q(t)$. We
then have $\omega(t)_j < \eta$ if and only if $H(t) < 0$, for $0<t
\leq t_0$. The polynomial $H$ is of degree at most $m+1$ and has
integer coefficients of magnitude at most
$k((m+1)2^{\tau+1})^{m+2}$. By \cref{LEM:Sep-bound} we have
$\omega(t)_j < \eta$ whenever $0<t \leq \min(t_0,t_2)$. Also from
\cref{LEM:closetolimit} we have $\abs{e_j - e(t)_j} < \eta$,
whenever $0<t \leq t_1$.

Putting all these observations together gives us the following
effective version of~\cite[Lemma 2.8]{AS:Kohlberg1974}.
\begin{lemma}
  There is an algorithm that given a $m \times n$ repeated game $A'$
  with absorbing states as above and $\eta = 1/k$ computes in
  polynomial time strategies $x$ and $x(t)$ that are optimal for
  Player~1 in $A(0,0)$ and $A(0,t_1)$, respectively, such that
\begin{enumerate}
\item If $\omega_j > 0$ then $\omega(t_2)_j > 0$ and $\abs{b_j - b(t_2)_j} < \eta$.
\item If $\omega_j = 0$ then $\omega(t_2)_j < \eta$ and $\abs{e_j - e(t_2)_j} < \eta$.
\end{enumerate}
where $t_2=(6k((m+1)2^{\tau+1})^{2(m+1)})^{-1}$.
\label{LEM:Kohlberg2.8}
\end{lemma}
\begin{proof}
  By \cref{PROP:Parametrized-Matrix-Game} if $B$ defines an
  optimal bfs for the matrix game $A'(0,t_2)$, then $B$ also defines an
  optimal bfs for $A'(0,t)$ for all $0<t'<t$. We may find such a $B$
  simply by solving the LP~(\ref{EQ:MatrixLP}) for $A'(0,t_2)$. Let now
  $P_i(t) =\det((M^{A'(0,t)}_B)_i)$ and $Q(t)=\det(M^{A'(0,t)}_B)$ as
  above. We can compute these polynomials, which are of degree at most
  $m+1$ by evaluating them on $m+2$ distinct points from the interval
  $(0,t_2)$ and interpolating. We can furthermore do this evaluation
  using LP~(\ref{EQ:MatrixLP}) since we already have found $B$. This
  further gives us $x_i(t)=P_i(t)/Q(t)$, and $x(t)$ is an optimal
  strategy in the matrix game $A(0,t)$ for all $0<t\leq t_2$. Let $x =
  \lim_{t\rightarrow 0^+} x(t)$ be the optimal limit strategy in
  $A'(0,0)$. As in the proof of \cref{LEM:closetolimit} we may
  compute each coordinate $x_i$ by considering the non-zero
  coefficients of the lowest degree of $P_i(t)$ and $Q(t)$. We
  conclude that $x(t_2)$ and $x$ together satisfy the required
  properties.
\end{proof}

We will now show how to reduce an arbitrary repeated games with
absorbing states to generalized Big Match games. By reduction we mean
that a generalized Big Match game $D$ is computed from a repeated game
with absorbing states $A$, such that a strategy $\sigma_1$ for $D$ can
be extended to a strategy $\tau_1$ for $A$. In case $\sigma_1$ is
$\eps'$ optimal for $D$ then $\tau_1$ is $\eps$-optimal for
$A$, and likewise, in case $\sigma_1$ is $\eps'$-supremum-optimal
for $D$ then $\tau_1$ is $\eps$-supremum-optimal for $A$, where
$\eps'$ depends on $\eps$ and $A$.

\begin{theorem}
  Let $A$ be a $m \times n$ repeated game with absorbing states and let
  $\eps=2^{-\ell}$. Assume the stage payoff $b_{ij}$ are rational
  numbers such that $\abs{b_{ij}} \leq 1$. Assume the stopping
  probabilities $\omega_{ij}$ are rational numbers with common
  denominator $\beta_1$ and the nominator and $\beta_1$ are of bit-size
  at most $\tau_1$. Then $A$ can be reduced in polynomial time to a
  generalized Big Match game satisfying \cref{matrixgameassumption}
  with integer entries of magnitude at most
  $(24(m+2)2^{\ell+\tau_1+1})^{20(m+2)^2 (2n+1)}$.
\end{theorem}
\begin{proof}
We will have 4 sources of error: In approximating the value of $A$, in
rounding the entries of $A$, from the strategy for the generalized Big
Match to which we reduce, and finally from additional strategies of
Player~2 that are not part of this. We shall allow $\eps/4$ to all
these.

First we use the Algorithm of \cref{PROP:ApproxValue} to
compute $u$ such that 
\begin{equation}
u + \frac{\eps}{2} \leq \val(A) < u + \frac{3\eps}{4} \enspace .
\label{EQ:u-valueapprox}
\end{equation}
Using $u$ we translate and round the entries of $A$ to obtain another
repeated game with absorbing states $A'$ with the same stopping
probabilities but with stage payoff $b'_{ij}$ given by
\[
b'_{ij} = \left\lfloor \frac{b_{ij}-u}{\eps/4} \right\rfloor \frac{\eps}{4} \enspace .
\]
Using \cref{EQ:u-valueapprox} we have that $\eps/4 \leq
\val(A') < 3\eps/4$. Also the rounded and translated stage payoffs
$b'_{ij}$ satisfy $-2 \leq b'_{ij} \leq 2$ and are rational numbers
with common denominator $\beta_2=4/\eps=2^{\ell+2}$ and nominators of
bit-size  at most $\tau_2=\ell+3$.

Since $\val(A') > 0$ from \cref{THM:ValueCharacterization} we
have that $\Delta_{A'}(0) > 0$, and this means that $\val(A'(0,t))$
can be bounded below by a linear function in an interval to the right
of 0. We shall make this explicit below, providing constants $\delta$
and $t_1$ such that
\begin{equation}
\label{KEQ:2.16}
\val(A'(0,t)) \geq \delta t
\end{equation}
whenever $0 \leq t \leq t_1$.

So $\Delta_{A'}(0) = \lim_{t \rightarrow 0^+}
\frac{\val(A'(0,t))}{t}>0$. We first fix $\delta>0$ and then determine
a corresponding $t_1$.  If $\val(A'(0,0))>0$ we may choose any
$\delta>0$. If $\val(A'(0,0))=0$ we should choose $\delta$ such that
$\delta < \Delta_{A'}(0)$.

Let $\tau=\tau_1+\tau_2$ and $\beta=\beta_1\beta_2$. Scaling the
entries of $A'(0,t)$ by $\beta$ and setting $t_0 =
(4((m+1)2^{\tau+1})^{2(m+1)})^{-1}$, whenever $B$ defines an optimal
bfs for $\beta A'(0,t_0)$ it also defines an optimal bfs for $A'(0,t)$
for all $0<t\leq t_0$ by \cref{PROP:Parametrized-Matrix-Game}. So let
$B$ define an optimal bfs for $A'(0,t_0)$. Let now $P(t)
=\det((M^{\beta A'(0,t)}_B)_{m+1})$ and $Q(t)=\det(M^{\beta
  A'(0,t)}_B)$. Then when $0<t\leq t_0$ we have $\val(A'(0,t)) =
\frac{P(t)}{\beta Q(t)}$. The polynomials $P$ and $Q$ are of degree at
most $m+1$ and having integer coefficients of magnitude at most
$((m+1)2^{\tau+1})^{m+1}$.

Suppose that $\val(A'(0,0))=0$. Then 
\[
\Delta_{A'}(0) = \margval{A'(0,t)}{t^+} = \lim_{t \rightarrow 0^+} \frac{\frac{d}{dt}P(t)Q(t)-P(t)\frac{d}{dt}Q(t)}{\beta(Q(t))^2} \enspace .
\]
Thus $\Delta_{A'}(0)$ is the ratio between the coefficients of
integers polynomial where the denominator has coefficients of maximum
magnitude $2^\tau((m+1)2^{\tau+1})^{2(m+1)}$. It follows that
\[
\frac{1}{2}\Delta_{A'}(0) \geq (((m+1)2^{\tau+1})^{2m+3})^{-1}
\]
so we let $\delta = (((m+1)2^{\tau+1})^{2m+3})^{-1}$. To determine
$t_1$ we need to ensure that $\val(A'(0,t)) \geq \delta t$. To this
end, define the polynomial $H(t)=P(t)/\delta- \beta t Q(t)$. This is
an integer polynomial of degree at most $m+2$ and $\norm{H}_\infty
\leq 2((m+1)2^{\tau+1})^{3m+4}$. By \cref{LEM:Sep-bound}, letting
$t_1=(4((m+1)2^{\tau+1})^{3m+4})^{-1}$ we obtain the desired
\cref{KEQ:2.16}.

Let $\eta=\delta/4$ and use the algorithm from
\cref{LEM:Kohlberg2.8} to compute strategies $x(t_2)$ and $x$,
where $t_2 = (24((m+1)2^{\tau+1})^{4(m+1)})^{-1}$. Note that $t_2 \leq
t_1$. Also note for later that $\delta \leq 2^{-\tau_2} = \eps/8$.

We may now proceed as in~\cite[Theorem 2.1]{AS:Kohlberg1974}. Player~1
will commit to at every stage playing either the strategy $x$ or the
strategy $x(t_2)$. In this way Player~1 becomes restricted to the $2
\times n$ repeated game with absorbing states $C=(c_{ij})$, where
\begin{equation}
c_{1j} = \begin{cases}\omega_j \absorb{b_j} & \text{if } \omega_j > 0\\
e_j & \text{if } \omega_j = 0
\end{cases}
\end{equation}
and similarly
\begin{equation}
c_{2j} = \begin{cases}\omega(t_2)_j \absorb{b(t_2)_j} & \text{if } \omega(t_2)_j > 0\\
e(t_2)_j & \text{if } \omega(t_2)_j = 0
\end{cases}
\end{equation}
Since $x$ is optimal in $A'(0,0)$ \cref{KEQ:2.16} gives for all $j$,
\begin{equation}
\omega_j b_j = \sum_{i=1}^m x_i \omega_{ij} b'_{ij} \geq 0
\label{KEQ:2.20}
\end{equation}
and similarly since $x(t_2)$ is optimal in $A'(0,t_2)$ \cref{KEQ:2.16} gives for all $j$,
\begin{equation}
\omega(t_2)_jb(t_2)_j+(1-\omega(t_2)_j)t_2e(t_2)_j=\sum_{i=1}^m x(t_2)_i \left( \omega_{ij}b'_{ij} + t_2(1-\omega_{ij}b'_{ij})\right) \geq \delta t_2
\label{KEQ:2.19}
\end{equation}
%Note that when $\omega_j=0$ we have $e_j=\sum_{i=1}^m x_i b_{ij}$.

Let $J = \{j \in \{1,\dots,n\} \mid \omega_j=0 \text{ and }
\omega(t_2)_j>0 \}$, and consider any $j \in J$. Since $\omega_j=0$,
\cref{LEM:Kohlberg2.8} gives $\omega(t_2)_j < \eta =
\delta/4$. Since $\abs{b'_{ij}}\leq 2$, we then get $\omega(t_2)_j
e(t_2)_j \leq \delta/2$, and \cref{KEQ:2.19} gives
\[
\omega(t_2)_jb(t)_j + t_2e(t_2)_j \geq \delta t_2/2 \enspace .
\]

Also from \cref{LEM:Kohlberg2.8} we have $\abs{e_j - e(t_2)_j} <
\eta \leq \delta/2$, which means $e(t_2)_j \geq e_j - \delta/2$, which
in turn means we have
\[
\omega(t_2)_jb(t_2)_j + t_2e_j \geq 0 \enspace .
\]
Let $\derivedgame{C}=(\widetilde{c}_{ij})$ be the derived matrix game
from $C$. For $j \in J$, $\widetilde{c}_{1j}=e_j$ and
$\widetilde{c}_{2j}=\omega(t_2)_jb(t_2)_j$. Thus, dividing by $1+t_2$ we get
\[
\frac{1}{1+t_2} \widetilde{c}_{2j} + \frac{t_2}{1+t_2}\widetilde{c}_{1j} \geq 0 \enspace ,
\]
which means that the value of the matrix game $\derivedgame{C}$
restricted to the columns of $J$ is at least 0. We define a $2 \times
\abs{J}$ repeated game with absorbing states $C'=(c'_{ij})$ by
restricting $C$ to the columns $J$ and subtracting a value from each
entry such the value of the derived matrix game $\derivedgame{C'}$ is
0. More precisely, let $v$ be the value of $\derivedgame{C}$ restricted
to the columns of $J$, and for $j \in J$ we let $c'_{1j} = e_j-v$ and
$c'_{2j} = \omega(t_2)_j \absorb{(b(t_2)-v/\omega(t_2)_j)}$.

Using the expressions previously obtained for $\omega(t)_j$, $b(t)_j$,
and $e(t)_j$, we have that for $0 < t \leq t_0$, each of
$\omega(t)_j$, $b(t)_j$, $e(t)_j$, and $\omega(t)_jb(t)_j$ can be
expressed as rational functions of integer polynomials of degree at
most $m+1$ and integer coefficients of magnitude at most
$((m+1)2^{\tau+1})^{m+2}$. This means that $e_j = \lim_{t \rightarrow
  0^+} e(t)_j$ is a rational number with nominator and denominator of
magnitude at most $((m+1)2^{\tau+1})^{m+2}$ as well. We can bound the
nominator and denominator of the numbers $\omega(t_2)_j$, $b(t_2)_j$,
and $\omega(t_2)_jb(t_2)_j$ by estimating the magnitudes after
substitution of $t_2$ in the corresponding rational functions. This
yield that they are rational numbers with nominator and denominator of
magnitude at most $((m+1)2^{\tau+1})^{m+1}
(24((m+1)2^{\tau+1})^{4(m+1)})^{m+2} \leq
(24(m+1)2^{\tau+1})^{4(m+2)^2}$.  Now the value $v$ is given by the
value of a $2 \times 2$ sub-game of the matrix game $\derivedgame{C}$
restricted to the columns of $J$. This in turn means that $v$ has
nominator and denominator of magnitude at most
$4(24(m+1)2^{\tau+1})^{16(m+2)^2}$.

We can now estimate the entries of
$\derivedgame{C'}=(\widetilde{c'}_{ij})$. These are just the entries from
$\derivedgame{C}$ subtracted $v$. Hence they all have nominator and
denominators of magnitude at most $8(24(m+1)2^{\tau+1})^{20(m+2)^2}
\leq (24(m+2)2^{\tau+1})^{20(m+2)^2}$.

We now scale the entries of $C'$ obtaining another repeated game with
absorbing states $D$ such that the entries of $\derivedgame{D}$ are
integers. We simply do this by multiplying by least common multiple
$M$ of all the denominators of the entries of $\derivedgame{C'}$. Note
that $M \leq (24(m+2)2^{\tau+1})^{40(m+2)^2 n}$, which makes the
entries of $\derivedgame{D}$ integers of magnitude at most
$K=(24(m+2)2^{\tau+1})^{20(m+2)^2 (2n+1)}$.

In case $\derivedgame{D}$ does not have a pure optimal strategy, then
$D$ satisfies \cref{matrixgameassumption}, and we let $\sigma_1$ be a
memory based strategy for Player~1 for $D$ with action map
$\sigma_1^a$ and update map $\sigma_1^u$ that is either
$\eps/(4M)$-optimal or $\eps/(4M)$-supremum optimal. In case that
$\derivedgame{D}$ has a pure optimal strategy we simple take
$\sigma_1$ to be the strategy that plays this pure action always.

From $\sigma_1$ we now construct a strategy $\tau_1$ for $A$. The
action map $\tau_1^a$ will sample an action from $\sigma_1^a$. In case
of a $\lef$ sample, $\tau_1^a$ will sample the final action from $x$
and in case of a $\rig$ sample, $\tau_1^a$ will sample the final
action from $x(t_2)$. The update map will be a simple filtering map
$\tau_1^u$ given as follows. Let $(m,j)$ be a pair of a memory state
$m$ and an action $j$ of Player~2. In case $j \in J$ we let
$\tau_1^u(m,j)=\sigma_1^u(m,j)$. But if $j \notin J$ we let
$\tau_1^u(m,j)$ stay in the memory state $m$, that is we let the next
state be $m$ with probability~1.

Looking at the rounds where $j \in J$, the strategy $\tau_1$ inherits
the performance of $\sigma_1$. Consider now $j \notin J$. Then we have
either (a) $\omega_j=0$ and $\omega(t_2)_j=0$ or (b) $\omega_j >0$. In
case (a) \cref{KEQ:2.19} gives $e(t_2) \geq \delta$ and from
\cref{LEM:Kohlberg2.8} follows $e_j \geq 0$. In case (b)
\cref{KEQ:2.20} gives $b_j \geq 0$. From \cref{LEM:Kohlberg2.8}
follows $\omega(t_2)_j > 0$ as well as $b(t_2)_j \geq -\eta =
-\delta/4 \geq -\eps/32$. Thus in each case the expected stage payoff
is at least $u-\eps/32 \geq \val(A)-\epsilon$.
\end{proof}

Note that $\log K = O(m^2n(\tau+\log m)) = O(m^2n(\log 1/\eps + \tau_1
+ \log m))$, which means that for each $\delta>0$ with probability at
least $1-\delta$ the resulting strategy $\eps$-supremum optimal
strategy will use space $O(f(T)+m^2n(\log 1/\eps + \tau_1 + \log m))$
and the resulting $\eps$-optimal strategy will use space
$O(\log\log T + m^2n(\log 1/\eps + \tau_1 + \log m))$.

\bibliographystyle{abbrv}
\bibliography{BigMatchSmallSpace}

\begin{thebibliography}{10}

\bibitem{AumannSurvey1981}
R.~J. Aumann.
\newblock Survey of repeated games.
\newblock In V.~Bohm, editor, {\em Essays in Game Theory and Mathematical
  Economics in Honor of Oskar Morgenstern}, volume~4 of {\em Gesellschaft,
  Recht, Wirtschaft}, pages 11--42. Bibliographisches Institut, Mannheim, 1981.

\bibitem{AMS:BlackwellFerguson1968}
D.~Blackwell and T.~S. Ferguson.
\newblock The big match.
\newblock {\em The Annals of Mathematical Statistics}, 39(1):159--163, 1968.

\bibitem{MOR:Coulomb1999}
J.~M. Coulomb.
\newblock Generalized ``big-match''.
\newblock {\em Mathematics of Operations Research}, 24(4):795--816, 1999.

\bibitem{BIT:Flajolet85}
P.~Flajolet.
\newblock Approximate counting: {A} detailed analysis.
\newblock {\em BIT}, 25(1):113--134, 1985.

\bibitem{AMS:Gillette1957}
D.~Gillette.
\newblock Stochastic games with zero stop probabilities.
\newblock In {\em Contributions to the Theory of Games III}, volume~39 of {\em
  Ann. Math. Studies}, pages 179--187. Princeton University Press, 1957.

\bibitem{HKM-LICS}
K.~Hansen, M.~Kouck\'y, and P.~Miltersen.
\newblock Winning concurrent reachability games requires doubly exponential
  patience.
\newblock In {\em Proc. of IEEE Symp. on Logic in Comp. Sci., LICS}, pages
  332--341, 2009.

\bibitem{STOC:HKLMT11}
K.~A. Hansen, M.~Kouck{\'y}, N.~Lauritzen, P.~B. Miltersen, and E.~P.
  Tsigaridas.
\newblock Exact algorithms for solving stochastic games.
\newblock In {\em {STOC} 2011}, pages 205--214. ACM, 2011.

\bibitem{JASA:Hoeffding63}
W.~Hoeffding.
\newblock Probability inequalities for sums of bounded random variables.
\newblock {\em Journal of the American Statistical Association},
  58(301):13--30, 1963.

\bibitem{RasmusThesis}
R.~Ibsen-Jensen.
\newblock {\em Strategy complexity of two-player, zero-sum games}.
\newblock PhD thesis, Aarhus University, 2013.

\bibitem{KalaiSurvey1990}
E.~Kalai.
\newblock Bounded rationality and strategic complexity in repeated games.
\newblock In T.~Ichiishi, A.~Neyman, and Y.~Tauman, editors, {\em Game Theory
  and Applications}, pages 131--157. Academic Press, 1990.

\bibitem{AS:Kohlberg1974}
E.~Kohlberg.
\newblock Repeated games with absorbing states.
\newblock {\em The Annals of Statistics}, 2(4):724--738, 1974.

\bibitem{AMS:Mills1956}
H.~D. Mills.
\newblock Marginal values of matrix games and linear programs.
\newblock In H.~W. Kuhn and A.~W. Tucker, editors, {\em Linear Inequalities and
  Related Systems}, volume~38 of {\em Annals of Mathematics Studies}, pages
  183--193. Princeton University Press, 1956.

\bibitem{CACM:Morris78a}
R.~Morris.
\newblock Counting large numbers of events in small registers.
\newblock {\em Communications of the ACM}, 21(10):840--842, 1978.

\bibitem{PNAS:Shapley53}
L.~Shapley.
\newblock Stochastic games.
\newblock {\em Proc. Natl. Acad. Sci. U. S. A.}, 39:1095--1100, 1953.

\bibitem{Sorin-chapter92}
S.~Sorin.
\newblock Repeated games with complete information.
\newblock In {\em Handbook of Game Theory with Economic Applications},
  volume~1, chapter~4, pages 71--107. Elsevier, 1 edition, 1992.

\bibitem{Sorin-FirstCourse2002}
S.~Sorin.
\newblock {\em A First Course on Zero Sum Repeated Games}.
\newblock Springer, 2002.

\bibitem{MA:vonNeumann28}
J.~von Neumann.
\newblock Zur theorie der gesellschaftsspiele.
\newblock {\em Mathematische Annalen}, 100:295--320, 1928.

\bibitem{Yap2000}
C.~K. Yap.
\newblock {\em Fundamental Problems of Algorithmic Algebra}.
\newblock Oxford University Press, New York, 2000.

\end{thebibliography}

\appendix

\section{Tail inequalities}

\begin{theorem}[Multiplicative Chernoff bound] 
  Let $X=\sum_{i=1}^n X_i$ where $X_1,\dots,X_n$ are random variables
  independently distributed in $[0,1]$. Then for any $\eps>0$
\begin{align*}
\Pr[X \geq (1+\eps)\Exp[X]] \leq
\left(\frac{e^\eps}{(1+\eps)^{(1+\eps)}}\right)^{\Exp[X]}\leq
\exp\left(-\frac{\eps^2}{2+\eps} \Exp[X]\right),
\end{align*}
and 
\[
\Pr[X \leq (1-\eps)\Exp[X]] \leq \left(\frac{e^{-\eps}}{(1-\eps)^{(1-\eps)}}\right)^{\Exp[X]} \leq \exp\left(-\frac{\eps^2}{2}\Exp[X]\right) \enspace .
\]
\label{THM:chernoff}
\end{theorem}

Hoeffding~\cite{JASA:Hoeffding63} gave the following bound for
sampling without replacement.
\begin{theorem}[Hoeffding]
  Let a population $C$ consist of $N$ values $c_1,\dots,c_N$, where
  $a_i \leq c_i \leq b_i$.  Let $X_1,\dots,X_n$ denote a random sample
  without replacement from $C$ and $X=\sum_{i=1}^n X_i$. Then
\[
\Pr[\abs{X-\Exp[X]} \geq t] \leq 2\exp\left(-\frac{2t^2}{\sum_{i=1}^n
    (b_i-a_i)^2}\right) \enspace .
\]
\label{THM:Hoeffding}
\end{theorem}

\end{document}